\documentclass[useAMS,usenatbib]{mnras}
\usepackage{graphicx}
\usepackage{longtable}
\usepackage{amssymb}
\usepackage{amsmath}
\title[High-redshift AGN in the \textit{Chandra}  Deep Fields]{High-redshift AGN in the \textit{Chandra} Deep Fields: the obscured fraction and space density of the sub-$L_*$ population}
\author[F. Vito et al.]
{F. Vito$^{1,2}$\thanks{E-mail: fvito@psu.edu},
W.N. Brandt$^{1,2,3}$,
G. Yang$^{1,2}$, 
R. Gilli$^{4}$,
B. Luo$^{5, 6}$,
C. Vignali$^{7,4}$,
Y.Q. Xue$^{8,9}$,
\newauthor
A. Comastri$^{4}$,
A.M. Koekemoer$^{10}$,
B.D. Lehmer$^{11}$,
T. Liu$^{8,9,12}$,
M. Paolillo$^{13,14,15}$,
\newauthor
P. Ranalli$^{16}$,
D.P. Schneider$^{1,2}$,
O. Shemmer$^{17}$,
M. Volonteri$^{18}$,
J. Wang$^{8,9}$
\\ \\
$^{1}$ Department of Astronomy \& Astrophysics, 525 Davey Lab, The Pennsylvania State University, University Park, PA 16802, USA\\
$^{2}$ Institute for Gravitation and the Cosmos, The Pennsylvania State University, University Park, PA 16802, USA\\
$^{3}$ Department of Physics, The Pennsylvania State University, University Park, PA 16802, USA\\
$^{4}$ INAF -- Osservatorio Astronomico di Bologna, via Gobetti 93/3. 40129 Bologna, Italy\\
$^{5}$ School of Astronomy and Space Science, Nanjing University, Nanjing 210093, China\\ 
$^{6}$Key Laboratory of Modern Astronomy and Astrophysics (Nanjing University), Ministry of Education, Nanjing, Jiangsu 210093, 
China\\
$^{7}$ Dipartimento di Fisica e Astronomia, Universit\`a degli Studi di Bologna, via Gobetti 93/2, 40129 Bologna, Italy \\
 $^{8}$ CAS Key Laboratory for Research in Galaxies and Cosmology, Department of Astronomy, University of Science and Technology of China,\\ Hefei, Anhui 230026, China\\
$^{9}$ School of Astronomy and Space Science, University of Science and Technology of China, Hefei 230026, China\\
$^{10}$ Space Telescope Science Institute 3700 San Martin Drive, Baltimore MD 21218, USA \\
$^{11}$ Department of Physics, University of Arkansas, 226 Physics Building, 825 West Dickinson Street, Fayetteville, AR 72701, USA\\
$^{12}$ Astronomy Department, University of Massachusetts, Amherst,MA 01003, USA\\
$^{13}$ Dip.di Fisica Ettore Pancini, University of Naples ``Federico II'', C.U. Monte SantAngelo, Via Cinthia, 80126, Naples, Italy\\
$^{14}$ INFN Sezione di Napoli, Via Cinthia, I-80126 Napoli, Italy\\
$^{15}$ Agenzia Spaziale Italiana - Science Data Center, Via del Politecnico snc, 00133 Roma, Italy\\
$^{16}$Lund Observatory, Department of Astronomy and Theoretical Physics,Lund University, Box 43, SE-22100 Lund, Sweden\\
$^{17}$ Department of Physics, University of North Texas, Denton, TX 76203\\
$^{18}$Institut d'Astrophysique de Paris, Sorbonne Universit\`{e}s, UPMC Univ Paris 6 et CNRS, UMR 7095, 98 bis bd Arago, 75014 Paris, France\\
}

\newcommand*{\chandra}{\textit{Chandra}}

\newcommand*{\chapt}[1]{\S~\ref{#1}}

\newcommand{\angstrom}{\mbox{\normalfont\AA}}

\defcitealias{Hsu14}{H14}
\defcitealias{Skelton14}{S14}
\defcitealias{Straatman16}{S16}
\defcitealias{Yang14}{Y14}
\defcitealias{Xue16}{X16}
\defcitealias{Luo17}{L17}
\begin{document}

\date{}

\graphicspath{{.}}

\pagerange{\pageref{firstpage}--\pageref{lastpage}} \pubyear{2014}

\maketitle

\label{firstpage}

\begin{abstract}
	We investigate the population of high-redshift ($3\leq z < 6$) AGN selected in the two deepest X-ray surveys, the 7 Ms \textit{Chandra} Deep Field-South and 2 Ms \textit{Chandra} Deep Field-North. Their outstanding sensitivity and spectral characterization of faint sources allow us to focus on the sub-$L_*$ regime (log$L_{\mathrm{X}}\lesssim44$), poorly sampled by previous works using shallower data, and the obscured population. 
	Taking fully into account the individual photometric-redshift probability distribution functions, the final sample consists of $\approx102$ X-ray selected AGN at $3\leq z < 6$. 
		The fraction of AGN obscured by column densities log$N_{\mathrm{H}}>23$ is \mbox{$\sim0.6-0.8$}, once incompleteness effects are taken into account, with no strong dependence on redshift or luminosity.
		We derived the high-redshift AGN number counts down to $F_{\mathrm{0.5-2\,keV}}=7\times10^{-18}\,\mathrm{erg\,cm^{-2}\,s^{-1}}$, extending previous results to fainter fluxes, especially at $z>4$.		
We put the tightest constraints to date on the low-luminosity end of AGN luminosity function at high redshift.
	 The space-density, in particular, declines at $z>3$ at all luminosities, with only a marginally steeper slope for low-luminosity AGN.
	 By comparing the evolution of the AGN and galaxy densities, we suggest that such a decline at high luminosities is mainly driven by the underlying galaxy population, while 
	 at low luminosities there are hints of an intrinsic evolution of the parameters driving nuclear activity. Also, the black-hole accretion rate density and star-formation rate density, which are usually found to evolve similarly at $z\lesssim3$, appear to diverge at higher redshifts. 
\end{abstract}


\section{Introduction}
Supermassive black holes (SMBH) and their hosting galaxies are broadly recognized to  influence the evolution of each other over cosmic time. This ``co-evolution" is reflected by the tight relations between the masses of SMBH and the properties of host galaxies in the nearby universe, such as masses and velocity dispersions of the bulges \citep[e.g.][]{Magorrian98,Ferrarese00, Marconi03} and the broadly similar evolution of the star formation and black hole accretion densities in the last $\sim10$~Gyr \citep[e.g.][]{Aird15}, although the details of this interplay are still not well known \citep[see e.g.][and references therein]{Kormendy13}. Studying galaxies and SMBH in the early universe, where these relations could be set in place, would boost our knowledge of how SMBH and galaxies formed and evolved.
However, while galaxy properties have been traced as far back in time as $z\sim8-10$ \citep[e.g.][]{Bouwens15}, our knowledge of SMBH is limited to later times. 

Only $\sim90$ accreting SMBH, shining as active galactic nuclei (AGN), have been identified at $z>6$ \citep[e.g.][]{Banados16, Wang17}, and are usually found to have masses of the order of $1-10$ billion solar masses \citep[e.g.][]{Mortlock11, Wu15}. 
The presence of such massive black holes a few $10^8$ years after the Big Bang challenges our understanding of SMBH formation and growth in the early universe, one of the major issues in modern astrophysics \citep[e.g.][and references therein]{Reines16}. Different classes of theories have been proposed to explain the formation of the BH seeds that eventually became SMBH, the two most promising ones involving ``light seeds" ($M\sim10^2M_\odot$), as remnants of the first Pop III stars, and ``heavy seeds" ($M\sim10^{4-6}M_\odot$), perhaps formed during the direct collapse of giant pristine gas clouds \citep[e.g.][and references therein]{Haiman13,Johnson16,Volonteri16b}. To match the masses of SMBH discovered at $z>6$, all such models require continuous nearly Eddington-limited or even super-Eddington accretion phases during which the growing SMBH is plausibly buried in material with large column densities, even exceeding the Compton-thick level \citep[e.g.][]{Pacucci15}.
However, these objects represent the extreme tail of the underlying distribution (in terms of both mass and luminosity) and are not representative of the overall population. 

X-ray surveys are the most suitable tools for investigating the evolution of the bulk of the AGN population up to high redshift: being less affected by absorption and galaxy dilution, they provide cleaner and more complete AGN identification with respect to optical/IR surveys \citep[and references therein]{Brandt15}. Over the last two decades, several works have focused on the properties and evolution of X-ray selected, $z>3$ AGN in wide \citep[e.g.][]{Brusa09, Civano11,Hiroi12, Marchesi16} and deep \citep[e.g.][]{Vignali02, Fiore12, Vito13, Giallongo15, Weigel15, Cappelluti16} surveys performed with \textit{Chandra} and \textit{XMM}-Newton, or using combinations of different surveys \citep[e.g.][]{Kalfountzou14,Vito14,Georgakakis15}. Common findings among such works are 1) a decline of the space density of luminous ($\mathrm{log}L_{\mathrm{X}}\gtrsim44$) AGN proportional to $(1+z)^d$ with $d\sim-6$ \citep[similar to the exponential decline of the space density of optically selected quasars, e.g.,][]{McGreer13}, and 2) a larger fraction of obscured AGN than that usually derived at lower redshifts, particularly at moderate-to-high luminosities \citep[e.g.][]{Aird15,Buchner15}. 

However, most of the low-luminosity ($\mathrm{log}L_{\mathrm{X}}\lesssim43$) and $z\gtrsim4$ AGN are missed even by the deepest surveys, leading to discrepant results among different studies. For instance, the evolution of the space density of low-luminosity, X-ray detected AGN is largely unconstrained: while \cite{Georgakakis15} reported an apparent strong flattening of the faint end of the AGN X-ray luminosity function (XLF) at $z>3$, \cite{Vito14} found that the decline of the space density of low-luminosity AGN is consistent with that of AGN with higher luminosities. Moreover, \cite{Giallongo15}, using detection techniques which search for clustering of photons in energy, space, and time, reported the detection of several faint AGN, resulting in a very steep XLF faint end \citep[see also][]{Fiore12}. These results also have strong impact on the determination of the AGN contribution to cosmic reionization \citep[e.g.][]{Madau15}. Moreover, the typical obscuration levels in these faint sources remain unknown, although hints of a decrease of the obscured AGN fraction with decreasing luminosity (for $\mathrm{log}L_{\mathrm{X}}\lesssim44$) at high-redshift have been found \citep[e.g.][]{Aird15, Buchner15, Georgakakis15}. This relation is the opposite trend to that found at low redshift \citep[e.g.][]{Aird15,Buchner15}, where the obscured AGN fraction shows a clear anti-correlation with AGN luminosity . Finally, the very detection of faint $z>5$ AGN in deep X-ray surveys is debated \citep[e.g.][]{Vignali02, Giallongo15, Weigel15, Cappelluti16, Parsa17}.

The recently-completed 7 Ms \textit{Chandra} Deep Field-South \citep[CDF-S;][]{Luo17} observations provide the deepest X-ray view of the early universe, reaching a flux limit of $F_{0.5-2\,\mathrm{keV}}=6.4\times10^{-18}\mathrm{erg\, cm^{-2}s^{-1}}$. 
   Moreover, the catalog of X-ray sources in the second deepest X-ray survey to date (limiting flux $F_{0.5-2\,\mathrm{keV}}=1.2\times10^{-17}\mathrm{erg\, cm^{-2}s^{-1}}$), the 2 Ms \textit{Chandra} Deep Field-North \citep[CDF-N;][]{Alexander03}, was recently re-analyzed by \cite{Xue16} with the same detection procedure applied to the CDF-S, which provides detections for more real sources. Therefore, the two deepest \textit{Chandra} fields allow us now to study high-redshift, faint AGN using homogeneous datasets.
In \cite{Vito16}, we applied a stacking technique to CANDELS \citep{Grogin11,Koekemoer11} selected galaxies to study the X-ray emission from individually-undetected sources in the 7 Ms CDF-S, finding that the emission is probably mostly due to X-ray binaries (XRB) rather than nuclear accretion, and concluding that most of the SMBH growth at $3.5<z<6.5$ occurred during bright AGN phases. In this paper, we combine the 7 Ms CDF-S and 2 Ms \mbox{CDF-N} data to study
 the X-ray properties of X-ray detected AGN at $z>3$, with a particular focus on low-luminosity sources (log$L_{\mathrm{X}}\lesssim44$), which are best sampled by deep, pencil-beam surveys. Taking fully into account the probability distribution functions (PDF) of the photometric redshifts for sources lacking spectroscopic identifications (see e.g. \citealt{Marchesi16} for a similar use of the photometric redshifts), the final sample consists of $\approx102$  X-ray detected AGN at $3\leq z<6$. The number of sources contributing to this sample with their PDF($z$) is 118. We performed a spectral analysis on our sample assuming the X-ray spectra are well represented by power-law emission subjected to Galactic and intrinsic absorption. The spectral analysis allowed us to take into account the full probability distribution of the intrinsic column densities. We also considered the probability distribution of the count rates of X-ray detected sources and applied a correction to mitigate the Eddington bias. The flux (and hence luminosity) probability distributions were derived by applying for each source the proper response matrices and the conversion factors between count-rate and flux, which depend on the observed spectral shape. We present the trends of the obscured AGN fraction with redshift and luminosity, the number counts, and the space density evolution of $3<z<6$ AGN. Throughout this paper we will use a $H_0=70\,\rmn{km\,s^{-1}Mpc^{-1}}$, $\Omega_m=0.3$, and $\Omega_\Lambda=0.7$ cosmology and we will assume Galactic column densities of $\mathrm{log}N_{\mathrm{H}}=0.9\times10^{20}$ and $1.6\times10^{20}\mathrm{cm^{-2}}$ along the line of sight of CDF-S and CDF-N, respectively. Errors and upper limits are quoted at the 68\% confidence level, unless otherwise noted.

\section{The sample}\label{sample_section}
\subsection{AGN parent sample and redshifts  in the 7 Ms CDF-S}\label{sample}
\begin{figure} 
	\centering
	\includegraphics[width=80mm,keepaspectratio]{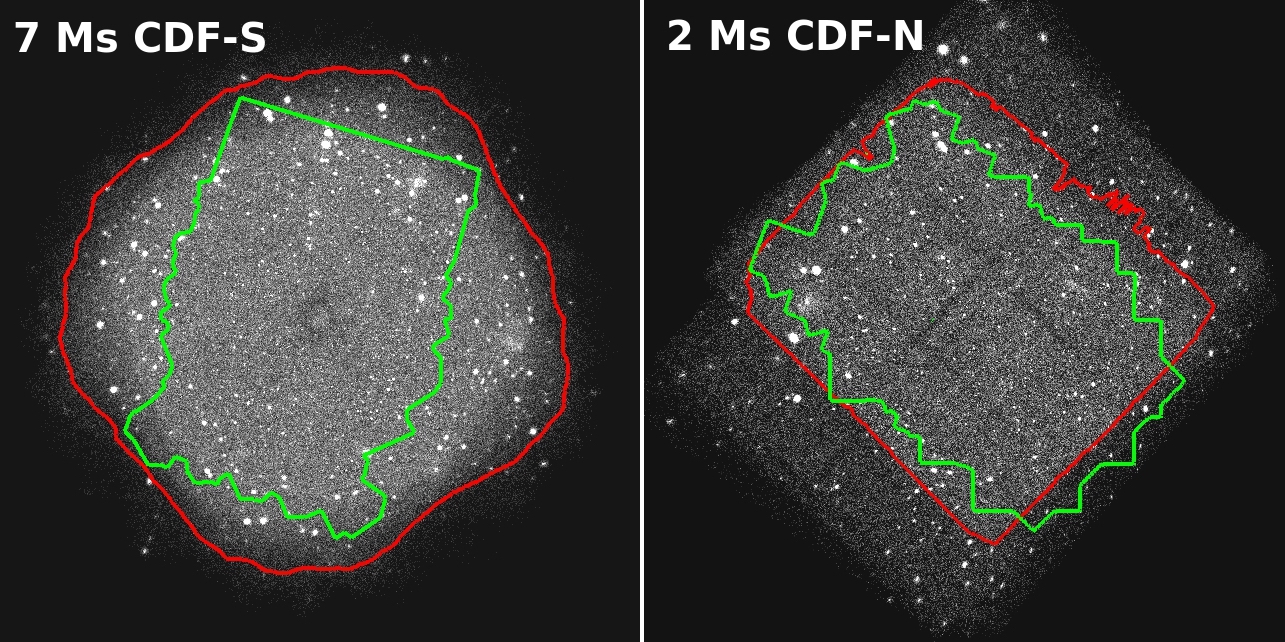}
	\caption{$0.45^\circ\times 0.45^\circ$ images of the 7 Ms CDF-S (left panel) and 2 Ms CDF-N (right panel) in the $0.5-2$ keV band. Red regions encompass the areas with effective exposure $>1$ Ms which are used in this work. Green polygons denote the CANDELS surveys in these fields.}
	\label{fields}
\end{figure}

We selected a sample of X-ray detected, $z>3$ AGN in the 7 Ms CDF-S\footnote{The integrated X-ray emission from high-mass and low-mass XRB in a galaxy can reach luminosities of log$L_{\mathrm{X}}\approx42$ \citep[i.e.][]{Lehmer16}.
	At $z>3$, the flux limit of the 7 Ms CDF-S corresponds to log$L_{\mathrm{X}}\gtrsim42$. Therefore we will consider all of the X-ray sources at $z>3$ to be AGN and discuss the possible level of contamination from XRB in \chapt{xlf_sect}.}, the deepest X-ray survey to date, from the \citet[hereafter \citetalias{Luo17}]{Luo17} catalog, which also provides multiwavelength identifications and spectroscopic and photometric redshifts for the X-ray sources. In particular, photometric redshifts were collected from \citet{Luo10}, \citet{Rafferty11}, \citet[hereafter \citetalias{Hsu14}]{Hsu14}, \citet[hereafter \citetalias{Skelton14}]{Skelton14}, \citet{Santini15}, and \citet[hereafter \citetalias{Straatman16}]{Straatman16}. %
Each X-ray source can therefore be associated with up to six different photometric redshifts. 
We considered only sources located in the area ($\sim330\,\mathrm{arcmin^2}$, red region in the left panel of Fig.~\ref{fields}) of the survey where the effective exposure is $\geq 1$ Ms, in order to exclude the outskirts of the field, where the PSF distortions and the effects of the vignetting affect the quality of the X-ray data and the optical identification rate and accuracy. Moreover, the inner region of the CDF-S is covered by the deepest optical/IR observations (green region), which are essential to derive highly-reliable spectroscopic and photometric redshifts. With this selection our parent sample in the CDF-S consists of 952 out of the 1008 X-ray sources in \citetalias{Luo17}.

We adopted the \citetalias{Luo17} definitions for the spectroscopic redshift quality.
Following \citetalias{Luo17}, we associated with each \mbox{X-ray} source a spectroscopic redshift if it is defined as ``secure" or ``insecure" but in agreement within \mbox{$\frac{|z_{\mathrm{phot}}-z_{\mathrm{spec}}|}{1+z_{\mathrm{spec}}}<15\%$} with at least one photometric redshift. Using more conservative criteria, such as requiring that the spectroscopic redshift agrees with at least 2 or $>50\%$ of the available photometric redshifts, would have no effect on the final sample of $z>3$ AGN. 
 The photometric redshifts used to validate the ``insecure" spectroscopic redshifts at $z>3$ are of good quality, with a $68\%$ confidence level $\Delta z=0.03-0.2$.

If the requirements for using the spectroscopic redshift are not satisfied, or if a source lacks spectroscopic identification, we assumed a photometric redshift from among those available. 
 \citetalias{Hsu14}, \citetalias{Skelton14} and \citetalias{Straatman16} provide the probability distribution functions (PDF) of their photometric redshifts. 
 We define a priority order among these catalogs by estimating the accuracy of the photometric redshifts
 \begin{equation}
  \frac{|\Delta z|}{1+z} = \frac{|z_{\mathrm{phot}}-z_{\mathrm{spec}}|}{1+z_{\mathrm{spec}}},
 \end{equation}
 where $z_{\mathrm{phot}}$ is the peak of the photometric-redshift PDF of each X-ray source in the \citetalias{Luo17} catalog with ``secure" spectroscopic redshift, and the normalized median absolute deviation, defined as 
\begin{equation}
 \sigma_{\mathrm{NMAD}}=1.48\times\mathrm{Med}(\frac{|\Delta z - \mathrm{Med(\Delta z)}|}{1+z_{\mathrm{spec}}}).
\end{equation}
We found median values of \mbox{$|\Delta z|/ (1+z_{\mathrm{spec}})$} of 0.009, 0.009, and 0.007, and $\sigma_{\mathrm{NMAD}}=0.010$, 0.011, and 0.008 using the photometric redshifts from \citetalias{Hsu14}, \citetalias{Skelton14}, and \citetalias{Straatman16}, respectively. A similar assessment of the photometric redshift accuracy for the sample of high-redshift sources is presented in \chapt{sample_tot}.

We also estimated the accuracy of the confidence intervals provided by the PDFs by computing the fraction of sources whose spectroscopic redshift is included in the $68\%$ confidence interval provided by its PDF (defined as the narrowest redshift interval where the integrated redshift probability is 0.68). If the PDFs provided accurate confidence intervals, that fraction would be 0.68, while we found $0.49$, $0.50$ and $0.63$ for \citetalias{Hsu14}, \citetalias{Skelton14}, and \citetalias{Straatman16}, respectively, reflecting a general mild underestimate of the confidence intervals, hence of the photometric-redshift errors. This effect could be due to underestimating the errors of the fitted photometric data (e.g., see \S~5.3 in \citealt{Yang14}). We found indeed that the most accurate confidence intervals are provided by \citetalias{Straatman16}, who addressed in detail this issue by employing an empirical technique to derive more accurate photometric errors than those usually provided by detection software such 
as SExtractor. The reported fractions refer to the particular comparative spectroscopic sample, i.e., X-ray selected galaxies, and are expected to be different considering the entire galaxy samples in those works.
The PDFs are usually derived by fitting the observed spectral energy distribution (SED) with models $M$ of galactic emission varying the redshift as $\mathrm{PDF}(z)\propto\mathrm{\mathrm{exp}}(-0.5\chi^2(z))$, where $\chi^2(z)=\sum_i\frac{(M_i(z)-SED_i)^2}{\sigma_i^2}$ is the test statistic of the fit, and the index $i$ represents the different photometric bands. If the photometric errors $\sigma_i$ are underestimated, the resulting PDFs will be too sharp and their confidence intervals will be underestimated as well. In this case, more accurate confidence intervals can be obtained by multiplying the photometric errors by a factor $\alpha$, which represents the average underestimating factor of the photometric errors among the used bands, or, equivalently, by using the ``corrected" distribution $\mathrm{PDF^{corr}}(z)=\mathrm{PDF^{input}}(z)^{\frac{1}{\alpha^2}}$, where $\alpha^2$ is computed empirically such that the $68\%$ considered interval provided by their PDFs encompasses the associated spectroscopic redshift in $68\%$ of the sample. This procedure is equivalent to empirically ``correcting'' (i.e., increasing) the  photometric errors following $\sigma^\mathrm{corr}=\alpha\times\sigma^\mathrm{input}$.  We obtained $\alpha^2=5.2$, 4.4, and 1.5 for \citetalias{Hsu14}, \citetalias{Skelton14}, and \citetalias{Straatman16}, respectively, and use the ``corrected" PDFs hereafter.

All of these tests led us to adopt the photometric redshift from \citetalias{Straatman16} as first choice. The photometric redshifts from \citetalias{Hsu14} and \citetalias{Skelton14} have similar accuracy, but \citetalias{Hsu14} provide the 
redshifts for the entire Extended CDF-S (E-CDF-S), while \citetalias{Skelton14} (as well as \citetalias{Straatman16}) is limited to the GOODS-S/CANDELS region. We therefore used the photometric redshifts from \citetalias{Hsu14} and \citetalias{Skelton14} as second and third choice, respectively.\footnote{With two exceptions, relevant to the purpose of this work: 1) XID 638 has no spectroscopic redshift, a photometric redshift from \citetalias{Straatman16} $z\sim3.10$ and a photometric redshift from \citetalias{Hsu14} \mbox{$z\sim2.64$}. A significant iron line is detected in the X-ray spectrum (see also \citealt{Liu17}). If it is produced by neutral iron at 6.4 keV, which is usually the strongest line, the observed line energy is consistent with the \citetalias{Hsu14} redshift. Even in the case of completely ionized iron, the line is consistent with a redshift $z<3$ at 90\% confidence level. 2) XID 341 has a photometric redshift $z=5.05$ in \citetalias{Straatman16}, which is inconsistent with the visible emission in the GOODS-S $B$ and $V$ bands. Therefore, for these two sources we adopted the \citetalias{Hsu14} redshifts instead of the \citetalias{Straatman16} solutions. } Among the 21 remaining sources with no entries in the above-considered photometric-redshift catalogs, two sources have a photometric redshift from \cite{Rafferty11}. We approximated their $\mathrm{PDF}(z)$, not provided by that work, as normalized Gaussian functions centered on the nominal redshift and with $\sigma$ equal to the $1\sigma$ error, separately for the positive and negative sides. Given the very low number of sources for which we adopted the \cite{Rafferty11} redshifts (two, and only one will be included in the final sample), the approximation we used to describe their PDF($z$) does not have any effect on the final results.

Nineteen sources out of the 952 X-ray sources in the parent sample ($1.9\%$) remain with no redshift information. Most of them (14/19) are not associated with any counterpart from catalogs at different wavelengths. 
The 5 sources with an optical counterpart but no redshift information have been assigned a flat PDF. We preferred not to use redshift priors based on optical magnitudes, as these sources are part of a very peculiar class of objects (i.e. extremely X-ray faint sources), whose redshift distribution is not well represented by any magnitude-based distribution derived for different classes of galaxies. We did not include the 14 \mbox{X-ray} sources with no optical counterparts in this work, as a significant number of them are expected to be spurious detections.
In fact, \citetalias{Luo17} estimated the number of spurious detections in the entire 7 Ms CDF-S main catalog to be $\sim19$ and argued that, given the superb multiwavelength coverage of the CDF-S and sharp \chandra\, PSF, resulting in high-confidence multiwavelength identifications, X-ray sources with multiwavelength counterparts are extremely likely to be real detections. Therefore, most of the spurious detections are expected to be accounted for by X-ray sources with no optical counterpart. This conclusion is especially true considering that most of the unmatched sources lie in the CANDELS/GOODS-S field, where the extremely deep multiwavelength observations would likely have detected their optical counterpart if they were true sources.  We also checked their binomial no-source probability ($P_B$, i.e. the probability that the detected counts of a source are due to a background fluctuation), provided by \citetalias{Luo17} for all the sources in their catalog. Thirteen out of the 14 sources with no counterparts have $P_B\gtrsim10^{-4}$, close to the detection threshold of $P_B=0.007$ (\citetalias{Luo17}). For comparison, $\sim80\%$ of the sources with optical counterparts have $P_B<10^{-4}$. The notable exception is XID 912, which has no optical counterpart, but $P_B\sim10^{-13}$ and $\approx77$ net counts. \citetalias{Luo17} suggested that XID 912 is a off-nuclear X-ray source associated with a nearby low-redshift galaxy, detected in optical/IR observations.

We fixed the PDFs of sources with spectroscopic redshift to zero everywhere but at the spectroscopic redshift where $\rmn{PDF(z_{\mathrm{spec}})=1}$. All the PDFs are normalized such as
 
\begin{equation}
 \int^{10}_{0} PDF(z)dz = 1
\end{equation}
where the redshift range was chosen to be the same as in \citetalias{Straatman16}. 
The fraction of the PDF for the $i$-th source at $z_1\leq z< z_2$, i.e., the probability the $i$-th source is at  $z_1\leq z< z_2$ is 
\begin{equation}\label{zfrac}
	P^i(z_1\leq z< z_2)= \int^{z_2}_{z_1}PDF^i(z)\, dz.
\end{equation} 
Tab.~\ref{nz} summarizes the number of X-ray sources in the \mbox{CDF-S} which are associated with spectroscopic, photometric, or no redshifts.

\subsection{AGN parent sample and redshifts in the 2 Ms CDF-N}\label{sample_N}

\citet[hereafter \citetalias{Xue16}]{Xue16} presented an updated version (683 sources in total) of the original 2 Ms CDF-N catalog (\citealt{Alexander03}), applying the same detection procedure used in the CDF-S \citep[\citetalias{Luo17}]{Xue11} and providing multiwavelength identifications and spectroscopic and photometric redshifts from the literature. In particular, photometric redshifts were collected from \citetalias{Skelton14} and \citet[hereafter \citetalias{Yang14}]{Yang14}. Both of these works provide the PDF of the photometric redshifts. We adopted the spectroscopic redshifts collected from \citetalias{Xue16}, as they considered only those marked as ``secure" in the original works.

For X-ray sources lacking spectroscopic redshift, we followed the procedure described in \chapt{sample} to define a priority order among the two used photometric catalogs. In particular, we derived median values of $|\Delta z |/ (1+z_{\mathrm{spec}})=0.026$ both for \citetalias{Skelton14} and \citetalias{Yang14}, $\sigma_{\mathrm{NMAD}}=0.025$ and 0.035  and $\alpha^2=9.2$ and 3.0 for \citetalias{Skelton14} and \citetalias{Yang14}, respectively. The $\alpha^2$ values mean that the photometric redshift PDFs from \citetalias{Yang14} account for the redshift uncertainties more realistically than the \citetalias{Skelton14} ones: this behavior can be explained again by considering that \citetalias{Yang14} applied an empirical method to estimate the photometric errors more reliably than those provided by standard detection software.
We therefore adopted the photometric redshifts from \citetalias{Yang14} and \citetalias{Skelton14} as first and second choice, respectively.

As in \chapt{sample}, we considered only the sources in the area covered by $\geq 1$ Ms effective exposure ($\sim215\,\mathrm{arcmin^2}$; red contour in the right panel of Fig.~\ref{fields}, almost coincident with the CANDELS survey in that field) as the parent sample (485 sources).
Among them, 35 sources ($\sim7\%$) have no redshift information.
This non-identification rate is mostly due to the method used in \citetalias{Xue16} 
to match the X-ray sources with the entries in the photometric-redshift catalogs. First, for each X-ray source they identified a primary multiwavelength counterpart using a likelihood-ratio procedure \citep{Luo10}. They then matched the coordinates of the primary counterparts with the photometric-redshift catalogs using a $0.5''$ radius. However, the positional uncertainties and offsets among the different photometric catalogs can be comparable to the utilized matching radius. We therefore visually inspected the positions of all the X-ray sources on the CANDELS/GOODS-N images\footnote{ https://archive.stsci.edu/prepds/candels/} \citep{Grogin11, Koekemoer11} and could match most of the X-ray sources  with no redshift information in \citetalias{Xue16} with an entry in one of the considered photometric-redshift catalogs. 
The visual multiwavelength matching allowed us also to associate a spectroscopic redshift from \citetalias{Yang14} (which included only high-quality spectroscopic redshifts) with several sources with only photometric or no redshifts in \citetalias{Xue16}.

The resulting number of X-ray sources with no redshift information is 12 ($\sim2.5\%$).
 As in \chapt{sources}, we excluded sources with no multiwavelength counterpart (8 sources) from this analysis, as most of them are expected to be spurious. Two out of the remaining 4 sources with counterparts but no redshifts in the considered catalogs have photometric-redshift entries in the CANDELS/GOODS-N catalog (Kodra D. et al. in preparation). Their PDF($z$) lie entirely at $z<3$, thus we excluded these sources from the high-redshift sample.
  Finally, we assigned flat $\mathrm{PDF}(z)$ over the range $z=0-10$ to the only two remaining sources with multiwavelength counterparts but no redshifts.
 Tab.~\ref{nz} summarizes the number of X-ray sources in the CDF-N which are associated with spectroscopic, photometric, or no redshifts.

\subsection{The sample of high-redshift  AGN in the Chandra Deep Fields}\label{sample_tot}
\begin{figure} 
	\centering
	\includegraphics[width=80mm,keepaspectratio]{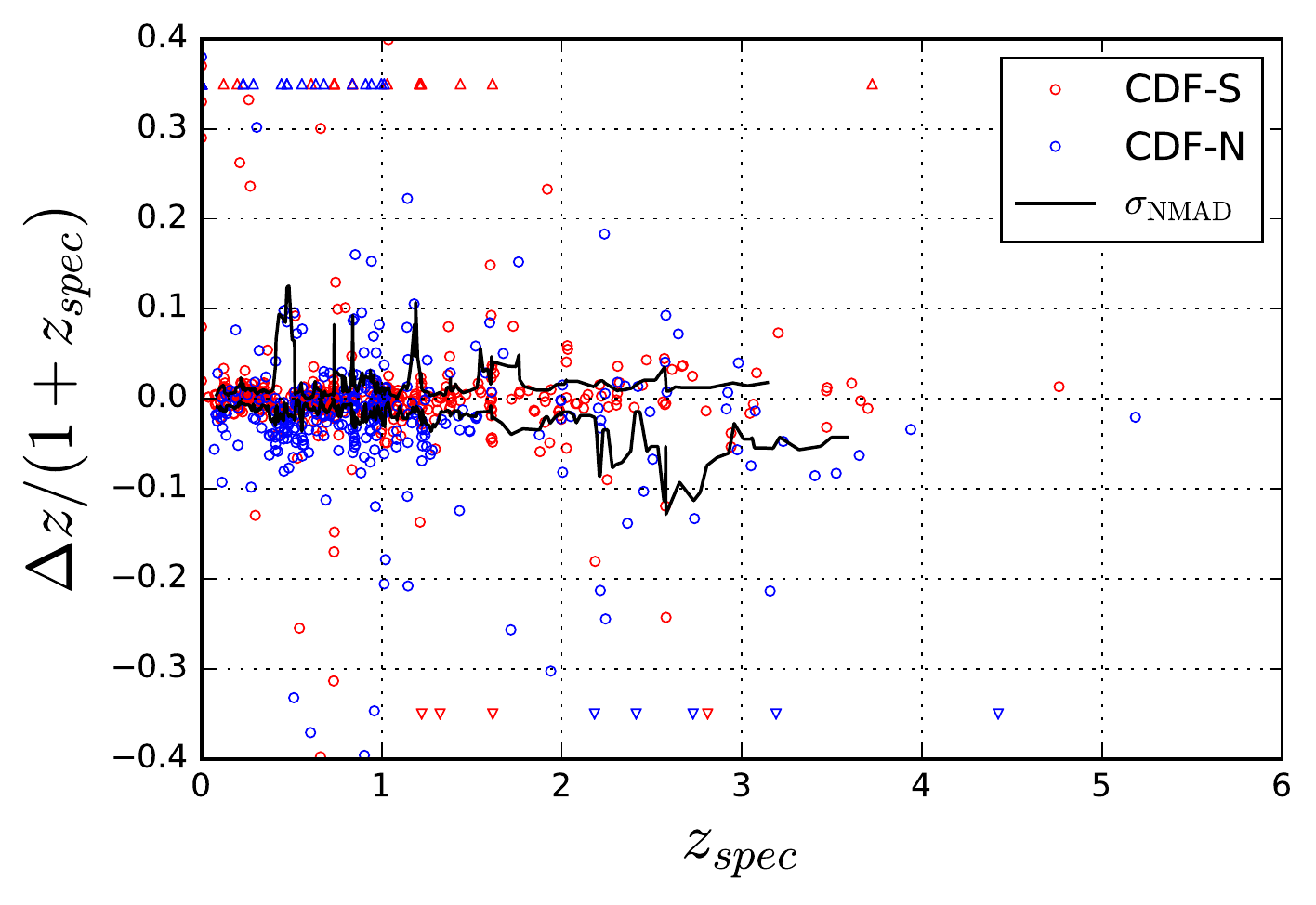}
	\caption{Photometric redshift accuracy for sources with secure spectroscopic redshift. The lines are the normalized median absolute deviations ($\sigma_{\mathrm{NMAD}}$) computed in a shifting redshift interval with variable width such to include 10 sources, separately for the positive and negative sides. Points at $\Delta z /(1+z_{\mathrm{spec}}) > 0.4$ ($<-0.4$) are plotted as upward-pointing (downward-pointing) triangles at $\Delta z /(1+z_{\mathrm{spec}})=0.35$ ($-0.35$).}
	\label{z_check}
\end{figure}

We checked the photometric redshift accuracy by plotting in Fig.~\ref{z_check} the $\Delta z / (1+z_{\mathrm{spec}})$ and the $\sigma_{\mathrm{NMAD}}$ for sources with secure spectroscopic redshifts (see also Tab.~\ref{nz}). The $\sigma_{\mathrm{NMAD}}$ is computed in a shifting interval of redshift with variable width such to include 10 sources (separately for the positive and negative sides). The photometric redshift for each AGN is chosen following the priority order described in \chapt{sample} and \chapt{sample_N}. We considered only sources within the area covered by $\geq1$ Ms exposure. The scatter increases slightly at \mbox{$z\sim2.5$} (by a factor of $\sim2-3$), but the photometric redshift accuracy does not appear to deteriorate dramatically at high redshift. 

\begin{figure} 
	\centering
	\includegraphics[width=80mm,keepaspectratio]{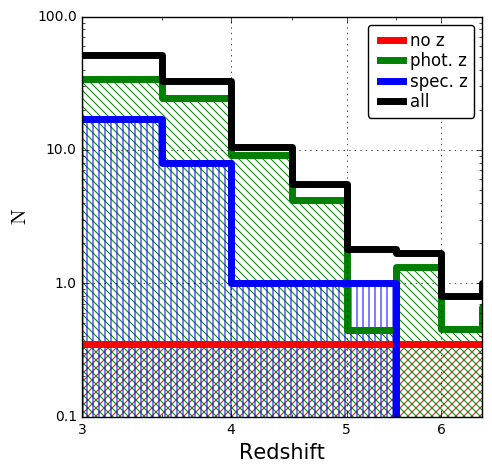}
	\caption{Redshift distribution of the X-ray sources in the \mbox{CDF-S} and \mbox{CDF-N} at $z>3$. Different classes of sources are represented by lines of different colors.}
	\label{z_distro}
\end{figure}

Fig.~\ref{z_distro} presents the redshift distributions of the sources at $z>3$ in the two deep fields, considering their PDF($z$). 
At $z>6$ the source statistics are poor and sources with no redshift information (red line), which carry little information, constitute a significant fraction of the total number of sources. We therefore chose to limit our investigation to the redshift range $3\leq z<6$. 
To prevent the inclusion of several sources whose PDFs show long tails of extremely low probability at high redshift, 
we also required that the probability of a source to be in the considered redshift range (computed as in Eq.~\ref{zfrac}) is $P(3\leq z<6)>0.2$.
The number of sources satisfying this criterion is $N_{\mathrm{hz}}=118$. Integrating their PDFs in the redshift range $3\leq z<6$, the number of ``effective" sources included in the sample is $N^{\mathrm{eff}}_{\mathrm{hz}}\simeq101.6$. The cut $P(3\leq z<6)>0.2$ results in the rejection of $\sim 2$  effective sources.  
 Tab.~\ref{nz} reports the breakdown of the redshift type for sources at $3\leq z <6$ .  Tab.~\ref{sources} provides the redshift along with the 68\% confidence interval, its origin and the probability of the source to be at $3\leq z <6$ according to its PDF for each source in the high-redshift sample. 
 
 In principle, Eq.~\ref{zfrac} measures only the uncertainty of the photometric redshift. The probability for a source to be at  $z_1\leq z< z_2$ can be computed by weighting Eq.~\ref{zfrac} with the X-ray luminosity function (XLF), in order to take into account the intrinsic variation of the AGN space density with redshift, that can introduce systematic errors  \citep[e.g.][]{Georgakakis15}. However, the generally high quality of the photometric redshifts we used (see \S~\ref{sample}, \ref{sample_N}, and Tab.\ref{nz}), due to the availability of deep multi-band data in the fields we considered, is reflected in narrow PDF$(z)$. In fact, the PDF$(z)$ of the spectroscopically unidentified sources included in our sample have a median $1\sigma$ uncertainty of $0.23$. Such narrow PDF$(z)$ would be at most sensitive to the ``local" XLF. For continuity reasons, the XLF cannot vary dramatically in a narrow redshift range and the applied weights are therefore expected to be negligible. This statement may not be true in cases of broader PDF$(z)$, which would be therefore sensitive to the space density variations over larger redshift intervals. 
 	
 	To quantify this effect in our case, we repeated the procedure followed to derive the redshift distribution of the sample, after weighting the PDF$(z)$ with the \cite{Ueda14} XLF. We also considered additional cases in which the \cite{Ueda14} XLF is modified at high redshift to match the results of \cite{Vito14} and \cite{Georgakakis15}. The new redshift distributions are barely distinguishable from Fig.~\ref{z_distro}, with $1.7-5.9\%$ (depending on the particular XLF used) of the sample shifting below $z=3$. Including in this computation sources with no redshift information (i.e., for which a flat PDF($z$) is assumed), we found that $2.5-7.7\%$ of the samples shift to low redshift. Moreover, the weighting is sensitive in particular to the assumed shape and evolution of the XLF faint end at high redshift, which is currently highly uncertain. For instance, using models with a steeper XLF faint end \citep[e.g.][]{Giallongo15} would move a fractional number ($<1$) of sources from $z\lesssim3$ to $z>3$, slightly increasing our sample size. Given the small effect obtained applying the XLF weighting and the source of additional uncertainty represented by the particular XLF assumed at high redshift, we will use Eq.~\ref{zfrac} to measure the intrinsic redshift probability distribution. 
 
 We also made a basic assessment of the effects of the expected presence of redshift outliers in our high-redshift sample (i.e., intrinsically low-redshift sources scattered to high redshift because of the catastrophic failure of their photometric redshifts, and hence PDF($z$)). We considered only X-ray sources with a high-quality spectroscopic redshift $z_{spec}<3$, assigned to them a photometric redshift following the priority order described in \S~\ref{sample} and \S~\ref{sample_N}, and computed the fraction of redshift outliers, defined as sources for which $|z_{spec}-z_{phot}|/(1+z_{spec})>0.15$ \citep[see, e.g.,][]{Luo10}, that would be counted in the high-redshift sample according to their photometric redshift (i.e., PDF($z$)). Considering this fraction and the number of X-ray selected, spectroscopically unidentified sources with photometric redshift $z_{phot}<3$, we estimated that $\approx4$ sources in our high-redshift sample ($\approx4\%$) are such low-redshift interlopers. The outlier fraction may be higher for spectroscopically unidentified sources, since these are typically fainter than the control sample we used, but our results are not qualitatively affected even considering an outlier fraction of $\approx10\%$.

Our sample includes a few sources studied in detail in previous, dedicated works, like the two Compton-thick candidate AGN in the CDF-S at $z=4.762$ \citep[ID 539 in Tab.~\ref{sources}]{Gilli14} and $z=3.70$ \citep[ID 551]{Comastri11}. We also include the three spectroscopically-identified $z>4$ AGN in the CDF-N analyzed by \citet{Vignali02}: ID 290 at $z=4.424$, ID 330 at $z=5.186$, which is also the X-ray source with the highest spectroscopic-redshift in the \textit{Chandra} deep fields, and ID 526 at $z=4.137$.\footnote{However, the quality of the optical spectrum of the latter source, presented in \cite{Barger02}, is low, and \citetalias{Yang14} and \cite{Xue16} discarded that spectroscopic redshift in favor of a photometric redshift $z=4.17^{+0.10}_{-0.13}$ from  \citetalias{Yang14}, which we adopted too, following the procedure described in \chapt{sample_N}.}

\begin{table}
	\tiny
	\caption{ (1) Redshift. (2) Number of X-ray sources which were assigned a spectroscopic redshift or (3) a photometric redshift from the different catalogs. At $3\leq z<6$ we considered only the fraction of PDF($z$) in that redshift interval. (4) Median accuracy of the photometric redshifts and (5) $\sigma_{\mathrm{NMAD}}$, both computed for X-ray sources with secure spectroscopic redshifts. (6) Number of sources with a multiwavelength counterpart and no redshift information, which are assigned a flat PDF($z$). (7) Number of sources with no multiwavelength counterpart, which are not included in the high-redshift sample, as most of them are expected to be spurious detections. (8) Total number of sources.  }\label{nz}
	\begin{tabular}{cccccccc}
		\hline
		\multicolumn{1}{c}{Sample} &
		\multicolumn{1}{c}{$z_\rmn{{spec}}$} &
		\multicolumn{1}{c}{$z_\rmn{{phot}}$}& 
		\multicolumn{1}{c}{$\mathrm{Med}(\frac{|\Delta z|}{1+z_{\mathrm{s}}})$}& 
				\multicolumn{1}{c}{$\sigma_{\mathrm{NMAD}}$}& 		
		\multicolumn{1}{c}{no $z$}&
			\multicolumn{1}{c}{no cp.}&
		\multicolumn{1}{c}{$N_{\mathrm{tot}}$}\\
		(1)       & (2) & (3) &(4) & (5)& (6) &(7)& (8)\\		
		\hline
		\multicolumn{7}{c}{CDF-S}\\		
		All        & 616 & 317 &0.008 & 0.009& 5 &14& 952\\
		$3\leq z<6$& 18 & 51.5 &0.015 & 0.014& 1.5 & --&71.0 \\
				\hline
		\multicolumn{7}{c}{CDF-N}\\		
		All        & 309 & 166 &0.026 & 0.031& 2 & 8&485\\
		$3\leq z<6$& 10 & 20.0&0.074 & 0.040& 0.6 &-- &30.6 \\	
				\hline	
		\multicolumn{7}{c}{CDF-S + CDF-N}\\		
		All        & 925 & 483 & 0.013&0.017 & 7&22 &1437\\
		$3\leq z<6$& 28  & 71.5& 0.032&0.039 & 2.1& --&101.6 \\			
		\hline
	\end{tabular}\\
\end{table}

\begin{table*}
\caption{Main properties of the high-redshift sources. }\label{sources}
\scriptsize
\begin{tabular}{cllllccccccc}	\hline
    ID & $z$                                  & Ref & $P$&Band    & $N_{H}$                                     & $\mathrm{log}F^{\mathrm{obs}}_{0.5-2\,\mathrm{keV}}$ &$\mathrm{log}F^{\mathrm{intr}}_{0.5-2\,\mathrm{keV}}$ & $\mathrm{log}L^{\mathrm{obs}}_{2-10\,\mathrm{keV}}$ &   $\mathrm{log}L^{\mathrm{intr}}_{2-10\,\mathrm{keV}}$ & \multicolumn{2}{c}{Net counts }\\
        &                                            &       &   $(3\leq z <6)$                         &            &   $(10^{22}\mathrm{cm^{-2}})$        &   $(\mathrm{erg\,cm^{-2}s^{-1}})$&  $(\mathrm{erg\,cm^{-2}s^{-1}})$ &   $(\mathrm{erg\,s^{-1}})$& $(\mathrm{erg\,s^{-1}})$ & $(0.5-2\,\mathrm{keV})$ & $(2-7\,\mathrm{keV})$ \\
   	(1)& (2)& (3)                                  & (4) & (5)                      &               (6)                   &                (7)                                          & (8) &  (9) & (10) & (11) & (12)\\ \hline
   \multicolumn{12}{c}{CDF-S}\\ \hline	
	8  &   $3.13_{-0.48}^{+3.28}$   &  P3 &   0.55                 &  S &     $59_{-18}^{+24}$  &   $-15.56_{-0.08}^{+0.07}$   &    $-15.59_{-0.08}^{+0.08}$ & $44.43_{-0.23}^{+0.20}$ &$44.41_{-0.27}^{+0.17}$  & 99.2 &  119.4        \\
	9  &    $1.93_{-0.20}^{+1.39}$  &  P3 &    0.44                & S &      $11_{-5}^{+8}$          &        $-15.31_{-0.08}^{+0.05}$    &  $-15.32_{-0.07}^{+0.06}$  & $44.04_{-0.13}^{+0.16}$ &$44.03_{-0.14}^{+0.15}$ & 140.7 & 95.4  \\
	10 &   $2.93_{-0.40}^{+0.85}$ &  P3 &   0.32                 &  S &    $3_{-2}^{+2}$            &      $-15.40_{-0.11}^{+0.06}$   &   $-15.43_{-0.09}^{+0.08}$   &  $43.76_{-0.13}^{+0.09}$  &$43.73_{-0.14}^{+0.09}$  & 70.1 &   38.9 \\
	12 &   $3.153$                         &   S2&    1.00                &  H &     $135_{-46}^{+69}$   &            $-16.59_{-0.75}^{+0.21}$     &   $<-17.28$      &   $44.06_{-0.18}^{+0.12}$ &$42.21_{-0.22}^{+1.32}$ &12.4 &   89.0     \\
	25 &     $-1$                          &   N&         0.30             &   S &   $<74$                      &      $-15.65_{-0.13}^{+0.08}$     &  $-15.69_{-0.15}^{+0.08}$ & $43.63_{-0.20}^{+0.44}$  &$43.60_{-0.18}^{+0.46}$   & 75.0 &     88.3  \\
	29 &   $3.33_{-0.29}^{+0.32}$  &  P4 &     0.99               &  S &    $<1$                       &         $-15.63_{-0.10}^{+0.05}$  &  $-15.65_{-0.11}^{+0.05}$  &  $43.47_{-0.13}^{+0.12}$ &$43.45_{-0.12}^{+0.13}$ & 119.0 &     44.4   \\
	35 &       $-1$                          &  N &       0.30                  &  S &   $25_{-17}^{+29}$     &     $-15.61_{-0.09}^{+0.09}$    &  $-15.63_{-0.11}^{+0.08}$ &  $43.97_{-0.20}^{+0.21}$  &$43.94_{-0.22}^{+0.20}$  &98.9 &   67.8  \\
	53 &  $3.202$                         & S1  & 1.00                     &  S &            $<22$              &         $-16.41_{-0.36}^{+0.18}$          &   $<-16.97$          & $42.74_{-0.40}^{+0.20}$   &$<42.30$   & 29.8& 40.7    \\
	84 &  $3.12_{-0.59}^{+0.35}$  & P1  & 0.70                     &  S &      $49_{-13}^{+14}$   &      $-16.07_{-0.11}^{+0.06}$ &  $-16.10_{-0.12}^{+0.06}$ &  $43.75_{-0.19}^{+0.14}$  &$43.71_{-0.19}^{+0.15}$   & 80.4 &      98.5   \\
	88 &  $-1$                               & N   & 0.30                     &  S &      $83_{-26}^{+19}$  &        $-15.98_{-0.08}^{+0.06}$  & $-16.01_{-0.09}^{+0.06}$ & $44.19_{-0.27}^{+0.12}$   &$44.16_{-0.25}^{+0.14}$  &99.3 & 195.6   \\
	99 &  $4.24_{-0.36}^{+0.05}$  &  P1&     0.99                 &  S &       $30_{-28}^{+34}$ &      $-15.91_{-0.19}^{+0.14}$               &  $<-16.19$                     & $43.55_{-0.26}^{+0.22}$    &$<43.32$  &41.3 &47.8 \\
	101& $4.34_{-0.61}^{+0.02}$  & P1  & 1.00                     &   S &     $68_{-22}^{+12}$   &       $-16.11_{-0.16}^{+0.09}$   &  $-16.20_{-0.27}^{+0.18}$ & $43.87_{-0.24}^{+0.24}$   &$43.78_{-0.37}^{+0.22}$   &51.2 & 110.8 \\
	109& $3.58_{-0.04}^{+0.03}$ & P1  & 1.0                        &  S &       $28_{-6}^{+5}$      &       $-16.48_{-0.23}^{+0.15}$                      &   $<-16.68$                       & $43.14_{-0.25}^{+0.16}$  &$<42.96$ &144.9 &150.9 \\
	110& $3.08_{-0.08}^{+0.38}$  & P1  & 0.87                     &   S &           $<1$                &        $-16.54_{-0.24}^{+0.15}$                       &  $<-16.74$                 & $42.55_{-0.24}^{+0.19}$  &$<42.40$   &22.8&  4.4\\
	121&  $3.42_{-0.05}^{+0.03}$  & P1  & 1.00                     &  S &      $68_{-21}^{+14}$    &     $-16.27_{-0.13}^{+0.09}$        & $-16.33_{-0.16}^{+0.11}$&  $43.69_{-0.22}^{+0.16}$  &$43.62_{-0.27}^{+0.16}$ &50.4&65.9   \\
	126&  $3.13_{-0.03}^{+0.07}$  & P1 & 0.95                     &  S &     $<28$                       &        $-15.84_{-0.23}^{+0.13}$             &  $<-16.55$                &  $43.32_{-0.22}^{+0.20}$     &$<42.67$   &39.9& 53.1  \\
	133& $3.474$                         & S1  & 1.00                     &   H &     $141_{-50}^{+73}$   &         $<-16.96$                               &  $<-17.39$  & $43.61_{-0.22}^{+0.13}$  &$42.23_{-0.15}^{+0.93}$ &4.0 & 56.3 \\
	165& $3.55_{-0.02}^{+0.02}$ & P1  & 1.00                     &  S &       $16_{-6}^{+6}$       &         $-15.94_{-0.09}^{+0.07}$      &   $-15.96_{-0.11}^{+0.06}$  &   $43.52_{-0.13}^{+0.09}$ &$43.5_{-0.15}^{+0.08}$  &71.9& 34.2  \\
	185& $3.09_{-0.03}^{+0.04}$ & P1  & 0.87                     & S &       $39_{-24}^{+24}$    &        $-16.70_{-0.19}^{+0.11}$      &  $-16.82_{-0.29}^{+0.16}$& $42.86_{-0.37}^{+0.22}$   &$42.71_{-0.45}^{+0.25}$    &18.7& 22.6  \\
	189& $3.65_{-0.04}^{+0.07}$ & P1  & 1.00                     & H &       $107_{-33}^{+25}$  &        $<-16.97$                                  &  $<-17.31$  &  $43.38_{-0.16}^{+0.09}$   &$43.30_{-0.19}^{+0.12}$    &7.1&  34.5  \\
	214& $3.74$                           & S2  & 1.00                     & S &       $26_{-2}^{+1}$     &     $-14.81_{-0.03}^{+0.00}$   &  $-14.82_{-0.03}^{+0.01}$  &  $44.81_{-0.03}^{+0.02}$   &$44.81_{-0.03}^{+0.02}$ & 1377.4& 1213.8   \\
	223& $3.76_{-0.17}^{+0.06}$ & P1  & 1.00                     &  S &            $<14$               &      $-16.46_{-0.15}^{+0.11}$       &  $-16.55_{-0.22}^{+0.13}$  &  $42.80_{-0.19}^{+0.11}$   &$42.71_{-0.23}^{+0.15}$   &27.9& 0.0   \\
	238&  $5.96_{-0.04}^{+0.01}$ & P2  & 1.00                     & S &             $<66$              &        $-16.29_{-0.17}^{+0.09}$   &  $-16.37_{-0.21}^{+0.14}$      &  $43.42_{-0.17}^{+0.15}$  &$43.35_{-0.24}^{+0.18}$    &40.5 & 7.1 \\
	248& $3.44_{-0.13}^{+0.22}$ & P1  & 1.00                     &   S &      $39_{-7}^{+5}$      &        $-15.55_{-0.06}^{+0.04}$    &  $-15.56_{-0.06}^{+0.04}$ & $44.18_{-0.10}^{+0.07}$   &$44.17_{-0.10}^{+0.07}$   &186.9&192.9 \\
	276&  $3.12_{-0.01}^{+0.02}$  & P1  & 1.00                     & S &        $15_{-3}^{+3}$      &       $-15.90_{-0.06}^{+0.04}$    & $-15.91_{-0.06}^{+0.04}$ & $43.51_{-0.08}^{+0.05}$    &$43.50_{-0.08}^{+0.05}$  & 122.9&  9.2   \\
	299& $3.79_{-0.05}^{+0.05}$ & P1  & 1.00                     &  S &       $10_{-9}^{+15}$    &         $-16.63_{-0.16}^{+0.10}$     &  $-16.71_{-0.19}^{+0.12}$  &  $42.67_{-0.18}^{+0.13}$   &$42.59_{-0.21}^{+0.15}$  &19.4& 21.0    \\
	329& $3.47_{-0.10}^{+0.10}$  & P1  & 1.00                     &  S &      $32_{-29}^{+33}$   &          $-16.69_{-0.20}^{+0.14}$                 &  $<-16.74$                   &  $42.61_{-0.25}^{+0.31}$   &$<42.77$     &17.2 &     17.1      \\
	337&  $3.660$                        & S1  & 1.00                     &  S &       $98_{-9}^{+7}$       &     $-16.08_{-0.09}^{+0.04}$     &  $-16.10_{-0.08}^{+0.05}$  &  $44.16_{-0.10}^{+0.07}$    &$44.14_{-0.09}^{+0.08}$ &78.3& 221.4  \\
	351&  $3.86_{-0.54}^{+0.14}$ & P1  & 0.97                     &  S &       $74_{-23}^{+9}$     &        $-16.28_{-0.14}^{+0.08}$    &  $-16.34_{-0.17}^{+0.10}$  &$43.71_{-0.19}^{+0.17}$     &$43.65_{-0.26}^{+0.15}$   &50.8& 73.2  \\
	356&   $3.38_{-0.19}^{+0.01}$ & P1  & 0.99                     & S &             $<8$                 &      $-16.38_{-0.14}^{+0.07}$    &   $-16.43_{-0.15}^{+0.08}$ &  $42.76_{-0.13}^{+0.10}$    &$42.71_{-0.15}^{+0.10}$   &31.9 & 31.4   \\
	366&  $3.67_{-0.18}^{+0.53}$ & P1  & 1.00                     &  S &      $81_{-15}^{+14}$     &       $-16.15_{-0.10}^{+0.06}$      &   $-16.18_{-0.10}^{+0.07}$ &  $43.96_{-0.14}^{+0.14}$    &$43.93_{-0.16}^{+0.13}$  &66.8 & 131.2 \\
	368& $3.21_{-0.08}^{+0.09}$ & P1  & 1.00                     &   S &           $<1$                 &             $-16.57_{-0.15}^{+0.10}$     &  $-16.65_{-0.19}^{+0.11}$ &  $42.52_{-0.19}^{+0.10}$     &$42.44_{-0.20}^{+0.11}$  &20.9&  0.0\\
	414& $3.58_{-0.10}^{+0.11}$   & P1  & 1.00                     &  H &            $-1^{**}$              &       $<-17.28$                                &   $<-17.41$   &   $42.71_{-0.36}^{+0.20}$    &$<42.61$   &0.3 &  9.7  \\
	416&  $3.470$                         & S1  & 1.00                     & S &             $<1$                 &    $-16.63_{-0.19}^{+0.10}$    &   $-16.75_{-0.27}^{+0.19}$  &  $42.52_{-0.19}^{+0.10}$     &$42.40_{-0.26}^{+0.15}$     &18.0 &0.4    \\
	433&  $4.11_{-0.12}^{+0.15}$   & P1  & 1.00                     & S &             $<1$                 &            $-16.80_{-0.23}^{+0.13}$            &   $<-16.86$                & $42.55_{-0.24}^{+0.14}$     & $<42.51$  &12.9& 0.3 \\
	459&  $3.91_{-0.04}^{+0.05}$ & P1  & 1.00                     &  H &      $166_{-64}^{+80}$  &     $<17.20$                                   &  $<-17.43$   &   $43.34_{-0.25}^{+0.17}$    &$<43.11$   & 0.3& 24.4  \\
	462&  $3.88_{-0.78}^{+0.41}$  &  P1 &   0.95                 &  H &       $51_{-21}^{+21}$     &      $-16.24_{-0.44}^{+0.19}$                &   $<-17.12$                     &  $43.92_{-0.25}^{+0.16}$    &$<42.69$   &25.6& 63.5\\
	464&  $4.76_{-0.10}^{+0.02}$ & P2  & 1.00                     & S &             $<16$               &      $-15.67_{-0.07}^{+0.04}$   &   $-15.68_{-0.06}^{+0.05}$  &  $43.81_{-0.08}^{+0.06}$   &$43.80_{-0.08}^{+0.06}$  &160.5 & 10.1 \\
	472& $4.16_{-0.17}^{+0.02}$  & P1  & 1.00                     &   H&    $110_{-94}^{+317}$  &        $<-16.97$                         &  $<-17.28$    &   $43.14_{-0.24}^{+0.17}$    &$<43.00$  &7.0&   19.2 \\
	490&  $4.73_{-0.08}^{+0.06}$ & P1  & 1.00                     & S &        $71_{-8}^{+9}$       &    $-16.05_{-0.07}^{+0.04}$    &  $-16.06_{-0.07}^{+0.04}$  &  $44.03_{-0.08}^{+0.07}$  &$44.01_{-0.10}^{+0.06}$   &86.8&     123.3\\
		500& $3.15_{-0.02}^{+0.04}$ & P1  & 1.00                     &S &        $40_{-38}^{+89}$   &     $<-16.79$                                    &  $<-17.19$    &   $<42.83$    &$<42.46$  &10.4& 16.6 \\
	503& $3.14_{-0.07}^{+0.03}$ & P1  & 0.99                     &    S &          $<5$                &               $<-16.93$                       &  $<-17.19$    &  $<42.20$    & $<41.95$     &7.0& 7.6  \\
	517& $3.256$                         & S2  & 1.00                     & S &             $<1$                &   $-16.24 _{-0.11}^{+0.06}$      &      $-16.27_{-0.12}^{+0.06}$ &$42.85_{-0.10}^{+0.07}$ &$42.82_{-0.11}^{+0.07}$  &44.1 &  24.0  \\%
	521&  $2.96_{-0.07}^{+0.07}$ & P1  & 0.41                     &  S &      $21_{-19}^{+60}$    &$-16.89 _{-0.33}^{+0.18}$          &      $<-17.11$            &  $42.34_{-0.36}^{+0.39}$     &$42.60_{-0.19}^{+0.15}$    & 13.0 & 27.7 \\%
	527&  $3.76_{-0.29}^{+0.05}$ & P1  & 1.00                     & S &             $<1$                 &   $-16.55 _{-0.15}^{+0.10}$      &     $-16.63_{-0.19}^{+0.12}$  & $42.68_{-0.17}^{+0.12}$ &$42.60_{-0.19}^{+0.15}$   & 22.5 & 0.0   \\%
	539& $4.762$                        & S1  & 1.00                     &  H &      $102_{-47}^{+49}$  &   $-16.48 _{-0.42}^{+0.18}$     &    $<-17.09$ &        $43.83_{-0.27}^{+0.13}$                &$<42.83$    &27.0 & 61.2 \\%
	551&  $3.700$                         & S1  & 1.00                   & S &        $76_{-5}^{+4}$      &   $-15.71 _{-0.05}^{+0.03}$  & $-15.72_{-0.05}^{+0.03}$   &  $44.37_{-0.05}^{+0.05}$  &$44.37_{-0.06}^{+0.04}$   &  209.8 & 441.4\\%
	580& $3.17_{-0.12}^{+0.19}$  & P1  & 0.97                     &  S &       $42_{-8}^{+10}$    & $-16.34 _{-0.08}^{+0.06}$     &     $-16.36_{-0.10}^{+0.05}$   & $43.40_{-0.13}^{+0.11}$  &$43.38_{-0.12}^{+0.12}$   & 45.1 & 66.6 \\%
	617& $3.58_{-0.13}^{+0.20}$ & P1  & 1.00                     &  S &            $<1$                  &   $-15.77 _{-0.07}^{+0.03}$ &    $-15.78_{-0.07}^{+0.03}$  & $43.41_{-0.08}^{+0.05}$    &$43.40_{-0.07}^{+0.06}$   &  136.0 & 69.1\\%
	619    & $3.67_{-0.09}^{+0.04}$ & P1  & 1.00  &  S &      $68_{-23}^{+18}$    &   $-16.85 _{-0.20}^{+0.10}$   &      $-16.97_{-0.22}^{+0.15}$   & $43.12_{-0.27}^{+0.19}$  &$42.99_{-0.32}^{+0.20}$  & 12.3 &  23.2\\%
	622    & $3.33_{-0.17}^{+0.09}$ & P1  & 1.00   &  S &            $<1$                  &      $- 16.50_{-0.19}^{+0.11}$    &     $-16.62_{-0.41}^{+0.18}$& $42.62_{-0.20}^{+0.14}$  &$42.49_{-0.44}^{+0.20}$    & 24.5 & 0.0 \\%
	623    & $3.58_{-0.03}^{+0.02}$ & P1  & 1.00  &  S &      $27_{-10}^{+10}$    &  $- 16.43_{-0.12}^{+0.07}$     &      $-16.47_{-0.12}^{+0.08}$  & $43.17_{-0.14}^{+0.13}$ &$43.13_{-0.18}^{+0.11}$    & 36.7 &  34.5\\%
	
  \\ \hline
\end{tabular}\\
(1) ID from \citetalias{Luo17} or \citetalias{Xue16} ;  (2) nominal redshifts (corresponding to the peaks of the PDFs) and uncertainties corresponding to the narrowest redshift intervals in which the integral of the PDFs is 0.68; (3) reference for the redshifts and PDFs: P1, P2, P3, P4, P5 refer to photometric redshifts from \citetalias{Straatman16}, \citetalias{Skelton14}, \citetalias{Hsu14},  \citet{Rafferty11}, and \citetalias{Yang14}, respectively. S1 and S2 stand for spectroscopic redshift from the collection presented in \citetalias{Luo17}, flagged as ``secure'' and ``insecure" (but upgraded as described in \chapt{sample}), respectively. S3 and S4 refer to spectroscopic redshift from \citetalias{Xue16} and \citetalias{Yang14} (for sources for which the multiwavelength matching has been improved via the visual inspection of X-ray and optical images, see \chapt{sample_N}), respectively. N refers to no redshift and, in such a case, column 2 is assigned a value of $-1$ and a flat PDF($z$) is assumed. (4) fraction of the PDFs at $3\leq z<6$ (see Eq.~\ref{zfrac}); (5) primary detection band (S: soft band, H: hard band; see \chapt{cr}): (6), (7), (8), (9) and (10): best estimates of the intrinsic column density, soft-band flux, and $2-10$ keV intrinsic luminosity, respectively, and 68\% confidence-level uncertainties. For fluxes and luminosities, both the observed (i.e. not applying the Eddington bias correction in Eq.~\ref{PDF_cr}; columns 7 and 9) and the intrinsic (i.e. for which the Eddington bias correction has been applied; columns 8 and 10) values are reported.  $^{**}$ for these sources we assumed a flat distribution in column density, as their spectral quality was too poor to perform a spectral analysis. (11) and (12): net counts in the soft and hard band, respectively. Continues. )
\end{table*}

\addtocounter{table}{-1}

\begin{table*}
	\scriptsize
	\caption{Continued.}
	\begin{tabular}{clllllccccccc}
		\hline
    ID & $z$                                  & Ref & $P$&Band    & $N_{H}$                                     & $\mathrm{log}F^{\mathrm{obs}}_{0.5-2\,\mathrm{keV}}$ &$\mathrm{log}F^{\mathrm{intr}}_{0.5-2\,\mathrm{keV}}$ & $\mathrm{log}L^{\mathrm{obs}}_{2-10\,\mathrm{keV}}$ &   $\mathrm{log}L^{\mathrm{intr}}_{2-10\,\mathrm{keV}}$ & \multicolumn{2}{c}{Net counts }\\
        &                                            &       &   $(3\leq z <6)$                         &            &   $(10^{22}\mathrm{cm^{-2}})$        &   $(\mathrm{erg\,cm^{-2}s^{-1}})$&  $(\mathrm{erg\,cm^{-2}s^{-1}})$ &   $(\mathrm{erg\,s^{-1}})$& $(\mathrm{erg\,s^{-1}})$ & $(0.5-2\,\mathrm{keV})$ & $(2-7\,\mathrm{keV})$ \\
(1)& (2)& (3)                                  & (4) & (5)                      &               (6)                   &                (7)                                          & (8) &  (9) & (10) & (11) & (12)\\ \hline
	640   & $3.77_{-0.03}^{+0.02}$ & P1  & 1.00   &  S &      $16_{-15}^{+18}$    &   $-16.68 _{-0.16}^{+0.10}$       &     $-16.75_{-0.19}^{+0.10}$ &  $42.65_{-0.18}^{+0.16}$ &$42.58_{-0.18}^{+0.18}$    &17.7 & 3.5  \\%
	657    & $3.58_{-0.10}^{+0.17}$  & P1  & 1.00   & S &       $115_{-19}^{+11}$   &   $- 16.59_{-0.14}^{+0.09}$      &    $-16.66_{-0.17}^{+0.10}$  &  $43.76_{-0.19}^{+0.13}$  &$43.69_{-0.20}^{+0.15}$  & 26.2 & 79.6 \\%
	662    & $4.84_{-0.12}^{+0.05}$ & P1  & 1.00   & S &             $<54$              &         $-16.82 _{-0.24}^{+0.14}$                &  $<-16.90$      &  $42.71_{-0.27}^{+0.18}$   &$<42.68$  &  13.3 & 2.6 \\%
	692    & $3.42_{-0.08}^{+0.13}$ & P1  & 1.00   &H &        $117_{-51}^{+56}$  &    $<-17.14$                          &    $<-17.37$  &          $42.84_{-0.31}^{+0.18}$              &$<43.22$    &3.6  &9.7 \\%
	714    & $3.48_{-0.09}^{+0.03}$ & P1  & 1.00   & S &  $<1$                           &   $- 16.25_{-0.10}^{+0.08}$     &  $-16.29_{-0.13}^{+0.07}$    &  $42.91_{-0.13}^{+0.07}$   &$42.87_{-0.13}^{+0.08}$  & 44.4 & 0.0 \\%
	723     & $3.045$                         & S1  & 1.00   &S &    $59_{-6}^{+4}$          &    $-16.02 _{-0.05}^{+0.05}$      &  $-16.03_{-0.05}^{+0.05}$& $43.88_{-0.10}^{+0.05}$    &$43.87_{-0.10}^{+0.05}$   &93.0 &  188.1  \\%
	746    & $3.064$                         & S1  & 1.00   &S &   $52_{-2}^{+1}$            &     $- 15.14_{-0.05}^{+0.02}$     &  $-15.15_{-0.04}^{+0.03}$ & $44.69_{-0.05}^{+0.03}$     &$44.68_{-0.05}^{+0.03}$    & 689.2 & 1281.4  \\%
	758    & $3.08_{-0.03}^{+0.01}$ & P1  & 1.00    &S &  $<36$                          &     $<-17.12$                            & $<-17.37$&          $<42.13$                 &$<41.92$                       & 8.7 &  13.5  \\%
	760    & $3.35$                           & S2  & 1.00  &S &    $74_{-5}^{+2}$            &  $- 15.111_{-0.03}^{+0.03}$ &  $-15.11_{-0.04}^{+0.02}$        & $44.94_{-0.05}^{+0.03}$    &$44.94_{-0.05}^{+0.03}$  & 763.5 & 999.8   \\%
	774     & $3.61$                            & S1  & 1.00   &S &    $7_{-1}^{+1}$              &   $-14.71 _{-0.04}^{+0.03}$     &  $-14.71_{-0.05}^{+0.02}$  & $44.63_{-0.03}^{+0.04}$ &$44.63_{-0.04}^{+0.03}$ &  1162.8 &  691.5\\%
	788     & $3.193$                          & S2  & 1.00  &S &  $2_{-1}^{+1}$                 & $- 15.11_{-0.03}^{+0.03}$   &   $-15.11_{-0.04}^{+0.02}$     &  $44.01_{-0.05}^{+0.02}$&$44.01_{-0.05}^{+0.02}$  &  646.8 &381.8  \\%
	811     & $3.471$                          & S1  & 1.00  & S &  $<1$                             &  $- 15.37_{-0.05}^{+0.02}$ &   $-15.37_{-0.05}^{+0.02}$     & $43.77_{-0.04}^{+0.03}$&$43.77_{-0.05}^{+0.02}$   &  312.9 &  194.9 \\%
	853    & $3.72_{-0.14}^ {+0.13}$ & P1  & 1.00   & S &  $<2$                            &       $- 16.21_{-0.29}^{+0.16}$                     & $<-16.81$     &  $43.05_{-0.35}^{+0.18}$ &$<42.60$    &37.0 &69.7  \\%
	859     & $2.88_{-0.09}^{+0.13}$ &  P1 &   0.22 &S &   $<3$                           &     $-16.71 _{-0.30}^{+0.17}$                      &  $<-16.93$       & $42.38_{-0.37}^{+0.21}$  &$<42.34$  &  16.4 &1.6 \\			%
	873     & $3.76_{-0.06}^{+0.04}$ & P1  & 1.00   & S & $<1$                            &   $- 16.25_{-0.12}^{+0.10}$        &  $-16.31_{-0.16}^{+0.11}$  & $42.98_{-0.13}^{+0.10}$ &$42.92_{-0.17}^{+0.11}$      &   41.4 &0.0  \\%
	876      & $3.470$                         & S2  & 1.00  &S &      $11_{-1}^{+1}$           &   $-14.57 _{-0.01}^{+0.01}$      & $-14.57_{-0.01}^{+0.01}$  &  $44.82_{-0.02}^{+0.01}$ &$44.82_{-0.02}^{+0.01}$   & 2211.9 &1482.2  \\%
	885    & $3.08_{-0.10}^{+0.02}$ & P1  & 0.78   & S &   $95_{-10}^{+10}$        &  $-16.18 _{-0.10}^{+0.07}$       &   $-16.21_{-0.13}^{+0.06}$  & $44.02_{-0.14}^{+0.10}$&$43.98_{-0.13}^{+0.12}$  & 64.4 &  201.9  \\%
	901      & $3.51_{-0.12}^{+0.25}$  & P1  & 1.00   &S &   $<43$                          &  $- 16.63_{-0.32}^{+0.17}$                        & $<-16.94$      & $42.65_{-0.34}^{+0.25}$  &$<42.43$  & 19.8 &14.7\\%
	908     & $3.60_{-0.20}^{+0.05}$ &  P1 &   1.00 &H &   $>23$                         &   $-16.94 _{-0.55}^{+0.61}$                          &  $<-17.20$     & $43.64_{-0.26}^{+0.27}$  &$<43.01$    & 7.5 &  52.5  \\			%
	921      & $3.082$                          & S1  & 1.00  &S &      $<4$                        &     $- 15.33_{-0.07}^{+0.02}$    &   $-15.34_{-0.06}^{+0.03}$ &$43.73_{-0.04}^{+0.05}$ &$43.73_{-0.07}^{+0.03}$   &347.6 & 204.8  \\%
	926     & $4.27_{-1.5}^{+0.06}$    & P1  & 0.53 &  S & $<7$                             &  $-15.61 _{-0.05}^{+0.05}$        &   $-15.62_{-0.07}^{+0.04}$ &$43.76_{-0.06}^{+0.06}$ &$43.75_{-0.08}^{+0.05}$  & 182.2 & 116.4 \\%
	939      & $-1$                               & N   & 0.30  &S &           $<1$                   &      $- 16.00_{-0.11}^{+0.10}$    &  $-16.05_{-0.16}^{+0.10}$  & $43.47_{-0.36}^{+0.12}$ &$43.38_{-0.36}^{+0.15}$   &  62.8 &15.0  \\%
	940      & $3.00_{-0.19}^{+0.26}$  & P1  & 0.63  &S &   $35_{-10}^{+7}$          &     $-16.00 _{-0.09}^{+0.06}$      &  $-16.03_{-0.09}^{+0.07}$& $43.64_{-0.16}^{+0.11}$ &$43.61_{-0.14}^{+0.13}$    & 91.7 & 97.4 \\%
	962      & $4.68_{-2.25}^{+2.25}$ & P3  & 0.45  &S & $141_{-48}^{+49}$        &    $- 16.40_{-0.31}^{+0.15}$                     &  $<-16.95$         & $44.01_{-0.43}^{+0.26}$  & $<43.54$    &35.2 &  130.2  \\%
	965      & $3.64_{-0.80}^{+0.11}$  & P3  & 0.65  &S &   $<1$                            &   $- 15.79_{-0.10}^{+0.06}$     &   $-15.81_{-0.11}^{+0.06}$  &  $43.36_{-0.13}^{+0.10}$ &$43.33_{-0.14}^{+0.10}$ &104.0 & 40.2   \\%
	971       & $2.17_{-0.59}^{+1.24}$   & P3  & 0.35  &S &   $36_{-13}^{+8}$           &   $- 15.61_{-0.05}^{+0.02}$    &    $-15.16_{-0.05}^{+0.02}$ & $44.73_{-0.28}^{+0.11}$& $44.73_{-0.29}^{+0.10}$   & 536.7 & 513.2  \\ %
	974      & $3.23_{-0.13}^{+0.18}$  &  P3 &   0.92 &S &   $9_{-8}^{+10}$            &   $-15.76 _{-0.10}^{+0.08}$      &     $-15.79_{-0.13}^{+0.07}$ &	$43.44_{-0.13}^{+0.11}$  &$43.41_{-0.14}^{+0.11}$    & 105.6 & 71.5 \\		%
	977$^*$& $-1$                               &  N  &   0.3   &S &    $89_{-29}^{+34}$       &      $-16.13 _{-0.18}^{+0.10}$     &     $<-16.20$  &       $44.04_{-0.29}^{+0.23}$         &$<44.03$      & 48.6 & 128.0  \\				%
	990      & $3.724$                         &  S1 &   1.00  &S &   $62_{-54}^{+64}$        &     $- 16.06_{-0.28}^{+0.17}$                  &  $<-16.99$        &  $43.35_{-0.24}^{+0.58}$  &$<42.86$   & 41.0 & 61.3 \\			%
\hline	\multicolumn{12}{c}{CDF-N}\\ \hline	
	31        & $3.45_{-0.36}^{+0.21}$  & P5 &    0.95 & S &  $<1$                          &    $-15.66 _{-0.12}^{+0.08}$        &  $-15.71_{-0.15}^{+0.08}$& $43.48_{-0.13}^{+0.12}$  &$43.43_{-0.17}^{+0.11}$  &44.4&  16.6\\%
	81         & $3.19$                            &  S3&     1.00 &S &   $<1$                          &   $- 15.34_{-0.10}^{+0.07}$   &     $-15.37_{-0.10}^{+0.07}$     & $43.72_{-0.10}^{+0.07}$&$43.69_{-0.10}^{+0.07}$ & 74.3 & 25.9 \\%
	112      & $3.15_{-0.18}^{+0.18}$   &P5  & 0.85    &  S & $<14$                         &   $-16.15 _{-0.14}^{+0.13}$       &    $-16.24_{-0.22}^{+0.13}$ & $43.00_{-0.19}^{+0.15}$ &$42.90_{-0.23}^{+0.18}$   & 20.7 &8.1     \\%
	129     & $3.938$                        & S3 &  1.00     & S & $<1$                           &     $-15.43 _{-0.09}^{+0.07}$      &   $-15.46_{-0.10}^{+0.07}$ & $43.84_{-0.09}^{+0.07}$&$43.81_{-0.11}^{+0.06}$    &63.7 & 24.3   \\%
	133       & $3.06_{-0.36}^{+0.46}$ &P5  & 0.63     &S &  $93_{-19}^{+16}$        &        $-16.34 _{-0.23}^{+0.15}$                     & $<-16.57$   &  $43.84_{-0.28}^{+0.21}$   &$<43.62$   & 15.0 & 41.0  \\%
	142       & $2.95_{-0.13}^{+0.08}$  & P5 & 0.23     &S &  $135_{-25}^{+24}$     &  $- 16.28_{-0.18}^{+0.13}$       &  $-16.41_{-0.23}^{+0.14}$     &$44.24_{-0.25}^{+0.23}$  &$44.11_{-0.31}^{+0.22}$   & 14.6 &55.0  \\%
	158     & $3.19_{-0.02}^{+0.01}$  & P2 & 1.00      &  S &$54_{-10}^{+8}$          &    $-15.74 _{-0.11}^{+0.08}$     &  $-15.68_{-0.12}^{+0.08}$ & $44.10_{-0.14}^{+0.12}$  &$44.06_{-0.16}^{+0.11}$  & 46.4 &  72.2 \\%
	177        & $3.44_{-0.34}^{+0.14}$ & P5 & 0.98      &S & $33_{-12}^{+14}$         & $- 16.09_{-0.15}^{+0.09}$       &  $-16.16_{-0.14}^{+0.11}$ &  $43.55_{-0.21}^{+0.17}$  &$43.48_{-0.21}^{+0.18}$  &  27.5 & 29.3 \\%
	195       & $-1$                               & N &  0.30      &S & $<2$                           &    $- 16.02_{-0.13}^{+0.11}$     & $-16.09_{-0.14}^{+0.11}$   &  $43.46_{-0.35}^{+0.13}$  &$43.38_{-0.33}^{+0.15}$  &28.3 &  12.9  \\%
	196       & $3.24_{-0.02}^{+0.02}$ & P2 &1.00       &S &  $85_{-9}^{+10}$          &    $- 16.06_{-0.13}^{+0.12}$      &    $-16.13_{-0.16}^{+0.10}$  &$44.08_{-0.19}^{+0.12}$ &$44.00_{-0.18}^{+0.13}$ & 29.9 &96.9   \\%
			201        & $4.43_{-1.33}^{+0.23}$ & P5 & 0.73     &  S &$-1^{**}$                     &     $<-16.78$                         &    $<-17.33$   &$<42.98$                  & $44.06_{-0.19}^{+0.12}$ & 3.4 &15.5\\%
207      & $3.652$                         &  S3 &   1.00  & S & $<1$                          &      $- 15.09_{-0.08}^{+0.05}$        &  $-15.11_{-0.08}^{+0.05}$&  $44.10_{-0.08}^{+0.05}$ &$42.21_{-0.15}^{+0.30}$  & 242.6 & 80.7 \\%
227      & $4.26_{-0.07}^{+0.03}$  & P2 & 1.00     & S & $-1^{**}$                     &         $-16.70 _{-0.31}^{+0.24}$                       &   $<-17.00$   & $42.71_{-0.55}^{+0.40}$ &$<43.18$   & 6.5 &3.9  \\%
229    & $3.413$                         &  S4 & 1.00     & S &$<1$                           &     $- 15.43_{-0.09}^{+0.06}$       &    $-15.46_{-0.09}^{+0.06}$& $43.70_{-0.09}^{+0.06}$ &$43.67_{-0.09}^{+0.06}$   &111.8  &48.8  \\%
257   & $2.95_{-0.44}^{+0.29}$ & P5 & 0.35     &S & $60_{-32}^{+37}$        &  $-16.56 _{-0.22}^{+0.14}$       &  $-16.74_{-0.30}^{+0.19}$   &  $43.24_{-0.43}^{+0.32}$ &$43.04_{-0.47}^{+0.37}$  & 9.6 &16.7  \\%
290       & $4.424$                         & S3  &  1.00    &S & $<1$                           &   $-16.24 _{-0.17}^{+0.10}$   &     $-16.33_{-0.18}^{+0.11}$   & $43.14_{-0.16}^{+0.11}$ &$43.05_{-0.17}^{+0.12}$   & 17.7 &9.1  \\%
293    & $3.96_{-0.30}^{+0.17}$  & P5 &  1.00     & S &$<8$                            &     $-15.32 _{-0.07}^{+0.06}$    &    $-15.33_{-0.07}^{+0.06}$& $43.98_{-0.12}^{+0.07}$  &$43.96_{-0.11}^{+0.08}$    &  115.0&66.9  \\%
297  & $3.23$                          &  S3 &    1.00  & S &$<1$                           &     $-16.09 _{-0.12}^{+0.12}$  &     $-16.16_{-0.16}^{+0.10}$   & $42.99_{-0.15}^{+0.10}$ &$42.02_{-0.15}^{+0.11}$   & 22.3 & 10.8 \\%
310 & $3.06_{-0.23}^{+0.36}$  &P5  &  0.70     & S & $<41$                         & $16.67 _{-0.32}^{+0.19}$                                &$<-16.93$   & $42.52_{-0.40}^{+0.30}$  &$<42.47$    &6.6  & 6.5 \\%
315 & $-1$                                &  N & 0.30     & S &   $<1$                        &       $- 16.61_{-0.25}^{+0.16}$      &   $- 16.86_{-0.51}^{+0.21}$     & $42.78_{-0.36}^{+0.23}$ &$42.46_{-0.51}^{+0.31}$    & 7.5 & 2.8 \\%
320 & $3.06_{-0.20}^{+0.26}$  & P5 & 0.69      & S & $<1$                          &   $- 16.73_{-0.28}^{+0.19}$                               &    $<-16.91$  & $42.37_{-0.31}^{+0.21}$ &$<42.23$   & 5.8 & 5.5 \\%
330 & $5.186$                          & S3 &  1.00     & S & $<1$                           &    $-15.53 _{-0.12}^{+0.07}$    &    $-15.57_{-0.12}^{+0.07}$    &$44.00_{-0.12}^{+0.07}$ &$43.96_{-0.12}^{+0.07}$   &  84.4& 31.4 \\%
		363 & $3.28_{-0.05}^{+0.05}$  & P2 &  1.00     & S & $<1$                           &   $-16.19 _{-0.17}^{+0.14}$         &    $-16.31_{-0.21}^{+0.14}$ &$42.92_{-0.17}^{+0.15}$ &$42.80_{-0.21}^{+0.15}$  & 15.8 & 3.9 \\%
373 & $3.36_{-0.29}^{+0.34}$  & P5 &  0.84     & S & $52_{-31}^{+31}$        &     $-16.76 _{-0.27}^{+0.19}$                          &$<-16.92$     &  $42.98_{-0.51}^{+0.25}$  &$<42.85$  &5.9  & 6.5 \\%
388 & $2.94_{-0.12}^{+0.08}$  & P5 &  0.31      &  S &$<1$                          &       $- 15.49_{-0.12}^{+0.07}$      &    $-15.53_{-0.13}^{+0.08}$& $43.56_{-0.13}^{+0.08}$ &$43.52_{-0.14}^{+0.09}$ & 56.5 & 0.0 \\%
390 & $3.19_{-0.29}^{+0.28}$  & P5 &  0.85     &  H & $204_{-46}^{+53}$    &      $<-16.89$                          &     $<-17.39$    & $43.38_{-0.13}^{+0.07}$ &$43.30_{-0.16}^{+0.09}$    &3.1  &31.5  \\%
404 & $3.15_{-0.13}^{+0.04}$  & P5 &  0.91       & S & $<1$                          &      $- 14.67_{-0.06}^{+0.03}$     &    $-14.67_{-0.06}^{+0.03}$  &$44.37_{-0.06}^{+0.04}$  &$44.36_{-0.06}^{+0.04}$     & 555.4 &  220.8\\%
424 & $4.43_{-0.26}^{+0.13}$  & P5 &  1.00       &H &  $166_{-40}^{+38}$   &   $-16.37 _{-0.36}^{+0.18}$                            &     $<-17.17$     & $43.34_{-0.25}^{+0.17}$ &$42.32_{-0.11}^{+0.58}$    & 13.1 &  62.2\\%
		\hline
	\end{tabular}\\
Continues.
$^*$ XID 977 has an insecure spectroscopic redshift in \citetalias{Luo17}, but no photometric redshifts. Here we conservatively discard that spectroscopic redshift and treat this source as spectroscopically unidentified. $^{**}$ for these sources we assumed a flat distribution in column density, as their spectral quality was too poor to perform a spectral analysis.
\end{table*}

\addtocounter{table}{-1}

\begin{table*}
		\scriptsize
	\caption{Continued.}
	\begin{tabular}{clllllccccccc}
		\hline
    ID & $z$                                  & Ref & $P$&Band    & $N_{H}$                                     & $\mathrm{log}F^{\mathrm{obs}}_{0.5-2\,\mathrm{keV}}$ &$\mathrm{log}F^{\mathrm{intr}}_{0.5-2\,\mathrm{keV}}$ & $\mathrm{log}L^{\mathrm{obs}}_{2-10\,\mathrm{keV}}$ &   $\mathrm{log}L^{\mathrm{intr}}_{2-10\,\mathrm{keV}}$ & \multicolumn{2}{c}{Net counts }\\
        &                                            &       &   $(3\leq z <6)$                         &            &   $(10^{22}\mathrm{cm^{-2}})$        &   $(\mathrm{erg\,cm^{-2}s^{-1}})$&  $(\mathrm{erg\,cm^{-2}s^{-1}})$ &   $(\mathrm{erg\,s^{-1}})$& $(\mathrm{erg\,s^{-1}})$ & $(0.5-2\,\mathrm{keV})$ & $(2-7\,\mathrm{keV})$ \\
(1)& (2)& (3)                                  & (4) & (5)                      &               (6)                   &                (7)                                          & (8) &  (9) & (10) & (11) & (12)\\ \hline
		428 & $3.03_{-0.11}^{+0.09}$  & P5 &  0.55       &S &  $18_{-3}^{+2}$         &  $-15.07 _{0.06-}^{+0.04}$     &     $-15.08_{-0.06}^{+0.04}$   & $44.36_{-0.08}^{+0.05}$ &$44.35_{-0.08}^{+0.05}$  & 258.7 & 188.1 \\%
		439 & $3.406$                         & S3 &  1.00       &S &  $27_{-26}^{+73}$     &    $-16.54 _{-0.28}^{+0.17}$     &     $-16.81_{-0.46}^{+0.27}$   & $42.76_{-0.35}^{+0.41}$  &$42.49_{-0.51}^{+0.46}$ &8.5  &15.8  \\%
		498 & $4.42_{-0.24}^{+0.31}$  & P5 &  0.99       &S &  $23_{-6}^{+6}$         & $-15.33 _{-0.09}^{+0.06}$      &   $-15.35_{-0.10}^{+0.05}$   &  $44.36_{-0.12}^{+0.09}$   &$44.34_{-0.12}^{+0.09}$ &149.2  &86.2  \\%
		502 & $3.157$                          & S3 &  1.00       &S &  $<1$                         &$- 15.56_{-0.13}^{+0.08}$    &    $-15.61_{-0.10}^{+0.10}$         & $43.50_{-0.13}^{+0.08}$&$43.45_{-0.13}^{+0.08}$   &  73.5&29.8  \\%
		504 & $3.36_{-0.09}^{+0.10}$  & P5 &  1.00       &H&  $831_{-537}^{+123}$ &            $<-16.31$                   &      $<-17.57$ &  $43.14_{-0.24}^{+0.17}$&$42.26_{-0.13}^{+0.59}$    & 9.8 & 30.8 \\%
		509 & $3.15_{-0.57}^{+0.12}$  & P5 &   0.47       &S &  $<15$                       &           $-16.26 _{-0.26}^{+0.18}$                       & $<-16.93$  &   $42.88_{-0.32}^{+0.21}$  &$<42.30$    & 15.1 & 3.2 \\%
		526 & $4.17_{-0.13}^{+0.10}$  & P5 &   1.00        &S &  $15_{-6}^{+7}$         &  $- 15.84_{-0.10}^{+0.07}$      &      $-15.88_{-0.10}^{+0.08}$  & $43.71_{-0.14}^{+0.09}$  &$43.67_{-0.15}^{+0.09}$   & 89.0 & 48.0 \\%
		545 & $3.524$                        & S3 &   1.00       &S &  $<11$                        &   $- 15.27_{-0.07}^{+0.05}$     &      $-15.28_{-0.07}^{+0.05}$  & $43.95_{-0.09}^{+0.07}$ &$43.93_{-0.08}^{+0.08}$ &  132.4& 46.8 \\%
		572 & $3.77_{-0.01}^{+0.03}$  &P2  &   1.00       &S &  $10_{-2}^{+2}$          &  $-15.49 _{-0.13}^{+0.08}$      &    $-15.54_{-0.17}^{+0.10}$    &  $43.94_{-0.14}^{+0.08}$ &$43.89_{-0.17}^{+0.11}$    &  288.9&  156.2\\%
	\end{tabular}\\

\end{table*}

\section{Spectral analysis and parameter distributions}\label{spec_analysis}
X-ray spectra of the sources in the $z>3$ sample were extracted with ACIS Extract \citep{Broos10}, as described in \citetalias{Luo17} and \citetalias{Xue16}. Following \cite{Vito13}, we analysed the individual X-ray spectra with XSPEC\footnote{https://heasarc.gsfc.nasa.gov/xanadu/xspec/} v12.9.0 \citep{Arnaud96} assuming an absorbed power-law model with fixed $\Gamma=1.8$ (a typical value for X-ray selected AGN, e.g. \citealt{Liu17}), Galactic absorption \citep{Kalberla05} and intrinsic absorption (XSPEC model $wabs \times zwabs \times powerlaw$), for a total of two free parameters (the intrinsic column density, $N_{\mathrm{H}}$, which assumes solar metallicity, and the power-law normalization), in the energy range $0.5-7$ keV. More complex parametrizations are precluded by the general paucity of net counts characterizing the sources in the sample, which, being at high-redshift, are typically faint. Appendix \ref{check_spec_analysis} describes our check through spectral simulations that the simplicity of the assumed spectral model does not affect the best-fitting column densities at log$N_{\mathrm H}\gtrsim23$ (see \S~\ref{absfrac} for discussion). In Appendix \ref{gamma} we study the photon index of a subsample of unobscured sources.
The $W$ statistic\footnote{https://heasarc.gsfc.nasa.gov/xanadu/\\xspec/manual/XSappendixStatistics.html}, based on the Cash statistic \citep{Cash79} and suitable for Poisson data with Poisson background, was used to fit the model to the unbinned spectra. If a spectroscopic redshift is assigned to a source, the fit was performed at that redshift, otherwise we used a 
grid of redshifts (with step $\Delta z =0.01$), obtaining for each source a set of best-fitting models as a function of redshift. In \chapt{cr} we derive the probability distribution of net count-rate for each source, including a correction for Eddington bias. In \chapt{NH} we convolve the set of best-fitting models with the redshift probability distribution function for each source to derive the probability distribution of best-fitting column density. The fluxes of the sources are derived from the observed count-rates in \chapt{flux}, where we also present the log$N$-log$S$ of the sample. In \chapt{luminosity} we combine the fluxes and column densities to derive the intrinsic X-ray luminosities. Results from spectral analysis are reported in Tab.~\ref{sources}. 

\subsection{Count rates, sky-coverage, and Eddington bias}\label{cr}
\begin{figure*} 
\centering
 \includegraphics[width=176mm,keepaspectratio]{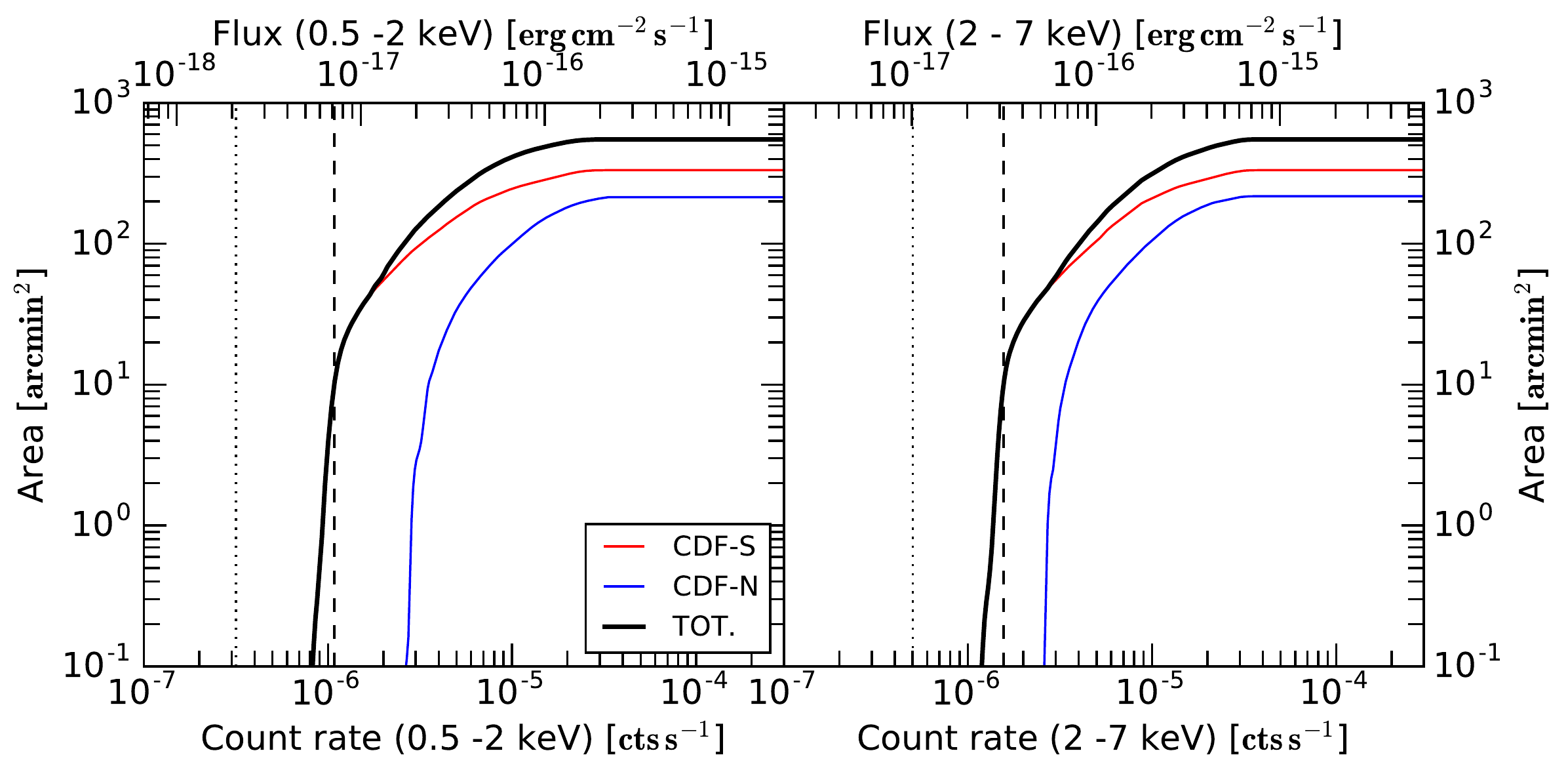}
\caption{Area covered by $\geq1$ Ms effective exposure sensitive to a given minimum count rate (lower $x$-axis) or flux (upper $x$-axis, assuming a power-law spectral shape with $\Gamma=1.4$) for the CDF-S (red curve), CDF-N (blue curve) and for the combination of the two surveys (black curve), in the soft band (left panel) and hard band (right panel). Different spectral assumptions would change only the flux axis. The dotted vertical line marks the minimum \textit{intrinsic} count rate we assumed for the sources in our sample (Eq.~\ref{cr_norm}). The vertical dashed lines indicate the minimum count rate corresponding to a $10\,\mathrm{arcmin^2}$ sensitive area, which we consider the observed count-rate limit for a source to be detected (see Eq.~\ref{omega}) to avoid large correction factors due to the small area sensitive at lower count rates.}
\label{skycov}
\end{figure*}

The fundamental quantity measured by an X-ray detector is the count rate of a source, which can then be converted into energy flux assuming a spectral shape and using the detector response matrices. Absorption affects strongly the spectral shape, in particular in the observed soft band ($0.5-2\,\mathrm{keV}$). Therefore different levels of absorption affecting sources with the same count rate result in different measured fluxes. Usually, when performing statistical studies of populations of astrophysical objects detected in X-ray surveys, weights are assigned to each source to take into account the different sensitive areas of the survey at different fluxes. The curves of the area sensitive to a given flux (namely, the sky-coverage) for the 7 Ms CDF-S and 2 Ms CDF-N  are shown in Fig.~\ref{skycov} (where we considered only the area covered by $\geq1$ Ms effective exposure, as per \chapt{sample}) for our two detection bands (see below). Since the source-detection algorithm used by \citetalias{Luo17} and \citetalias{Xue16} operates in the count-rate space, sources with the same fluxes in the detection energy band can be detected over different areas, if their spectral shapes are different (i.e. the same count rate can correspond to different fluxes). This is 
usually not taken into account, and an average spectral shape (usually a power-law with $\Gamma=1.4$) is assumed for all the sources, which can introduce errors up to a factor of several in the weights associated with each source \citep{Vito14}.  

We will therefore remain in the count-rate space, transforming count rates to energy fluxes, when needed, by assuming every time the specific spectral shape and response files suitable for a source. In doing so we do not assume any \textit{a priori} spectral shape to compute the weights related to the sky-coverage for our sources. Following \cite{Lehmer12}, we derived the probability distribution of the count rate for one source as

\begin{equation}\label{PDF_cr}
 P(CR)\propto (\frac{T}{T+b})^s\,(1-\frac{T}{T+b})^b\,\frac{dN}{dCR}
\end{equation}
where 
\begin{equation}
T=t_{\mathrm{exp}}\, CR \, f_{\mathrm{PSF}} + b\, \frac{A_{\mathrm{s}}}{A_{\mathrm{b}}}.
\end{equation}
\noindent
In these equations, $s$ and $b$ are the numbers of counts detected in the source and background regions, which have areas of $A_{\mathrm{s}}$ and $A_{\mathrm{b}}$, respectively, $t_{\mathrm{exp}}$ is the exposure time and $f_{\mathrm{PSF}}$ is the fraction of \textit{Chandra} PSF corresponding to $A_{\mathrm{s}}$ at the position of the source. The parameter $T$ is therefore the expected total (i.e. source plus background) number of counts detected in the $A_{\mathrm{s}}$ region for a source with a net count rate of $CR$. The first two factors of Eq.~\ref{PDF_cr} derive from the binomial probability of observing $s$ counts in the $A_{\mathrm{s}}$ region given the expected number $T$ and $b \frac{A_{\mathrm{s}}}{A_{\mathrm{b}}}$ background counts. The last factor is the X-ray source number counts (differential number of \mbox{X-ray} sources per unit count rate) derived by \citetalias{Luo17} using the 7 Ms CDF-S data set. This factor includes our 
knowledge of the intrinsic distribution in 
flux (or, in this case, count rate) of the AGN population, and accounts for the Eddington bias (see \citealt{Lehmer12} for discussion), expected for our sample of faint sources. Count-rate probability distributions are preferentially computed in the soft band, where the \textit{Chandra}/ACIS sensitivity peaks. For sources undetected in the soft band and detected in the hard band ($2-7\, \mathrm{keV}$) in the X-ray catalogs, we computed the count rates in the hard band. \citetalias{Luo17} provide AGN number counts for both bands. The number of sources in our sample detected in the soft (hard) band is 108 (14), as flagged in Tab.~\ref{sources}.\footnote{Two sources in our sample, XID 500 and 503, in the 7 Ms \mbox{CDF-S} catalog are reported as full-band detections only. Given the small number of such objects, for the purposes of this work we consider them as soft-band detections.}

The $\frac{dN}{dCR}$ factor diverges for very faint fluxes, which can be explained by considering that the \citetalias{Luo17} model is constrained only down to the 7 Ms CDF-S flux limit.  For this reason, we set a hard lower limit to the \textit{intrinsic} count rate, $CR^{\mathrm{intr}}_{\mathrm{lim}}=10^{-6.5}$ and $10^{-6.3}\mathrm{cts\,s^{-1}}$ for the soft and hard bands, respectively, a factor of $\sim0.5$ dex fainter than the nominal count-rate limit of the survey (dotted vertical lines in Fig.~\ref{skycov}) in those bands and normalized 
$P(CR)$ to unity above that limit:
\begin{equation}\label{cr_norm}
 \int_{CR^{\mathrm{intr}}_{\mathrm{lim}}}^\infty P(CR) \, dCR=1;
\end{equation}
i.e., we assume that the intrinsic count rate of a detected source is at most $\sim0.5$ dex lower than the 7 Ms \mbox{CDF-S} count-rate limit.
This approach allows us to apply the Eddington bias correction and obtain convergent probabilities.
The small area covered by the 7 Ms CDF-S at fluxes close to that limit would result in unreasonably large weights for very small count rates. To prevent this bias, we applied another cut, $CR^{\mathrm{obsc}}_{\mathrm{lim}}$, corresponding to the count rate at which the sensitive area is, following \cite{Lehmer12}, $\geq10\,\mathrm{arcmin^2}$ in the two bands (dashed vertical lines in Fig.~\ref{skycov}).
Finally, we define the weighting factor

\begin{equation}\label{omega}
 \Omega(CR)=\begin{cases}
\frac{\mathrm{MAX}(skycoverage)}{skycoverage(CR)} & \mathrm{if\, CR > CR^{\mathrm{obsc}}_{\mathrm{lim}}}\\
0.& \mathrm{otherwise}
            \end{cases}
\end{equation}
where $skycoverage(CR)$ is the black curve in Fig.~\ref{skycov} in each band, i.e., the sum of the sky-coverages of the CDF-S and CDF-N (computed by \citealt{Luo17} and \citealt{Xue16}, respectively), and the maximum sensitive area covered by the survey with $\geq1$ Ms effective exposure is $\mathrm{MAX}(skycoverage)\sim551\,\mathrm{arcmin^2}$. Using the combined sky-coverage, we assume that each source could in principle be detected in both surveys, i.e., we adopt the coherent addition of samples of \cite{Avni80}.

\subsection{Intrinsic column-density distribution}\label{NH}
Once the best-fitting model for a source at a given redshift had been obtained, we derived the observed best-fitting $\mathrm{log}N_{\mathrm{H}}$ distribution, $P(\mathrm{log}N_{\mathrm{H}}|z)$, through the XSPEC $steppar$ command. The probabilities are normalized such that

\begin{equation}\label{NH_z}
\int^{25}_{\mathrm{log}N_{\mathrm{H}}=20}  P(\mathrm{log}N_{\mathrm{H}}|z)\, \mathrm{dlog}N_{\mathrm{H}} = 1.
\end{equation}

The intrinsic column density probability \textit{at a given redshift} is then weighted by the PDF($z$) at high redshift to derive the intrinsic $N_H$ probability distribution for each source:

\begin{equation}\label{NH_int}
P(\mathrm{log}N_{\mathrm{H}}) = \frac{\int^{6}_{z=3}  P(\mathrm{log}N_{\mathrm{H}}|z) \, \mathrm{PDF}(z) \, dz}{P(3\leq z<6)},
\end{equation}
where $P(3\leq z<6)$ is defined in Eq.~\ref{zfrac}. Note that Eq.~\ref{NH_int} is normalized to unity, in order to compute confidence intervals and estimate uncertainties.

\begin{figure*} 
	\centering
	\includegraphics[width=180mm,keepaspectratio]{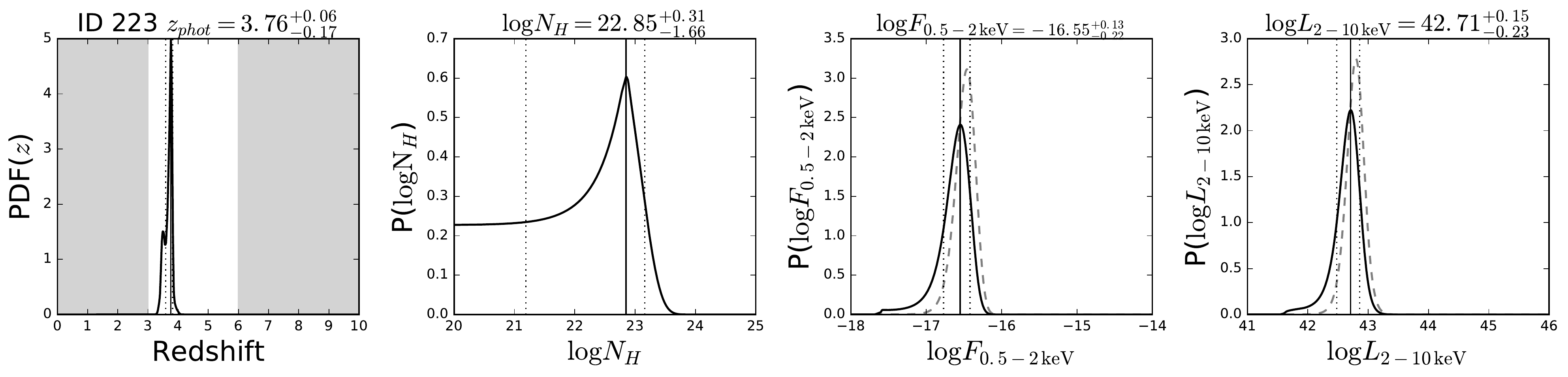}
	\caption{From left to right: probability distribution functions of redshift, column density (Eq.~\ref{NH_int}), flux (Eq.~\ref{P_F}), and luminosity (Eq.~\ref{P_L}) for ID 223. The area not shadowed in the first panel shows the redshift range of this work and the part of the PDF($z$) used to compute the probability distributions of the other parameters. In each panel the solid and dotted vertical lines mark the position of the nominal value and the 68\% confidence interval of the relative parameter. Dashed grey curves in the last two panels are the probability distribution of flux and luminosity if the correction for Eddington bias is not applied in Eq.~\ref{PDF_cr}. Probability distributions for all sources will be made available online. Most of the distributions are well behaved like those shown here, while a few have more complex (i.e., broad or with double peaks) shapes. The complex functions are produced by the combined effects of the measured PDF($z$), and, especially for faint sources, the uncertainties on $N_{\mathrm{H}}$ and flux, and the area-correction (i.e. the $\Omega$) factors.}
	\label{PDF_param}
\end{figure*}

\begin{figure} 
	\centering
	\includegraphics[width=80mm,keepaspectratio]{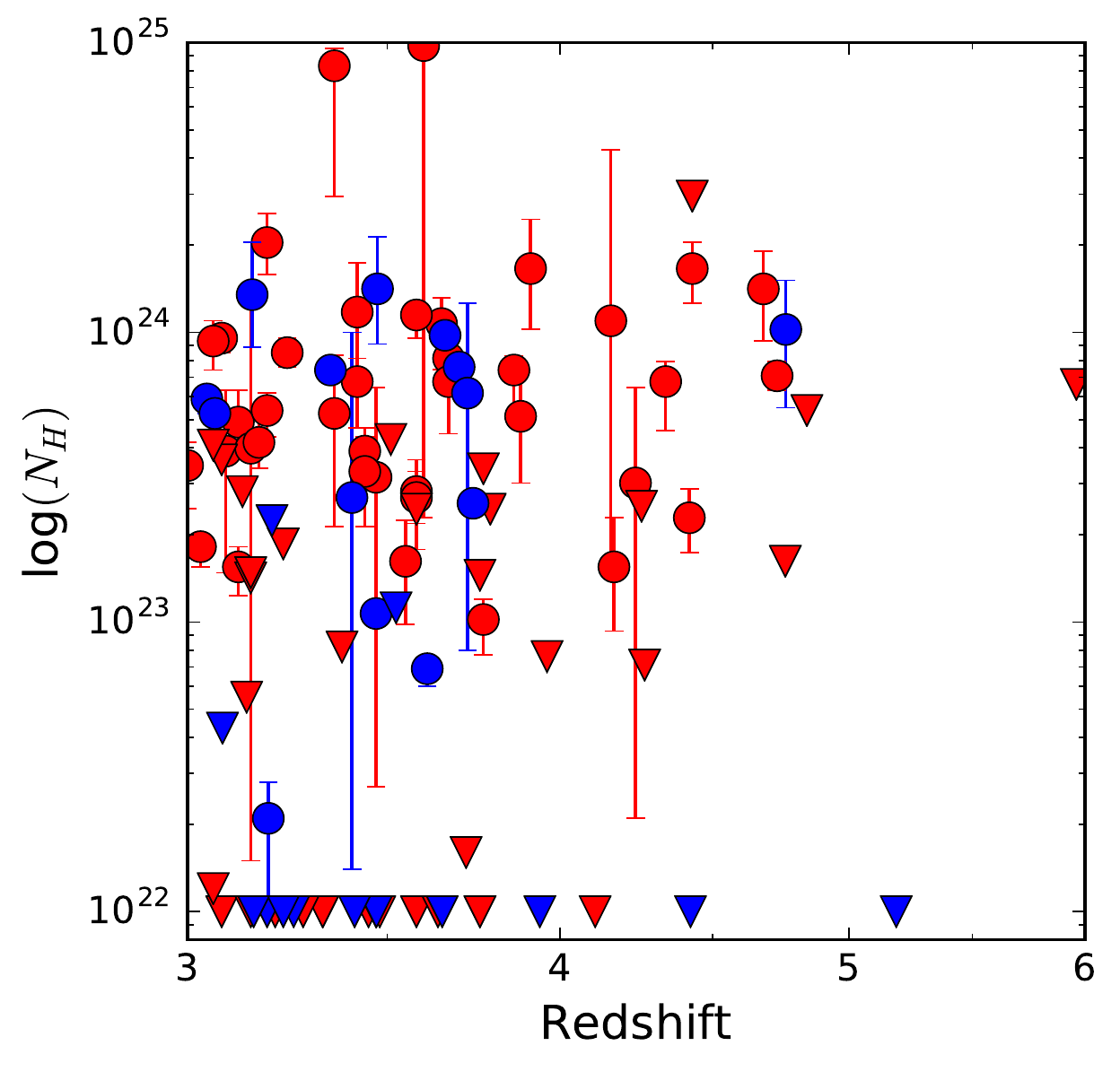}
	\caption{Best-fitting column density versus redshift for sources with spectroscopic (blue symbols) and photometric (red symbols) redshift. Upper limits are displayed as downward-pointing triangles.}
	\label{nh_z}
\end{figure}

  For each source we report in Tab.~\ref{sources} the $\mathrm{log}N_{\mathrm{H}}$ that maximizes Eq.~\ref{NH_int} (i.e., our best estimate of the intrinsic column density) and the 68\% confidence level uncertainties,  corresponding to the narrowest $\mathrm{log}N_{\mathrm{H}}$ interval in which the integrated $P(\mathrm{log}N_{\mathrm{H}})$ is 68\%. 
Fig.~\ref{PDF_param} presents the $\mathrm{PDF}(z)$ and $P(\mathrm{log}N_{\mathrm{H}})$ for a source as an example. The PDF of redshift, column density, flux (see \S~\ref{flux}), and luminosity (see \S~\ref{luminosity}) for every source in the sample are made available with the online version of this paper. Fig.~\ref{nh_z} presents the best-fitting column density plotted against the nominal redshift of each source (i.e., the spectroscopic redshift or the redshift corresponding to the peak of PDF($z$)).

We then derived the intrinsic distribution of column density of our sample as

\begin{equation}\label{NH_distro_eq}
 N^{\mathrm{tot}}(\mathrm{log}N_{\mathrm{H}}) = \sum_{i=1}^{N_{\mathrm{hz}}} P^i(\mathrm{log}N_{\mathrm{H}})\,\Omega^i\,P^i(3\leq z<6),
\end{equation}

where 
\begin{equation}
 \Omega^i=\int_{CR^{\mathrm{intr}}_{\mathrm{lim}}}^\infty\Omega(CR)\, P^i(CR)\, dCR.
\end{equation}
and the sum is performed over the $N_{\mathrm{hz}}=118$ sources which contribute to the high-redshift sample. $CR$ and $CR_{\mathrm{lim}}^{\mathrm{intr}}$ are defined in the soft or hard bands for sources detected in the respective bands, as discussed in \chapt{cr}.
The $P^i(3\leq z<6)$ term in Eq.~\ref{NH_distro_eq} keeps track of the actual probability of the $i$-th source to be at high redshift.

\begin{figure} 
\centering
\includegraphics[width=88mm,keepaspectratio]{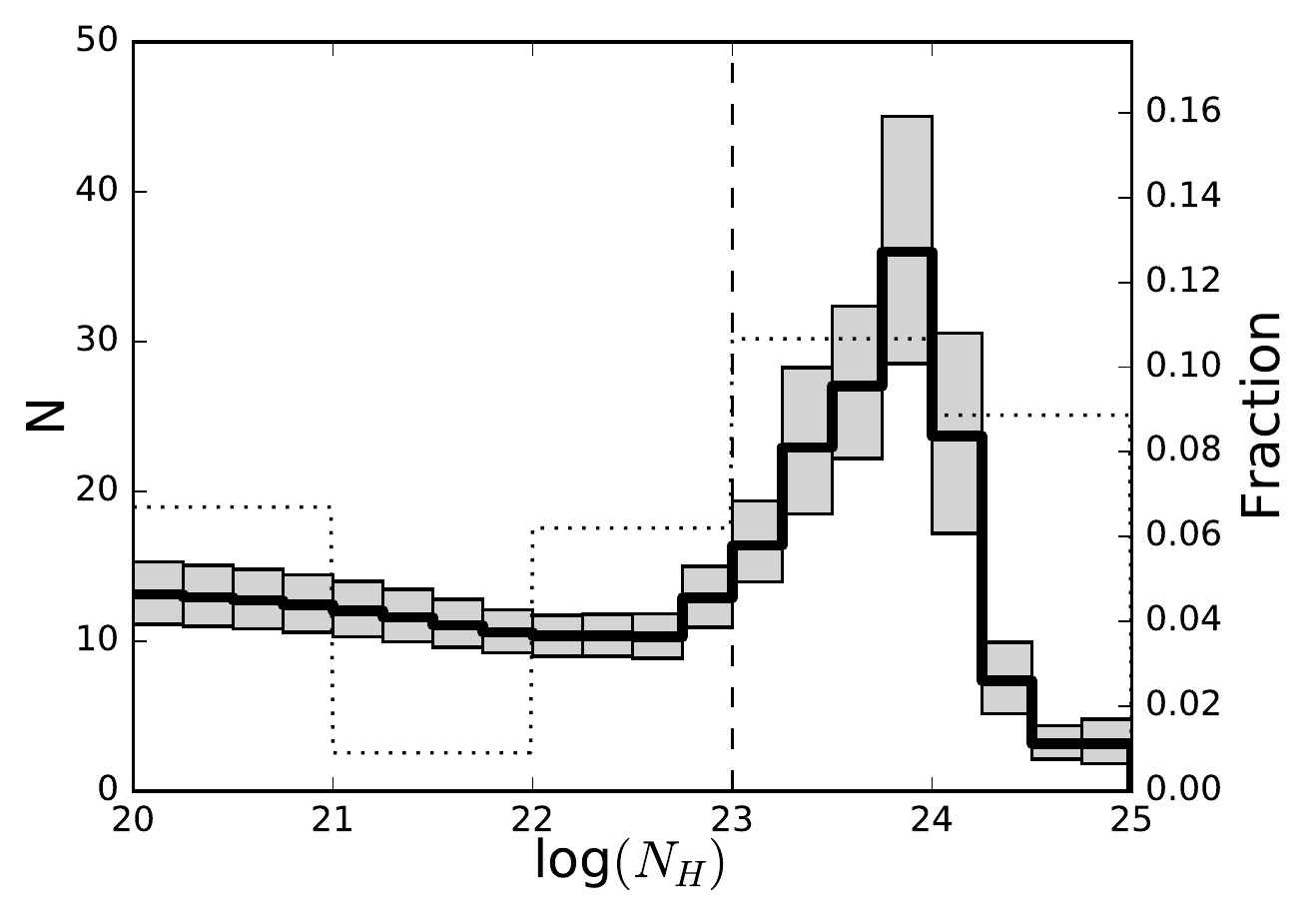}
\caption{Binned intrinsic (i.e., weighted by the redshift probabilty distribution functions and sky-coverages associated with the individual sources) distribution of column density of our sample (Eq.~\ref{NH_distro_eq}). Uncertainties (at 68\% confidence level; grey regions) have been computed through a bootstrap procedure. The vertical dashed line represents the threshold used in this work to separate obscured and unobscured AGN at high redshift. The flatness of the distribution below log$N_H\approx23$ is due to the lack of information in X-ray spectra at high redshift useful to constrain such low levels of obscuration, as discussed in \S~\ref{NH}. The dotted histogram is the column-density distribution assumed by \citet{Gilli07}, which will be used in \chapt{incompl}.}
\label{NH_distro}
\end{figure}

Fig.~\ref{NH_distro} shows the binned $N^{\mathrm{tot}}(\mathrm{log}N_{\mathrm{H}})$. Errors have been computed through a bootstrap procedure: we created a list of $N_{\mathrm{hz}}$ sources randomly chosen among the $N_{\mathrm{hz}}$ sources in our sample, allowing for repetition (i.e., one source can appear in the new list 0, 1, or multiple times). We then recomputed $N^{\mathrm{tot}}(\mathrm{log}N_{\mathrm{H}})$. Repeating this procedure 1000 times, we derived 1000 distributions of $N^{\mathrm{tot}}(\mathrm{log}N_{\mathrm{H}})$ or, equivalently, at each $\mathrm{log}N_{\mathrm{H}}$ we derived 1000 values of probability. The 68\% confidence interval at each $\mathrm{log}N_{\mathrm{H}}$ has been computed by sorting the corresponding 1000 probabilities and selecting the 16th and 84th quantile values. \cite{Liu17} presented an X-ray spectral analysis of the 7 Ms CDF-S sources with $>80$ net counts.
Fig.~\ref{NH_distro} is consistent with the column-density distribution presented in that work in an overlapping redshift bin ($z=2.4-5$), in spite of the different assumed spectral models and performed analysis. Also, the priority we assigned to the different photometric-redshift catalogs is different than that used by \cite{Liu17}, resulting in some sources having different redshifts.

 The flattening at $\mathrm{log}N_{\mathrm{H}}\lesssim23$ is  due to the photoelectric cut-off detection limit. Indeed, the determination of the best-fitting $N_{\mathrm{H}}$ for an X-ray spectrum is largely driven by the detection of the photoelectric cut-off, which is a function of both redshift and column density. In particular, the photoelectric cut-off shifts to higher energies for increasing $N_{\mathrm{H}}$ and to lower energies for increasing redshift. Thus, at high redshift, unless a source is heavily obscured, the photoelectric cut-off is shifted outside the \chandra\, bandpass and, for \mbox{small-to-moderate} numbers of net counts, it is virtually impossible to determine the intrinsic column density. In \cite{Vito13} we estimated the minimum column density which could be constrained for a typical ($\sim100$ net counts) AGN in the 4 Ms CDF-S at $z\sim3.5$ to be $\rmn{log}N_{\mathrm{H}}\approx23$. 
 
 We stress that the flattening at $\mathrm{log}N_{\mathrm{H}}\lesssim23$, due to the lack of information in X-ray spectra useful to constrain such low values of obscuration at high redshift, should not be considered real. We can derive the probability of a source having $\mathrm{log}N_{\mathrm{H}}\lesssim23$, but not the exact probability distribution in column density bins below this threshold, resulting in a flat behavior.
 According to these considerations, we will adopt $\mathrm{log}N_{\mathrm{H}}=23$ as the threshold between obscured and unobscured sources. This choice is further reinforced by the spectral simulations we performed in Appendix~\ref{check_spec_analysis}.

\subsection{Flux distribution and number counts}\label{flux}
The observed flux in one band of a source at a given redshift and characterized by a given intrinsic column density is defined as
\begin{equation}\label{F_X}
 F_{\mathrm{X}}(CR, \mathrm{log}N_{\mathrm{H}},z)=\frac{CR}{ f_{\mathrm{X,CR}}(\mathrm{log}N_{\mathrm{H}},z)}
\end{equation}
where $ f_{\mathrm{X,CR}}(\mathrm{log}N_{\mathrm{H}},z)$, computed with XSPEC, is the conversion factor between count rate and flux, which depends on the response files associated with the source, and its observed-frame spectral shape (i.e., $N_{\mathrm{H}}$ and $z$, given that the photon index is fixed). $CR$ is the count rate in the same energy band.
According to the change of variables in Eq.~\ref{F_X} \citep[e.g.,][]{Bishop06}, we can derive the probability distribution density of flux from the known $P(CR)$ (Eq.~\ref{PDF_cr}):
\begin{equation}\label{change_var}
\begin{split}
	&P(F_{\mathrm{X}}(CR,\mathrm{log}N_{\mathrm{H}},z))=\\ &P(F_{\mathrm{X}}(CR,\mathrm{log}N_{\mathrm{H}},z)  f_{\mathrm{X,CR}}(\mathrm{log}N_{\mathrm{H}},z))  f_{\mathrm{X,CR}}(\mathrm{log}N_{\mathrm{H}},z)=\\
	&P(CR) f_{\mathrm{X,CR}}(\mathrm{log}N_{\mathrm{H}},z),
	\end{split}
\end{equation}
where the probability density in the first term is defined in the flux space, while the probability densities in the second and third term are defined in the count-rate space. The Jacobian of the transformation, $f_{\mathrm{X,CR}}(\mathrm{log}N_{\mathrm{H}},z)$, conserves the integrated probability during the change of variable, i.e.,
\begin{equation}
	\int_{F_{\mathrm{X,1}}}^{F_{\mathrm{X,2}}} P(F_{\mathrm{X}}) \mathrm{d}F_{\mathrm{X}}=\int_{CR_1}^{CR_2} P(CR) f_{\mathrm{X,CR}}\mathrm{d}CR,
\end{equation}
where $F_{\mathrm{X,1}}$ and $F_{\mathrm{X,2}}$ are linked to $CR_1$ and $CR_2$, respectively, through Eq.~\ref{F_X}.

\begin{figure} 
\centering
\includegraphics[width=88mm,keepaspectratio]{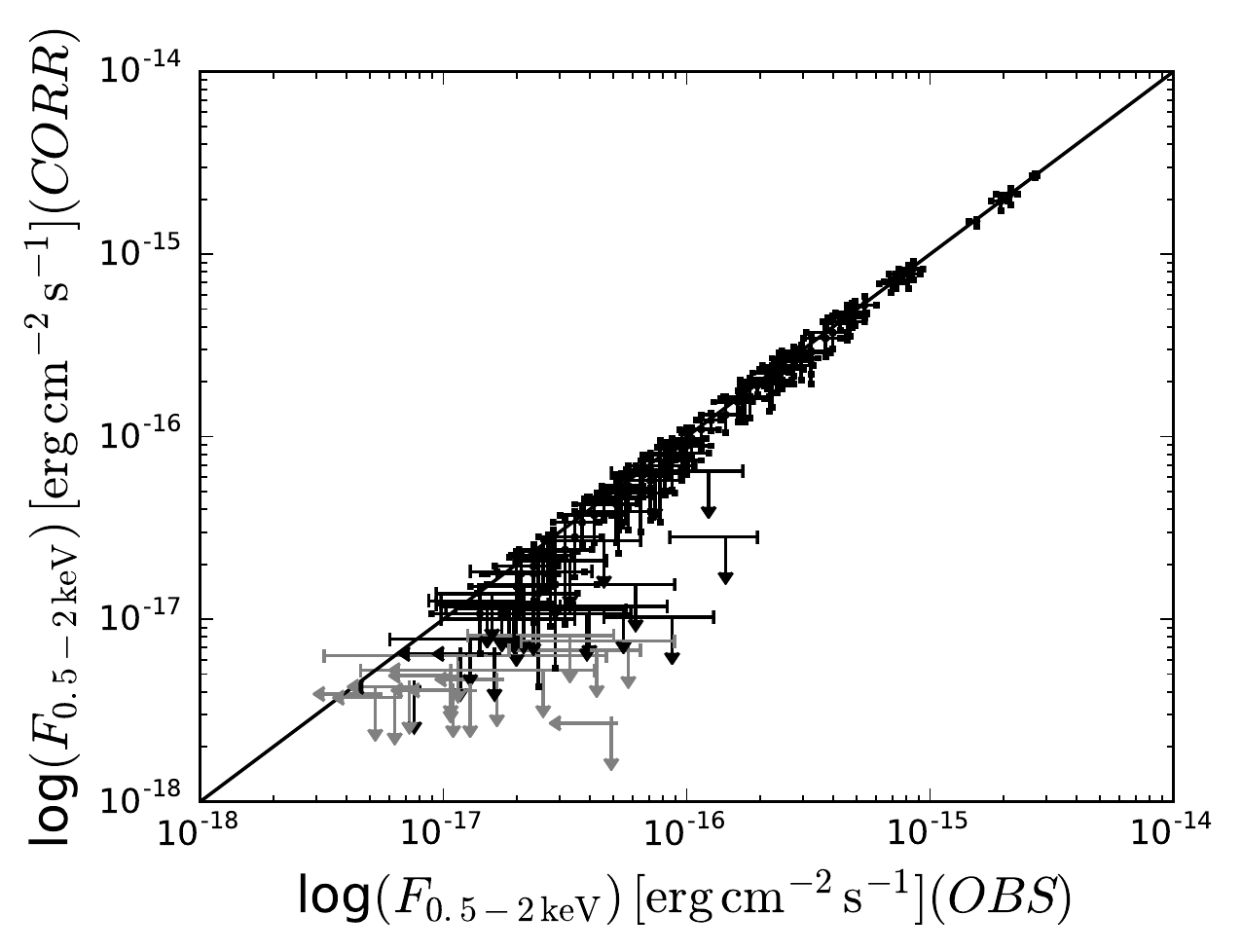}
\caption{Best estimates of the soft-band flux obtained by integrating Eq.~\ref{P_F}, with relative 68\% confidence level uncertainties, corrected for Eddington bias (see Eq.~\ref{PDF_cr}), compared to the uncorrected (i.e. observed) values. Grey symbols represent soft-band undetected sources, which are shown for completeness, for which the count rate (and therefore flux) in the hard band has been used everywhere else in this work.}
\label{F_comp}
\end{figure}

We can therefore define the probability density distribution in the $\left\{  F_{\mathrm{X}},\mathrm{log}N_{\mathrm{H}},z \right\} $ parameter space as
\begin{equation}\label{P_F}
\small
	 P(F_{\mathrm{X}},\mathrm{log}N_{\mathrm{H}},z)= \frac{P(F_{\mathrm{X}}(CR, \mathrm{log}N_{\mathrm{H}},z)) \, P(\mathrm{log}N_{\mathrm{H}}|z) \,  PDF(z) }{P(3\leq z<6)}
\end{equation}
 Eq.~\ref{P_F}  includes the correction for Eddington-bias applied in Eq.~\ref{PDF_cr} and $P(F_{\mathrm{X}},\mathrm{log}N_{\mathrm{H}},z)$ is normalized to unity. 
The flux probability distribution ($P(F_{\mathrm{X}})$) of each source is derived by integrating Eq.~\ref{P_F} over the considered redshift and column density ranges. Fig.~\ref{PDF_param} displays the flux probability distribution for a source as an example.
 Tab.~\ref{sources} reports for each source the soft-band flux value that 
maximizes $P(F_{\mathrm{X}})$ and the 68\% confidence level uncertainties, corresponding to the narrowest interval containing 68\% of the total probability. If less than 68\% of $P(CR)$ of a source lies above the count-rate limit, we report the upper limit on $P(F_{\mathrm{X}})$.
 For consistency, Tab.~\ref{sources} lists the soft-band flux also for sources detected in the hard-band only, for which, however, the hard-band flux has been used to derive the luminosity in \chapt{luminosity}.
 Fig.~\ref{F_comp} presents the $F_{0.5-2\,\mathrm{keV}}$ estimated by applying and not applying the Eddington-bias correction (i.e., the last factor in Eq.~\ref{PDF_cr}), which, as expected, causes the slight bend at faint fluxes in Fig.~\ref{F_comp}.

Considering only soft-band detected sources, we derived the intrinsic soft-band flux distribution (which includes the sky-coverage correction) as

\begin{equation}\label{F_distro_eq}
\small
\begin{split}
& N^{\mathrm{tot}}(F_{0.5-2\,\mathrm{keV}}) = \sum_{i=1}^{N_{\mathrm{hz}}^{\mathrm{sb}}}[P^i(3\leq z<6)\\ & \int_{z=3}^{6}\!\int_{\mathrm{log}N_{\mathrm{H}}=20}^{25}\!\!P^i(F_{\mathrm{X}},\mathrm{log}N_{\mathrm{H}},z)\,\Omega(F_{\mathrm{X}},\mathrm{log}N_{\mathrm{H}},z)\,\mathrm{dlog}N_{\mathrm{H}} \, dz]
 \end{split}
\end{equation}
where $N_{\mathrm{hz}}^{\mathrm{sb}}=108$ is the number of soft-band detected sources contributing to the high-redshift sample (see \chapt{cr}). In Eq.~\ref{F_distro_eq} the weighting factor is expressed as a function of the integrating variables, following standard rules for a change of variables, as:
\begin{equation}\label{omega_F}
\small
\Omega(F_{\mathrm{X}},\mathrm{log}N_{\mathrm{H}},z)=\Omega(F_{\mathrm{X}}(CR, \mathrm{log}N_{\mathrm{H}},z)\,f_{\mathrm{X,CR}}(\mathrm{log}N_{\mathrm{H}},z))=\Omega(CR)
\end{equation}
This change of variables allows us to account for the specific spectral shape of each source, as discussed in \S~\ref{cr}, since the same observed flux can correspond to different count rates for different values of $N_H$ and $z$, which, according to our simple spectral model, describe completely the spectral shape.

\begin{figure} 
	\centering
	\includegraphics[width=88mm,keepaspectratio]{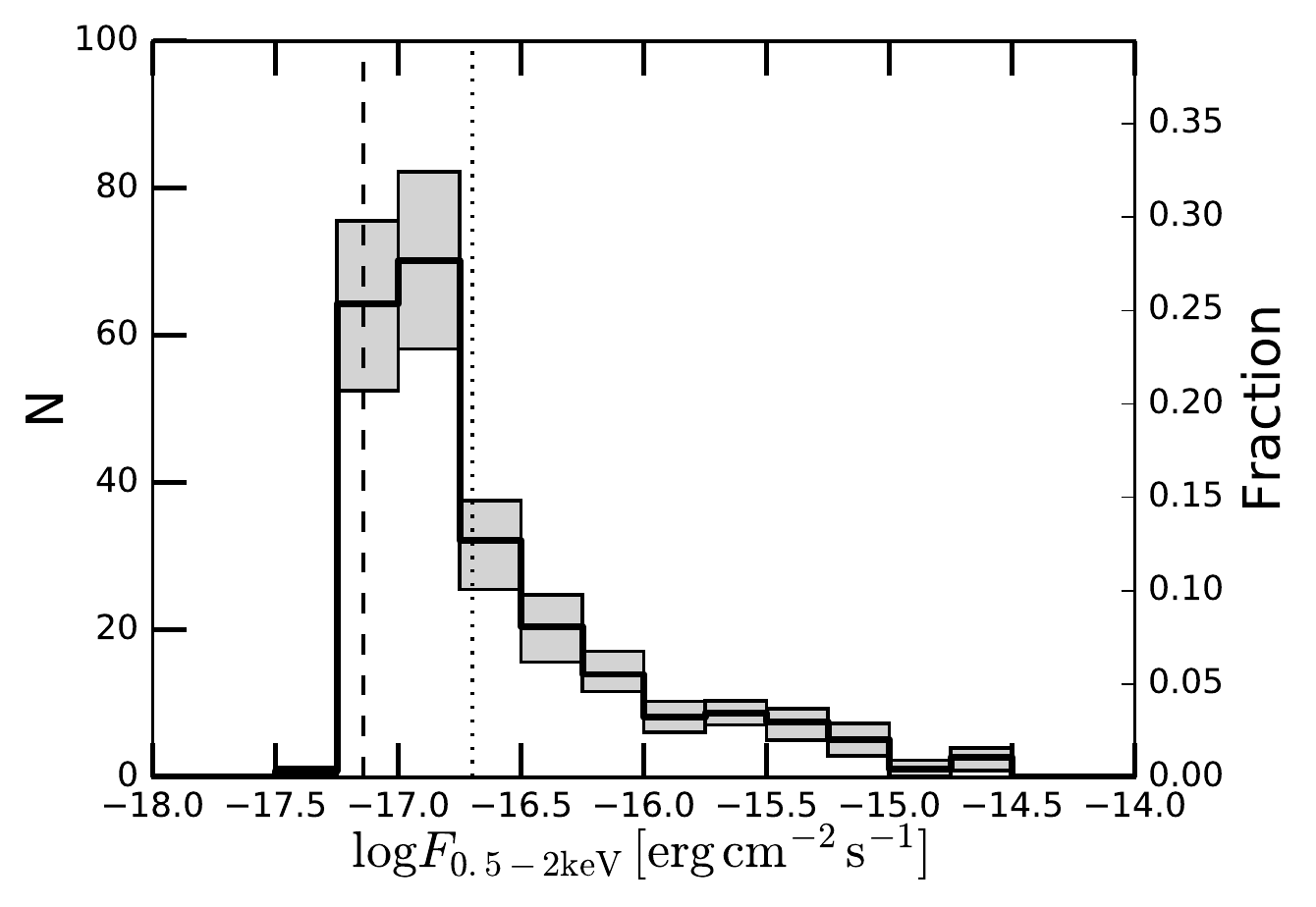}
	\caption{Binned intrinsic (i.e., weighted by the redshift probabilty distribution functions and sky-coverages associated with the individual sources) distribution of flux for our subsample of soft-band detected AGN (Eq.~\ref{F_distro_eq}) and relative uncertainties (at 68\% confidence level; grey bars). The dashed and dotted vertical lines represent the flux limits (assuming $\Gamma=1.4$) of the CDF-S (corresponding to a sensitive area $A=10\,\mathrm{arcmin^2}$, i.e., dashed line in Fig.~\ref{skycov}) and the CDF-N, respectively.}
	\label{F_distro}
\end{figure}

Fig.~\ref{F_distro} shows the binned  $N^{\mathrm{tot}}(F_{0.5-2\,\mathrm{keV}})$  with 68\% confidence intervals derived with a bootstrapping procedure, similarly to what was done in \chapt{NH} and Fig.~\ref{NH_distro}.
 We note that, as we remained in the count-rate space, the distributions extend slightly below the nominal flux-limit, since the assumed count-rate limit corresponds to slightly different fluxes for different observed spectral shapes and response matrices (see \chapt{cr}).

\subsubsection{AGN logN-logS at $3\leq z< 6$}\label{lnls_text}
\begin{figure} 
	\centering
	\includegraphics[width=80mm,keepaspectratio]{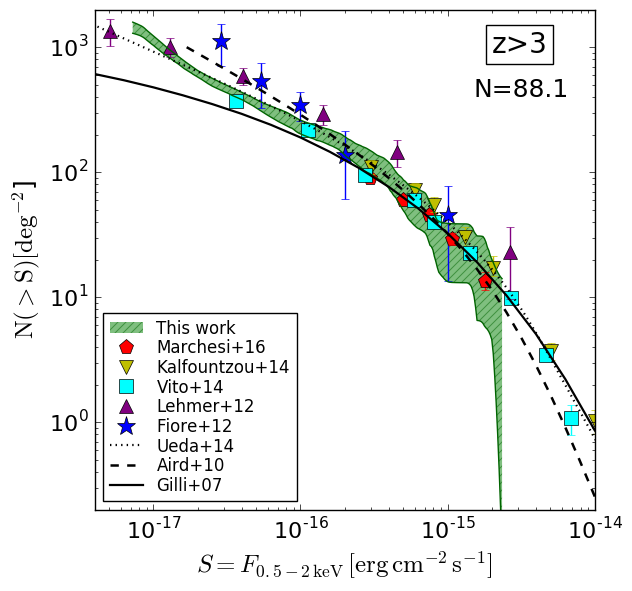}
	\includegraphics[width=80mm,keepaspectratio]{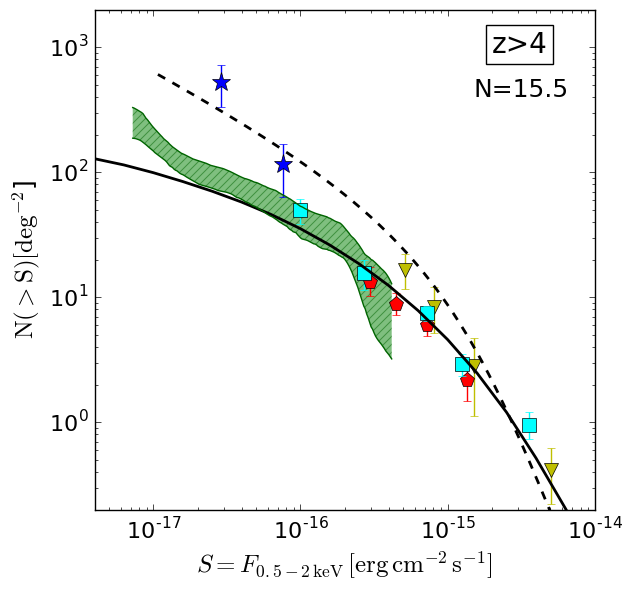}
	\caption{Confidence region (at the 68\% level) of the log$N$-log$S$ of our subsample of soft-band detected, $z>3$ (top panel) and $z>4$ (bottom panel) AGN, compared with results from previous works. The numbers of soft-band detected sources, weighted by the PDF($z$), included in these figures are also reported. }
	\label{lnls}
\end{figure}

The cumulative version of Eq.~\ref{F_distro_eq}, divided by the total area in $\mathrm{deg^2}$, is the log$N$-log$S$ relation of our sample. The 68\% confidence region derived through the bootstrap procedure is compared with results from previous works at $z>3$ and $z>4$ in Fig.~\ref{lnls}. We derived the $z>4$ log$N$-log$S$ by repeating the procedures described in this section using $z=4$ as the lower limit for the considered redshift range.
Bright and rare sources are not sampled properly by the pencil-beam, deep surveys such as those used in this work. Therefore, Fig.~\ref{lnls} displays the curves up to the flux at which the median expected number of sources from the bootstrap analysis is 1 in the area considered in this work ($\sim550\,\mathrm{arcmin^2}$, see Fig.~\ref{skycov}, corresponding to $\sim6.5\,\mathrm{deg^{-2}}$ sources).

Our results are in good agreement with previous measurements at log$F_{0.5-2\,\mathrm{keV}}\gtrsim-16$, which is the flux regime better sampled by wide surveys. At fainter fluxes, where we can exploit the excellent sensitivity of the \textit{Chandra} deep fields, our results are consistent with previous results which used data from the 4 Ms CDF-S \citep{Lehmer12, Vito14} and with the \cite{Ueda14} curve. 
\cite{Lehmer12} presented the number counts derived in the 4 Ms CDF-S down to slightly fainter fluxes than those reached in this work. This is due to the less-conservative approach they used to compute the sensitivity of the 4 Ms CDF-S compared to that used by \cite{Luo17} for the 7 Ms CDF-S, which we adopted here.

At $z>4$ we can push the AGN number counts down to log$F_{0.5-2\,\mathrm{keV}}\approx-17$. While our curve agrees with previous results at bright fluxes and with the \cite{Gilli07} log$N$-log$S$ reasonably well, it excludes very steep number counts such as those reported by \cite{Fiore12} at faint fluxes. The selection in  \cite{Fiore12} made use of pre-determined optical positions of galaxies and adaptive X-ray detection bands, and it is therefore less conservative than a blind X-ray detection. However, about half of the 4 Ms CDF-S sources reported in that work at $z>4$ were not detected even in the deeper 7~Ms exposures, leaving doubts about their detection significance (especially considering that the \cite{Fiore12} analysis is plausibly affected by Eddington bias). This issue, together with the different, more recent photometric redshifts we used, that shift some of the \cite{Fiore12} to lower redshifts, accounts completely for the discrepancies in Fig.~\ref{lnls}.

\begin{figure} 
	\centering
	\includegraphics[width=80mm,keepaspectratio]{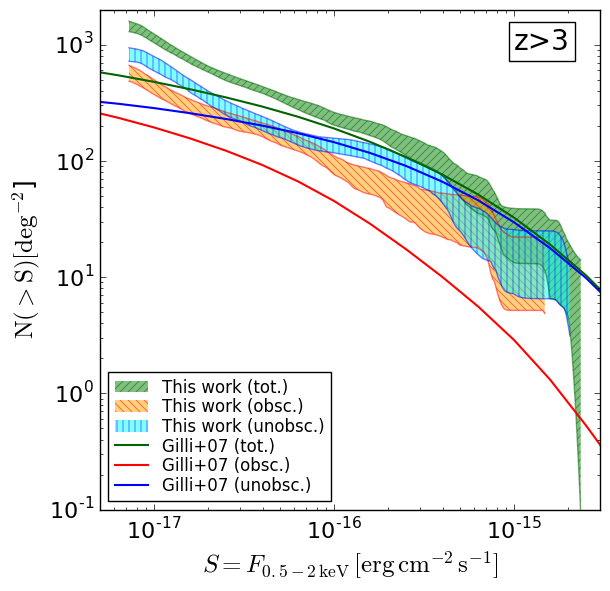}
	\includegraphics[width=80mm,keepaspectratio]{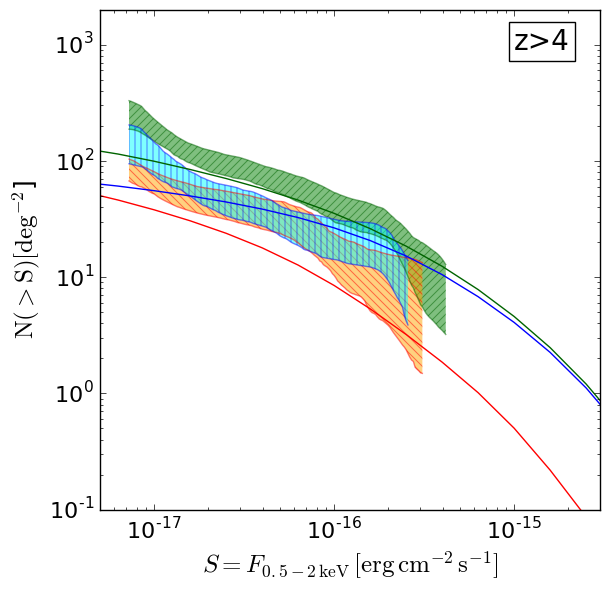}
	\caption{Confidence region (at the 68\% level) of the log$N$-log$S$ of our subsample of soft-band detected, $z>3$ (top panel) and $z>4$ (bottom panel) AGN, for the total sample (green stripes), obscured (orange stripes) and unobscured (cyan stripes) AGN. The results are compared to the expectations of the \citet{Gilli07} X-ray background synthesis model, using the same $N_{\mathrm{H}}$ threshold to define the obscuration-based subsamples.}
	\label{lnls_abs}
\end{figure}

By changing the limits of the integral of column density in Eq.~\ref{F_distro_eq}, we derived separately the log$N$-log$S$ for obscured ($23\leq\mathrm{log}N_{\mathrm{H}}<25$) and unobscured ($20\leq\mathrm{log}N_{\mathrm{H}}<23$) AGN (Fig.~\ref{lnls_abs}).
The \cite{Gilli07} X-ray background synthesis model underestimates the number of obscured AGN at high redshift
 and the steepening of our number counts of unobscured sources at log$F_{0.5-2\,\mathrm{keV}}<-16.5$ (see Fig.~\ref{lnls_abs}). Such steepening could be due to the population of star-forming galaxies, which begin providing a significant contribution to the number counts at the faintest fluxes probed by this work \citep[e.g.][]{Lehmer12}.

\subsection{Luminosity distribution}\label{luminosity}
The intrinsic (i.e., absorption-corrected) rest-frame \mbox{$2-10\,\mathrm{keV}$} luminosity ($L_{\mathrm{X}}$) can be derived from the observed flux as
\begin{equation}
L_{\mathrm{X}}(F_{\mathrm{X}},\mathrm{log}N_{\mathrm{H}},z)=\frac{F_{\mathrm{X}}}{l_{\mathrm{X}}(\mathrm{log}N_{\mathrm{H}},z)}
\end{equation}
where the $l_{\mathrm{X}}(\mathrm{log}N_{\mathrm{H}},z)$ factor, computed with XSPEC for each source, depends on the observed-frame spectral shape (i.e., $N_{\mathrm{H}}$ and $z$, since the photon index is fixed). $F_{\mathrm{X}}$ is the flux in the soft or hard band, as derived in \chapt{flux}, for sources detected in the soft band or only in the hard-band, respectively, as marked in col. 5 of Tab.~\ref{sources}. The $l_{\mathrm{X}}$ factors are computed in both bands and applied accordingly for each source. 

Similarly to \chapt{flux},
we derived for each source the hard-band intrinsic  luminosity probability density distribution as:

\begin{equation}\label{P_L}
P(L_{\mathrm{X}},\mathrm{log}N_{\mathrm{H}},z)=\frac{P(L_{\mathrm{X}}(F_{\mathrm{X}},\mathrm{log}N_{\mathrm{H}},z))}{P^i(3\leq z<6)} 
\end{equation}
where

\begin{equation}
\begin{split}
&P(L_{\mathrm{X}}(F_{\mathrm{X}},\mathrm{log}N_{\mathrm{H}},z))=\\&P(L_{\mathrm{X}}(F_{\mathrm{X}},\mathrm{log}N_{\mathrm{H}},z)\,l_{\mathrm{X}}(\mathrm{log}N_{\mathrm{H}},z))\,{l_{\mathrm{X}}(\mathrm{log}N_{\mathrm{H}},z)}
\end{split}
\end{equation}

$P(L_{\mathrm{X}},\mathrm{log}N_{\mathrm{H}},z)$ is normalized to unity. Eq.~\ref{P_L}  includes the correction for Eddington-bias applied in Eq.\ref{PDF_cr}. However, we checked that neglecting this correction results in a similar luminosity distribution, with a slight decrease in the number of log$L_X<42.5$ sources, balanced by an increase at higher luminosities, as expected.
The luminosity probability distribution ($P(L_{\mathrm{X}})$) of each source is derived by integrating Eq.~\ref{P_L} over the considered redshift and column-density ranges. Fig.~\ref{PDF_param} presents $P(L_{\mathrm{X}})$ for a source as an example.
 Tab.~\ref{sources} reports for each source the luminosity value that 
 maximizes $P(L_{\mathrm{X}})$ (i.e. our best estimate of 
 the luminosity of each source) and the 68\% confidence level uncertainties, corresponding to the narrowest interval containing 68\% of the total probability. 
 If less than 68\% of $P(CR)$ of a source lies above the count-rate limit, we report the upper limit of $P(L_{\mathrm{X}})$.

We derived the luminosity distribution of our sample as

\begin{equation}\label{L_distro_eq}
\small
\begin{split}
 &N^{\mathrm{tot}}(L_{\mathrm{X}}) = \\ &\sum_{i=1}^{N_{\mathrm{hz}}}\int_{z=3}^{6}\!\int_{\mathrm{log}N_{\mathrm{H}}=20}^{25}\!\!P^i(L_{\mathrm{X}},\mathrm{log}N_{\mathrm{H}},z)\,\Omega(L_{\mathrm{X}},\mathrm{log}N_{\mathrm{H}},z)\,\mathrm{dlog}N_{\mathrm{H}} \, dz\\
 \end{split}
\end{equation}
where, similarly to Eq.~\ref{omega_F},
\begin{equation}
\small
\begin{split}
\Omega(L_{\mathrm{X}},\mathrm{log}N_{\mathrm{H}},z)&=\Omega(L_{\mathrm{X}}(F_{\mathrm{X}}, \mathrm{log}N_{\mathrm{H}},z)\,l_{\mathrm{X}}(\mathrm{log}N_{\mathrm{H}},z))=\\  &=\Omega(F_{\mathrm{X}},\mathrm{log}N_{\mathrm{H}},z)=\Omega(CR)
\end{split}
\end{equation}

\begin{figure} 
\centering
\includegraphics[width=88mm,keepaspectratio]{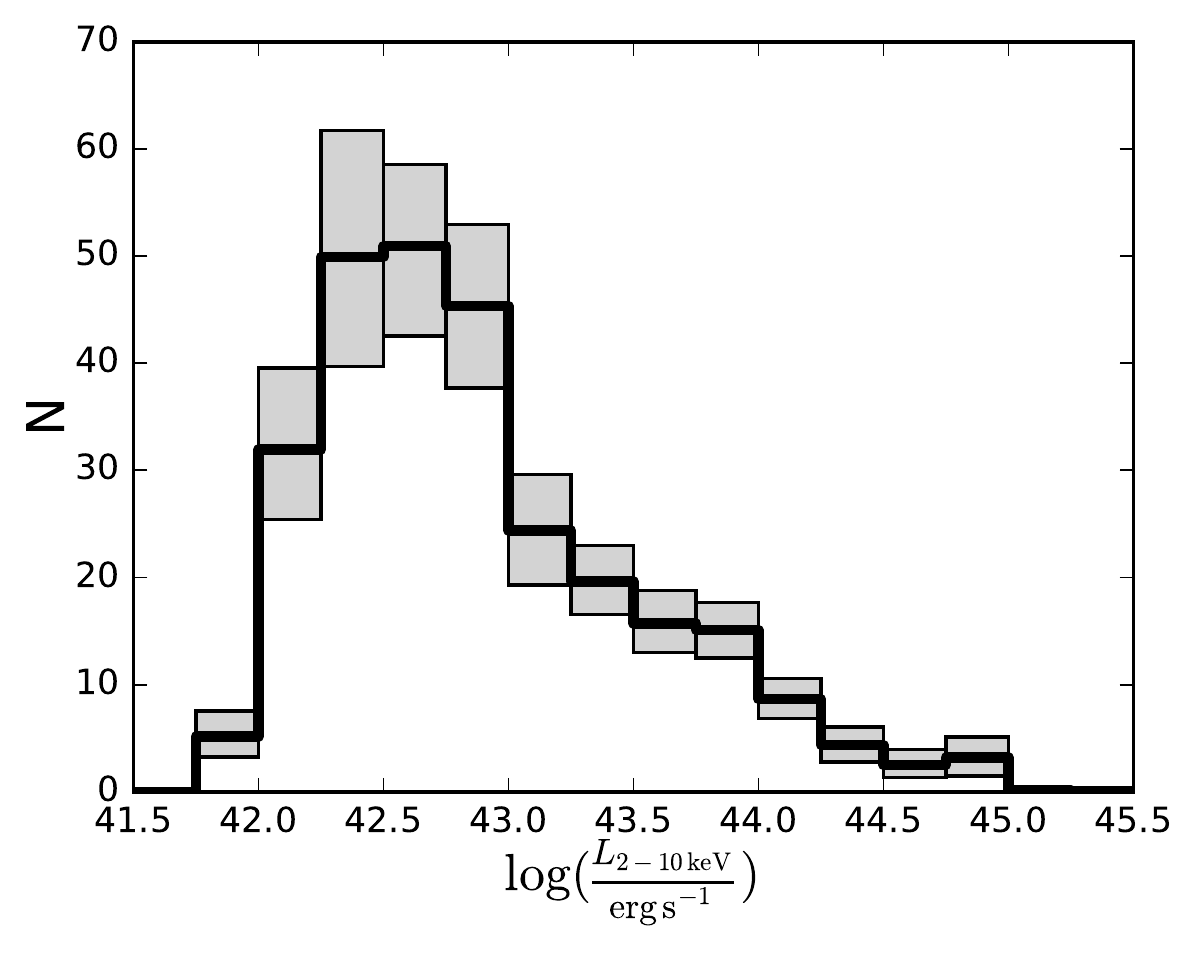}
\caption{Binned intrinsic (i.e., weighted by the redshift probabilty distribution functions and sky-coverages associated with the individual sources) distribution of hard-band luminosity of our sample (Eq.~\ref{L_distro_eq}) and relative uncertainties (at 68\% confidence level, grey bars). }
\label{L_distro}
\end{figure}

Fig.~\ref{L_distro} shows the binned $N^{\mathrm{tot}}(L_{\mathrm{X}})$ with 68\% confidence intervals derived from a bootstrapping procedure, similarly to what was done in \chapt{NH} and Fig.~\ref{NH_distro}.

\section{Obscuration-dependent incompleteness}\label{incompl}

The flux (or count-rate) limit of a survey is effectively a limit in luminosity, redshift, and column density, as these are the physical parameters defining the flux (assuming our simple spectral model). Therefore a low-luminosity AGN at a given redshift can be detected only if its column density is below a certain level, causing an obscuration-dependent incompleteness that affects the low-luminosity regime by favoring the detection of unobscured AGN (see also\S~5.1.2 of \citealt{Liu17}). Such incompleteness usually appears in X-ray studies of flux-limited AGN samples as a lack of detected sources in the low-luminosity and heavy-obscuration region of the $L_X-N_H$ parameter space.

It is worth noting that this effect is not corrected by the $\Omega(CR)$ factor, which compensates for the different areas sensitive to different observed fluxes for \textit{detected} sources, irrespective of their intrinsic properties  (e.g., obscuration). In fact, while representing a potentially significant fraction of the population, low-luminosity and heavily obscured AGN are not detected at all, and are therefore not accounted for when deriving intrinsic properties (e.g., space density) of the low-luminosity population. This effect is particularly important when two obscuration-based subsamples are defined to compute the obscured AGN fraction, $F_{\mathrm{obsc}}$, which would be biased toward lower values, and its trends with redshift and luminosity \citep[e.g.][]{Gilli10b}. Therefore, before deriving $F_{\mathrm{obsc}}$ as a function of redshift and luminosity in \chapt{absfrac}, in this section we assess the effect of the obscuration-dependent incompleteness and derive suitable corrections.

In order to evaluate this bias, we computed the observed count rate (in both the soft and hard bands) corresponding to the parameters $(z, L_{\mathrm{X}}, N_{\mathrm{H}})$:
\begin{equation}
CR(z,L_{\mathrm{X}},N_{\mathrm{H}})=L_{\mathrm{X}}\,l_{\mathrm{X}}(z, N_{\mathrm{H}})\,f_{\mathrm{X,CR}}(z, N_{\mathrm{H}})
\end{equation} 
where the factors $f_{\mathrm{X,CR}}$ and $l_{\mathrm{X}}$ are defined in \chapt{flux} and \chapt{luminosity}, respectively.
We then assigned a value of 1 (detection) or 0 (non-detection) to this triplet of coordinates if the resulting count rate is larger or smaller, respectively, than the count-rate limit in the 7 Ms CDF-S in at least one of the two considered bands. By comparing the observed count rate with the count-rate limit of the survey we are assuming that all the sensitivity dependence on position is properly taken into account by the $\Omega$ factor. 

\begin{figure} 
	\centering
	\includegraphics[width=80mm,keepaspectratio]{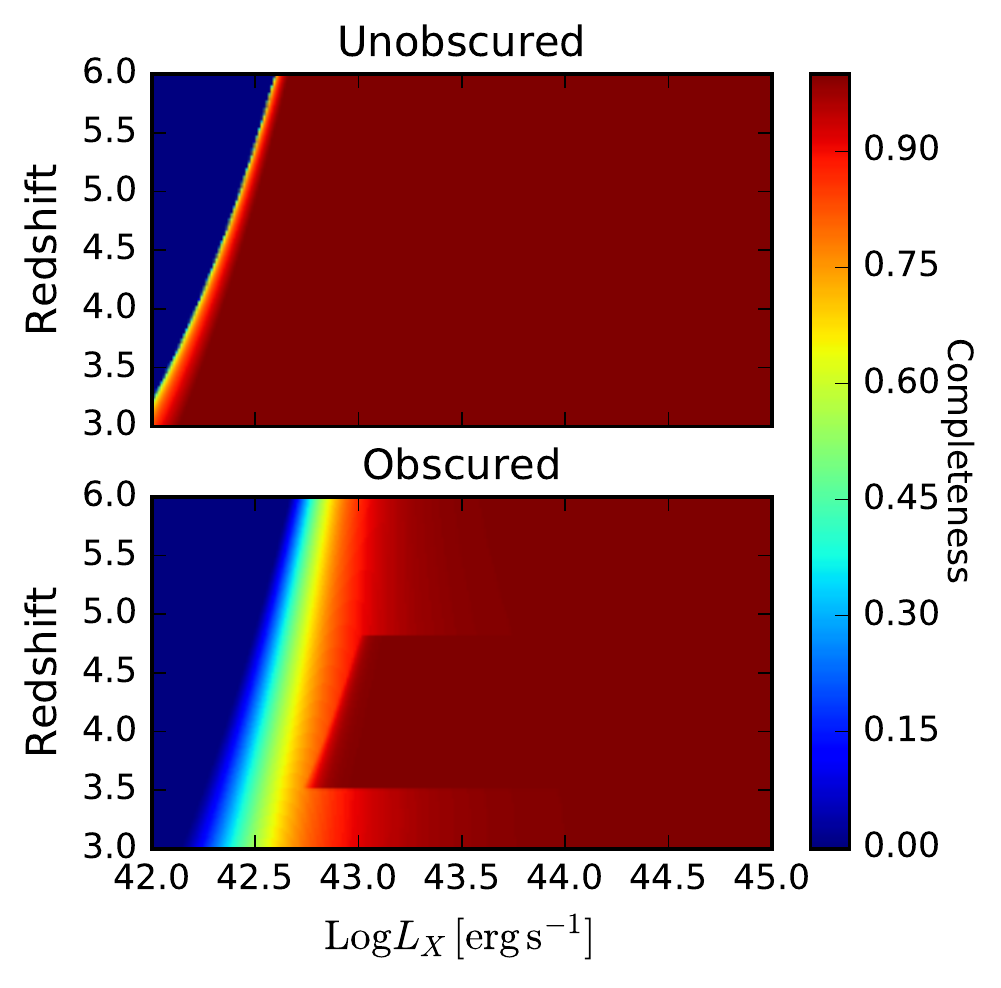}
	\caption{Detection completeness at the 7 Ms CDF-S depth as a function of redshift and luminosity for obscured (lower panel) and unobscured (upper panel) sources. An intrinsic column density distribution similar to Fig.~\ref{NH_distro} is assumed, as described in \chapt{incompl}, where the dark thick stripe at  $\mathrm{log}L_{\mathrm{X}}\sim43-43.5$ and $z\sim3.5-4.5$ is also discussed.}
	\label{z_L_compl}
\end{figure}

The completeness level in the $z-L_{\mathrm{X}}$ plane is shown in Fig.~\ref{z_L_compl} by computing  the mean over the $N_{\mathrm{H}}$ axis separately for $\mathrm{log}N_{\mathrm{H}}=23-25$ and $\mathrm{log}N_{\mathrm{H}}=20-23$, weighted over an assumed intrinsic column-density distribution $P(\mathrm{log}N_{\mathrm{H}})$, which we here considered to be represented by Fig.~\ref{NH_distro}. At $\mathrm{log}L_{\mathrm{X}}\lesssim43$ the two classes of objects are characterized by very different completeness levels: in the upper panel of Fig.~\ref{z_L_compl}, the detection of unobscured sources is complete up to $z\sim6$ for $\mathrm{log}L_{\mathrm{X}}\gtrsim42.5$, while lower-luminosity AGN can be detected only at lower redshifts. Since the photoelectric cut-off for $\mathrm{log}N_{\mathrm{H}}\lesssim23$ shifts close to or below the lower-limit of the \textit{Chandra} bandpass, for unobscured sources the completeness is almost independent of the column-density distribution. For this reason, the transition between 0\% and 100\% completeness is sharp, as it depends only on redshift and luminosity.\footnote{ This result again confirms that at $z>3$ column densities of \mbox{$\mathrm{log}N_{\mathrm{H}}\lesssim23$ } do not sensibly affect X-ray spectra, and therefore cannot be correctly identified. As a consequence, a column density threshold larger than the usual $\mathrm{log}N_{\mathrm{H}}=22$ value must be used to define obscured AGN.} The lower panel shows that for $\mathrm{log}N_{\mathrm{H}}=23-25$ at a given redshift the transition is smoother and occurs at larger luminosities. In fact, contrary to the case of unobscured sources, the flux of an obscured source depends strongly on the particular $N_{\mathrm{H}}$ value: more heavily obscured sources can be detected only at higher luminosities than sources affected by milder obscuration. The darker stripe at $\mathrm{log}L_{\mathrm{X}}\sim43-43.5$ and $z\sim3.5-4.5$ is due to the inclusion of the hard band; heavily-obscured sources in those redshift and luminosity intervals are more easily detected in the hard band than in the soft band.

\begin{figure} 
	\centering
	\includegraphics[width=88mm,keepaspectratio]{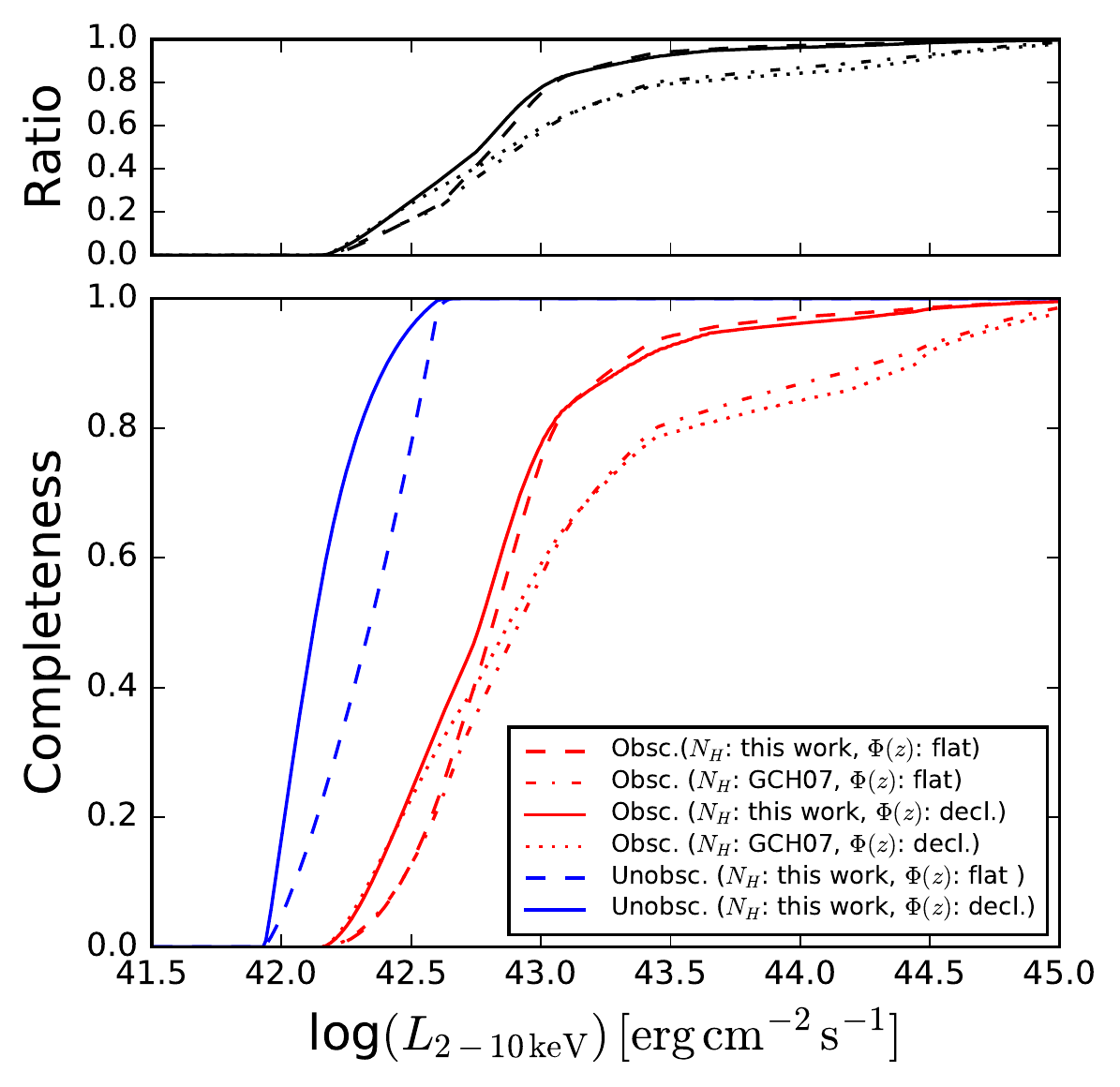}
	\caption{Detection completeness at the 7 Ms CDF-S depth as a function of luminosity for obscured (red lines) and unobscured (blue lines) sources, assuming different intrinsic distributions of column density (the one presented in this work in Fig.~\ref{NH_distro} and the one used in \citealt{Gilli07}) and redshift (flat or more realistically declining with redshift as $\propto10^{-6(1+z)}$; e.g. \citealt{Hiroi12, Vito13}), as described in \chapt{incompl}. For the unobscured case, the curve corresponding to the two column-density distributions are almost indistinguishable and therefore are not plotted for clarity. The solid lines have been used to derive the correction factors applied to Fig.~\ref{absfrac_z_fig} and \ref{absfrac_L_fig}. The upper panel presents the ratio of the completeness of obscured and unobscured sources.}
	\label{L_completeness}
\end{figure}

By projecting Fig.~\ref{z_L_compl} over the luminosity axis (i.e., averaging over the redshift range), we derived the completeness curve as a function of luminosity (solid lines in Fig.~\ref{L_completeness}). The averaging is weighted assuming an intrinsic redshift distribution characterized by a decline in the space density of high-redshift AGN proportional to $(1+z)^{-6}$ \citep[e.g.,][solid lines]{Hiroi12,Vito14}. The unweighted projection corresponds to a flat distribution in redshift and is shown for completeness as dashed lines in Fig.~\ref{L_completeness}. The specific redshift distribution has a small effect on the displayed curves: this behavior is due to the negative curvature of absorbed X-ray spectra, that causes inverted K-corrections at increasing redshift.

To assess the dependence on the particular choice of column-density distribution, we also assumed the one from \cite{Gilli07}, in which, in particular, Compton-thick AGN are more numerous (see Fig.~\ref{NH_distro}). While the completeness curve for unobscured sources is not sensitive to the particular choice of $P(\mathrm{log}N_{\mathrm{H}})$ (we do not show the curve corresponding to the \citealt{Gilli07} case for clarity), the completeness curve for obscured AGN is more severely affected by the \cite{Gilli07} $P(\mathrm{log}N_{\mathrm{H}})$. 

The bias affecting the obscured AGN fraction is linked to the relative completeness levels of the obscured and unobscured subsamples as a function of redshift and luminosity. To visualize this effect, the upper panel in Fig~\ref{L_completeness} show the ratio of the completeness characterizing obscured and unobscured subsamples as a function of luminosity. The ratio varies rapidly from log$L_{\mathrm{X}}=42$ to 43, rising from zero to $\approx0.8$. This is the regime in which the incompleteness effects are strongest: at low intrinsic luminosities the presence or absence of even moderate obscuration is crucial for observing a flux above or below the survey sensitivity limit. At higher luminosities, only the most-obscured systems cannot be detected, and incompleteness is less severe.

\begin{figure} 
	\centering
	\includegraphics[width=88mm,keepaspectratio]{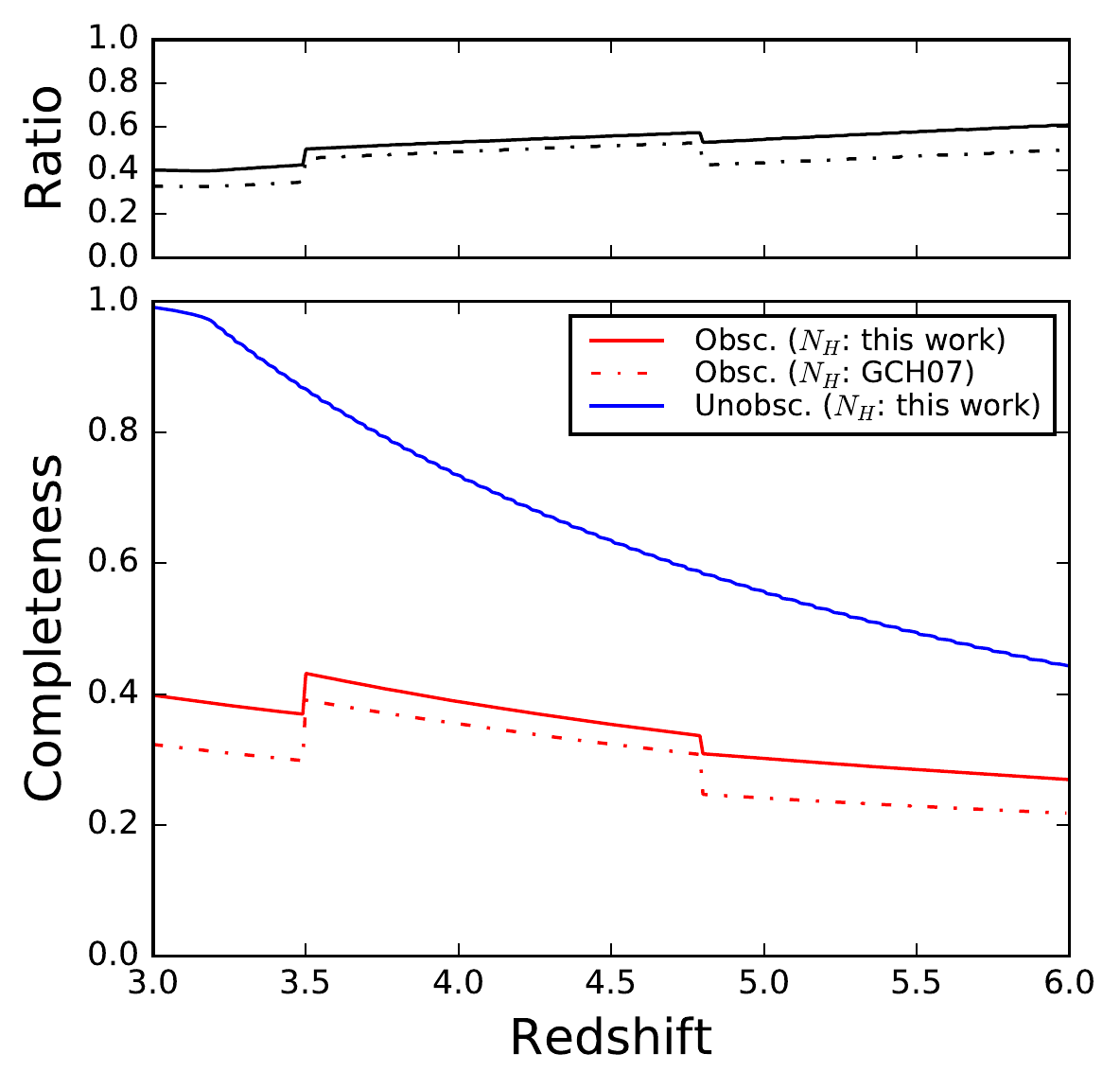}
	\caption{Detection completeness at the 7 Ms CDF-S depth as a function of redshift for obscured (red lines) and unobscured (blue line) sources, assuming different intrinsic distributions of column density and the luminosity function of \citet{Vito14}. The solid line has been used to derive the correction factors applied to Fig.~\ref{absfrac_z_fig} and Fig.~\ref{xlf_abs}. The upper panel shows the ratio of the completeness of obscured and unobscured sources.}
	\label{z_completeness}
\end{figure}

Similarly, we derived the completeness as a function of redshift (Fig.~\ref{z_completeness}) by projecting  Fig.~\ref{z_L_compl} over the redshift axis and assuming an intrinsic luminosity distribution (i.e. luminosity function). We used the pure density evolution model of \cite{Vito14}. 
The upper panel of Fig.~\ref{z_completeness} shows that the relative strength of incompleteness for obscured and unobscured sources does not significantly vary with redshift, but the evolution is slightly stronger for unobscured sources: this behavior is due to the inverse K-correction characterizing obscured X-ray sources, which helps the detection of these systems at higher redshifts.

\section{Obscured AGN fraction}\label{absfrac}
In this section we derive the obscured AGN fraction $F_{\mathrm{obsc}}$ as a function of redshift, flux, and luminosity for our sample of high-redshift AGN. 
We define the obscured AGN fraction as a function of a parameter $x$, where $x$ is redshift, flux, or luminosity in the next sections, as

 \begin{equation}\label{fabs}
  F_{\mathrm{obsc}}(x)=\frac{N_{\mathrm{obsc}}(x)}{N_{\mathrm{tot}}(x)}
 \end{equation}
where $N_{\mathrm{tot}}(x)=N_{\mathrm{obsc}}(x)+N_{\mathrm{unobs}}(x)$ is the total number of observed sources, and $N_{\mathrm{obsc}}(x)$ and $N_{\mathrm{unobs}}(x)$ are the numbers of obscured and unobscured sources, respectively, as a function of $x$. Since all of these are intrinsic numbers, as shown in the following sections, Eq.~\ref{fabs} is equivalent (modulo a volume term) to computing the ratio of space densities of obscured and unobscured AGN, as is usually done to derive $F_{\mathrm{obsc}}$ \citep[e.g.][]{Buchner15, Aird15, Georgakakis15}.

As discussed in \chapt{NH} and following \cite{Vito13,Vito14}, throughout this work we used log$N_{\mathrm{H}}=23$ as the column-density threshold dividing obscured and unobscured AGN at $z>3$. This value is higher than the commonly adopted threshold of log$N_{\mathrm{H}}=22$ in X-ray studies of the AGN population \citep[e.g.,][]{Ueda14,Aird15,Buchner15}, which is also more in agreement with optical classification \citep[e.g.,][]{Merloni14}. However, such low levels of obscuration are extremely difficult to detect at high redshift, where the photoelectric cut-off in X-ray spectra shifts below the energy band probed by \mbox{X-ray} observatories.
The redshifted cut-off, together with the typical limited spectral quality of high-redshift sources, leads to a general overestimate of obscuration for low values of $N_{\mathrm{H}}$, as can be seen in Appendix~\ref{check_spec_analysis}. This effect is especially important when considering the obscured AGN fraction and its trends with other quantities, such as redshift and luminosity.
We therefore prefer to define log$N_{\mathrm{H}}=23$ as the minimum column density of (heavily) obscured AGN at high redshift and discuss the dependence of (heavily) obscured AGN on redshift and luminosity. Unfortunately, this choice complicates the comparison with previous works utilizing more standard definitions. 
In Appendix~\ref{nominal_parameters} 
we show that neglecting the full probability distributions of redshift and spectral parameters would lead to different results than those presented in the following subsections.

\subsection{Obscured AGN fraction versus redshift}\label{absfrac_z}
\begin{figure} 
\centering
\includegraphics[width=88mm,keepaspectratio]{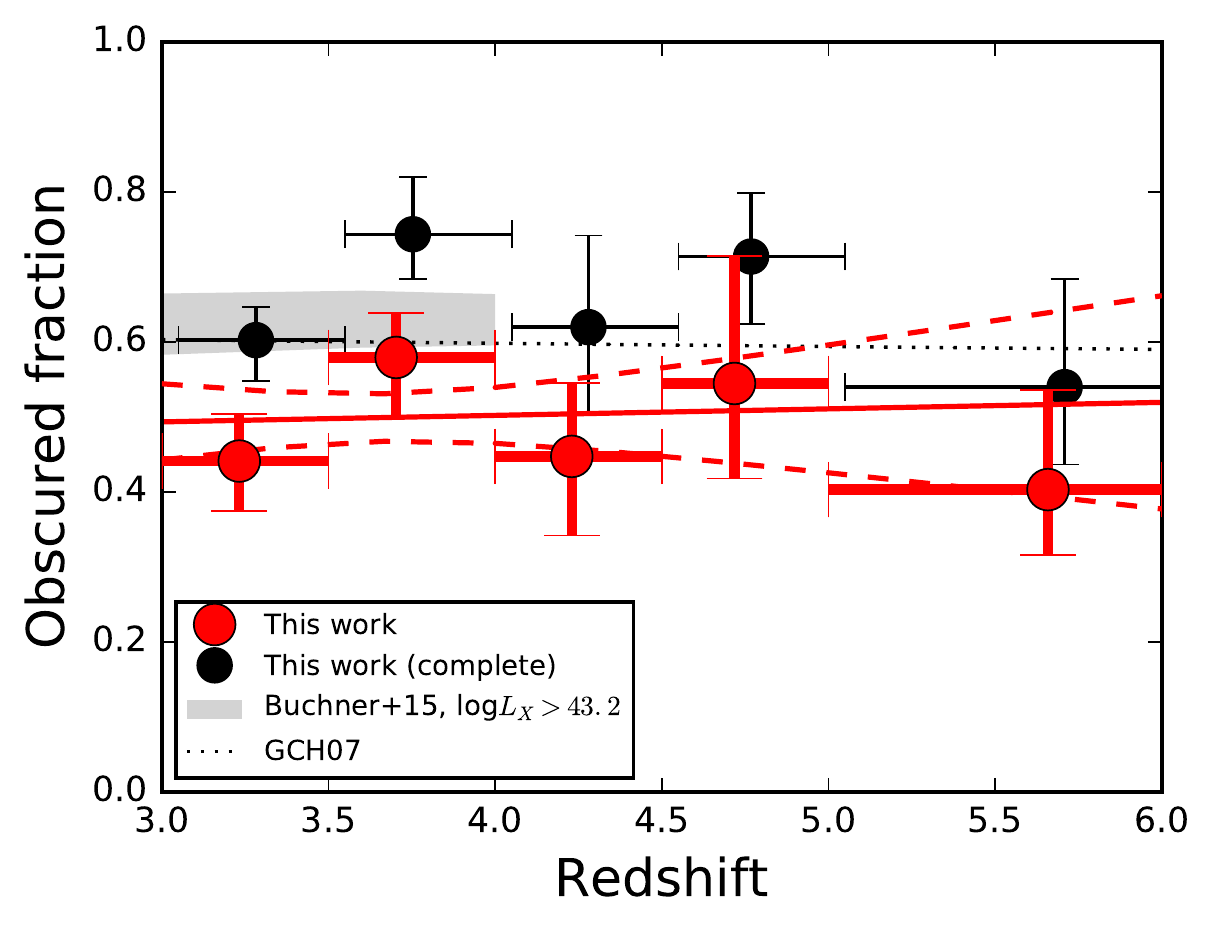}
\caption{Binned estimates of the obscured AGN fraction as a function of redshift ($F_{\mathrm{obsc}}(z)$; red circles). The red solid and dashed lines show the best-fitting linear model and 68\% confidence level region, respectively. The continuous line is not the fit to the points: line and points are different estimates of the obscured AGN fraction as a function of redshift derived using different (parametric and non-parametric) methods. Grey regions encompass the $F_{\mathrm{obsc}}(z)$ from \citet{Buchner15} for luminous AGN. Black circles, slightly shifted for visual purposes, indicate the non-parametric estimate of  $F_{\mathrm{obsc}}(z)$ where a correction for the obscuration-dependent incompleteness has been applied (see \chapt{incompl}). The dotted line is the predictions from the X-ray background synthesis model of \citet{Gilli07}}
	
\label{absfrac_z_fig}
\end{figure}

For each source $i$ in the high-redshift sample, we define
\begin{equation}\label{N_obsc_z}
\small
 \begin{split}N_{\mathrm{obsc}}(z)=\sum_{i=1}^{N_{\mathrm{hz}}}\int_{23}^{25}P_i(\mathrm{log}N_{\mathrm{H}}|z)\, PDF_i(z) \, \Omega^i \, d\mathrm{log}N_{\mathrm{H}}\\
  N_{\mathrm{unobs}}(z)=\sum_{i=1}^{N_{\mathrm{hz}}}\int_{20}^{23}P_i(\mathrm{log}N_{\mathrm{H}}|z)\, PDF_i(z) \, \Omega^i \, d\mathrm{log}N_{\mathrm{H}}
 \end{split}
\end{equation}
Red circles in Fig.~\ref{absfrac_z_fig} indicate the obscured AGN fraction as a function of $z$ in five redshift bins with errors computed through a bootstrapping procedure. We also report the results by \cite{Buchner15}, derived from their space densities of luminous (log$L_{\mathrm{X}}>43.2$) AGN with column densities log$N_{\mathrm{H}}=20-23$ and log$N_{\mathrm{H}}=23-26$, and show the predictions of the X-ray background synthesis model of \cite{Gilli07}. 
 
To derive an independent and parametric estimate of the dependency of $F_{\mathrm{obsc}}$ on redshift,
we performed a Bayesian analysis\footnote{We used the EMCEE Python package (http://dan.iel.fm/emcee/current/), which implements the Affine Invariant Markov chain Monte Carlo (MCMC) Ensemble sampler by \cite{Goodman10}.} of $F_{\mathrm{obsc}}(z)$, assuming  a linear model of the form \mbox{$F^{\mathrm{m}}_{\mathrm{obsc}}=q_{z} + m_{z}(z-3)$}.
We applied a flat prior to $q_{z}$ and a minimally informative prior to $m_{z}$, such that $P(m_{z})\propto(1+m_{z}^2)^{3/2}$ (e.g., \citealt{Vanderplas14} and references therein). The latter is preferred over a flat prior on $\beta$, which would weight steep slopes more.\footnote{This effect can be visualized considering that, while $\beta$ can be any real number, half of the plane is covered by $-1<\beta<1$. Giving equal weight to values in and out of this range would in principle preferentially select $|\beta|>1$, hence steep slopes. However, we checked that using a flat prior on $m_{z}$ does not affect significantly the results.} We also associated a null probability to parameter pairs resulting in unphysical $F_{\mathrm{obsc}}$ values (i.e., negative or larger than unity) in the redshift interval probed by the data.

By construction, our sources can be assigned to one of the two obscuration-defined classes (obscured or unobscured AGN) only with a certain probability, since a probability distribution of $N_{\mathrm{H}}$ is associated with each source. The outcome of this approach is that the number of sources in the two classes (i.e.,  the number of ``successes" and ``failures") is fractional (see Eq.~\ref{N_obsc_z}). Therefore, the use of the Binomial distribution for describing the data likelihood in this case is not formally correct. However, we can use the Beta distribution

\begin{equation}\label{beta_distro}
B(x; \alpha, \beta)\propto x^{\alpha-1}(1-x)^{\beta-1}
\end{equation}

\noindent as a formally correct expression of the data likelihood in one redshift bin, with $\alpha-1=N_{\mathrm{obsc}}(z)$, $\beta-1=N_{\mathrm{unobs}}(z)$, and $x=F^{\mathrm{m}}_{\mathrm{obsc}}(z)$.\footnote{An assessment of the covariance between $z$ and $L_X$ is discussed in \S~\ref{absfrac_l}.}
The total likelihood of the data under a particular set of model parameters is therefore derived multiplying Eq.~\ref{beta_distro} over the entire redshift range.

The resulting best-fitting values are $q_{z}=0.50_{-0.05}^{+0.05}$ and $m_{z}=0.00_{-0.06}^{+0.06}$.
The red dashed lines in Fig.~\ref{absfrac_z_fig} enclose the $68\%$ confidence interval of $10^6$ MCMC realizations derived using this parametric approach, and the solid line is the median value. The parametric and non-parametric representations, derived independently, agree well, showing the robustness of our approach.
The observed obscured AGN fraction is found in Fig.~\ref{absfrac_z_fig} to be flat at $z=3-5$ at $F_{\mathrm{obsc}}\approx0.5$.
$F_{\mathrm{obsc}}$ is usually found to increase with redshift up to $z\approx2-2.5$, and then saturates at higher redshifts (e.g., \citealt{Treister06}, \citealt{Hasinger08}, \citealt{Iwasawa12}, \citealt{Buchner15}, \citealt{Liu17}).
We probe such behavior up to $z\sim6$. 
Black circles in Fig.~\ref{absfrac_z_fig} are derived by correcting the red points using the solid lines in Fig.~\ref{z_completeness}, causing $F_{\mathrm{obsc}}(z)$ generally to increase slightly, with no strong dependence on redshift, to values $\approx0.65$, consistent with the results by \citealt{Buchner15} at high luminosities (log$L_\mathrm{X}>43.2$) up to $z\sim4$. We could push the investigation of $F_{\mathrm{obsc}}(z)$ down to lower luminosities.

\subsection{Obscured AGN fraction vs soft-band flux}\label{absfrac_f}
Considering only the soft-band detected sources, we defined
$N_{\mathrm{unobs}}(F_{\mathrm{X}})$ and $N_{\mathrm{obsc}}(F_{\mathrm{X}})$ from Eq.~\ref{F_distro_eq} by limiting the integral over column density to the ranges $\mathrm{log}N_{\mathrm{H}}=20-23$ and $23-25$, respectively, and summing only the soft-band detected sources.

\begin{figure} 
\centering
\includegraphics[width=88mm,keepaspectratio]{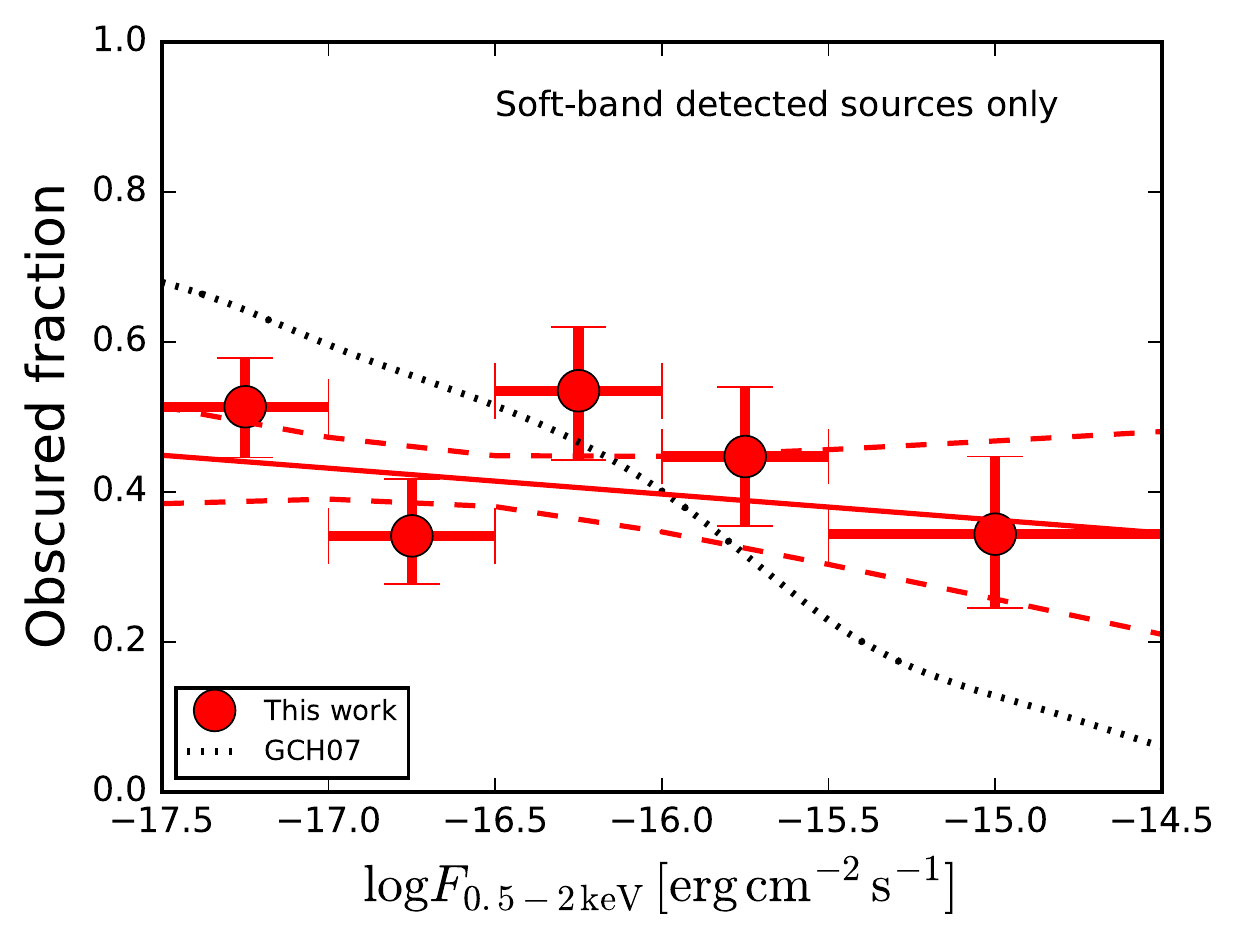}
\caption{Binned estimates of the obscured, soft-band detected AGN fraction as function of flux (red circles).  Red continuous and dashed lines are the best-fitting linear model and 68\% confidence level region, respectively. The prediction of the \citet{Gilli07} X-ray background synthesis model are indicated by the dotted black line.}
\label{absfrac_F_fig}
\end{figure}

Similar to the process described in \chapt{absfrac_z}, we derived the parametric (assuming a linear model \mbox{$F_{\mathrm{obsc}}=q_{F_{\mathrm{X}}} + m_{F_{\mathrm{X}}}(\mathrm{log}F_{\mathrm{X}}-(-17))$} and non-parametric obscured AGN fraction as a function of flux. The results are shown in Fig.~\ref{absfrac_F_fig} and compared with the predictions of the \cite{Gilli07} X-ray background synthesis model. The resulting best-fitting values are $q_{F_{\mathrm{X}}}=0.43_{-0.04}^{+0.04}$ and $m_{F_{\mathrm{X}}}=-0.04_{-0.06}^{+0.06}$.

\subsection{Obscured AGN fraction vs luminosity}\label{absfrac_l}
\begin{figure*} 
	\centering
	\includegraphics[width=120mm,keepaspectratio]{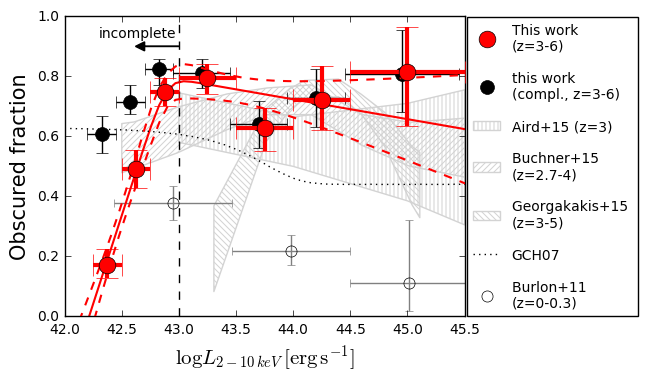}
	\caption{Binned estimates of the obscured AGN fraction as a function of intrinsic hard-band luminosity($F_{\mathrm{obsc}}(L_{\mathrm{X}})$, red circles). Solid and dashed lines represent the best-fitting model (as defined in Eq.~\ref{Fobs_L_model}) and 68\% confidence level region, respectively, as derived through a Bayesian analysis. Grey dashed regions encompass the $F_{\mathrm{obsc}}(L_{\mathrm{X}})$ derived by \citet{Aird15}, \citet{Buchner15}, and \citet{Georgakakis15}, as labelled in the figure. The curves are plotted for slightly different redshift ranges, as reported in the legend. Obscured AGN are defined as those with log$N_{\mathrm{H}}=22-24$ in \citet{Aird15} and \citet{Buchner15}, while we require log$N_{\mathrm{H}}>23$. \citet{Georgakakis15} derived their obscured AGN fraction by comparing X-ray and UV luminosity functions. The dotted black line indicates the prediction of the X-ray background synthesis model of \citet{Gilli07}. We also plot the results derived by \citet{Burlon11} in the local Universe as grey empty circles to show the strong evolution of $F_{\mathrm{obsc}}$ from low to high redshift. The \citet{Burlon11} points use the same column-density threshold as we assumed to define obscured and unobscured sources, and are therefore directly comparable with our points. Black circles (slightly shifted for visual purposes) show the non-parametric estimate of  $F_{\mathrm{obsc}}(L_{\mathrm{X}})$ where a correction for the obscuration-dependent incompleteness has been applied (see \chapt{incompl}).}
	\label{absfrac_L_fig}
\end{figure*}

We defined $N_{\mathrm{unobs}}(L_{\mathrm{X}})$ and $N_{\mathrm{obsc}}(L_{\mathrm{X}})$ from Eq.~\ref{L_distro_eq} by limiting the integral over column density to the ranges $\mathrm{log}N_{\mathrm{H}}=20-23$ and $23-25$, respectively.
Fig.~\ref{absfrac_L_fig} presents the obscured AGN fraction as a function of intrinsic, $2-10$ keV luminosity with errors computed with a bootstrapping procedure (red symbols). As this non-parametric description cannot be well parametrized by a simple linear model, in this section we allowed the slope to change above a characteristic luminosity, i.e.:
\begin{equation}\label{Fobs_L_model}
 F_{\mathrm{obsc}}(L_{\mathrm{X}})=\begin{cases}q_L+ m_L(\mathrm{log}L-\mathrm{log}L_*)& \mathrm{if}\, \mathrm{log}L\leq \mathrm{log}L_*\\
 q_L+ n_L(\mathrm{log}L-\mathrm{log}L_*)& \mathrm{if} \,\mathrm{log}L>\mathrm{log}L_*\\
 \end{cases}
\end{equation}
where all the parameters ($q_L$, $m_L$, $n_L$ and $\mathrm{log}L_*$) are allowed to vary. The best-fitting model derived with the Bayesian analysis is shown in Fig.~\ref{absfrac_L_fig}.  The most noticeable feature is the drop of $F_{\mathrm{obsc}}(L_{\mathrm{X}})$ at low luminosities. However, this trend can be ascribed to a selection effect described in detail in \chapt{incompl}, where tentative corrections are also provided. The best-fitting values describing the relation at high-luminosities are $q_L=0.79_{-0.06}^{+0.06}$, $m_L=-0.06_{-0.09}^{+0..08}$, and $\mathrm{log}L_*=42.89_{-0.09}^{+0.09}$.

We compare the high fraction of heavily obscured AGN ($F_{\mathrm{obsc}}\approx0.7-0.8$) we derived at high-luminosities ($\mathrm{log}L_{\mathrm{X}}>43$) with previous findings by \cite{Aird15} and \cite{Buchner15}, where obscured AGN are defined in a more ``standard" way as those characterized by  log$N_{\mathrm{H}}=22-24$.
The column-density threshold used to separate obscured and unobscured AGN in their works, log$N_{\mathrm{H}}=22$, while useful for comparison purposes with results at lower redshift, is not suitable at high redshift, where the photoelectric cut-off shifts at low energies, close to or even below the lower energy boundary of the \textit{Chandra} bandpass and is therefore poorly constrained, as discussed in \chapt{NH_distro} and Appendix~\ref{check_spec_analysis}. For completeness, we also show the results of \cite{Georgakakis15}, where the obscured AGN fraction is derived by comparing X-ray and UV luminosity functions, and therefore is not directly comparable to the other purely X-ray defined curves.

While at log$L_{\mathrm{X}}>43$ incompleteness effects are negligible, they dominate the observed trend of $F_{\mathrm{obsc}}$ at lower luminosities in Fig.~\ref{absfrac_L_fig}. 
Applying the corrections discussed in \chapt{incompl} results in values, represented by the black circles in Fig.~\ref{absfrac_L_fig}, consistent with those at log$L_{\mathrm{X}}>43$ (i.e., $F_{\mathrm{obsc}}\approx0.7-0.8$), although  at log$L_{\mathrm{X}}<43$
a slightly decreasing trend of  $F_{\mathrm{obsc}}$, down to $\approx0.6$, with decreasing luminosity remains visible.
However, we caution that the applied corrections depend quite strongly on the particular intrinsic column density and luminosity distributions. Also, the possible detection of star-forming galaxies, which have typically steep spectra, at the lowest luminosities probed by this work (log$L_\mathrm{X}\approx42-42.5$) may decrease $F_{\mathrm{obsc}}$ in such regime. We therefore do not consider such trend to be significant.

The curve of \cite{Buchner15} appears to suffer less from similar issues at low luminosities, although they reported a decrease of $F_{\mathrm{obsc}}(L_{\mathrm{X}})$ at low luminosities. The difference with our observed result is probably due to the different procedure used; \cite{Buchner15} applied a Bayesian procedure which disfavors strong variations of $F_{\mathrm{obsc}}$ over close redshift and luminosity bins. At high redshift, the faint regime is not well sampled, probably causing the obscured AGN fraction at low luminosities to be  dominated by the priors, i.e., to be similar to the value at lower redshift and/or higher luminosity, where incompleteness issues are less severe, alleviating this issue. The slight decrease of $F_{\mathrm{obsc}}$ toward low luminosity, which they ascribed to different possible physical mechanisms, may be at least partly due to incompleteness effects.

A strong anti-correlation between $F_{\mathrm{obsc}}$ and luminosity is usually found at low redshifts. For instance, grey empty symbols in Fig.~\ref{absfrac_L_fig} represent the fraction of AGN with log$N_{\mathrm{H}}>23$ derived by \citealt{Burlon11} at $z<0.3$. This behavior appears not to hold at high-redshift, or at least to be much less evident.
Comparing our points with the  \citealt{Burlon11}, we note the positive evolution of $F_{\mathrm{obsc}}$ from the local universe to high redshift, which is stronger at high luminosities.
In \cite{Vito14}, where we derived similar results from a combination of different X-ray surveys, we ascribed the larger fraction of luminous obscured AGN at high redshift than at low redshift to the larger gas fractions of galaxies at earlier cosmic epochs \citep[e.g.][]{Carilli13}, which can cause larger covering factors and/or longer obscuration phases. This description is especially true if luminous AGN are preferentially triggered by wet-merger episodes \citep[e.g.][]{DiMatteo05, Menci08, Treister12}, whose rate is expected to be higher in the early universe. In this case, chaotic accretion of gas onto the SMBH can produce large covering factors.

Fig.~\ref{absfrac_z_fig} and ~\ref{absfrac_L_fig} do not account for the covariance between the $F_{\mathrm{obsc}}-z$ and $F_{\mathrm{obsc}}-L_{\mathrm{X}}$ trends, due to our dataset being flux-limited. This could in principle bias the results when investigating $F_{\mathrm{obsc}}$ separately as a function of one of the two parameters.
To check for this possible effect, we derived the trend of $F_{\mathrm{obsc}}$ with redshift separately for two luminosity bins, log$L_{\mathrm{X}}=43-43.5$ and $43.5-45$, chosen not to suffer significantly from incompleteness and to include approximately the same number of sources. For both subsamples $F_{\mathrm{obsc}}$ is consistent with being flat at similar values, reassuring us that this result does not strongly depend on the intrinsic trend with redshift. We also repeated the procedure dividing the sample in two redshift bins and deriving the trend of $F_{\mathrm{obsc}}$ with luminosity in each of them. The results are consistent with those reported above in this section within the (larger) uncertainties.

\section{Evolution of the high-redshift AGN population}\label{xlf_sect}
\subsection{The AGN XLF at high redshift}
\begin{figure*} 
 	\centering
 	\includegraphics[width=88mm,keepaspectratio]{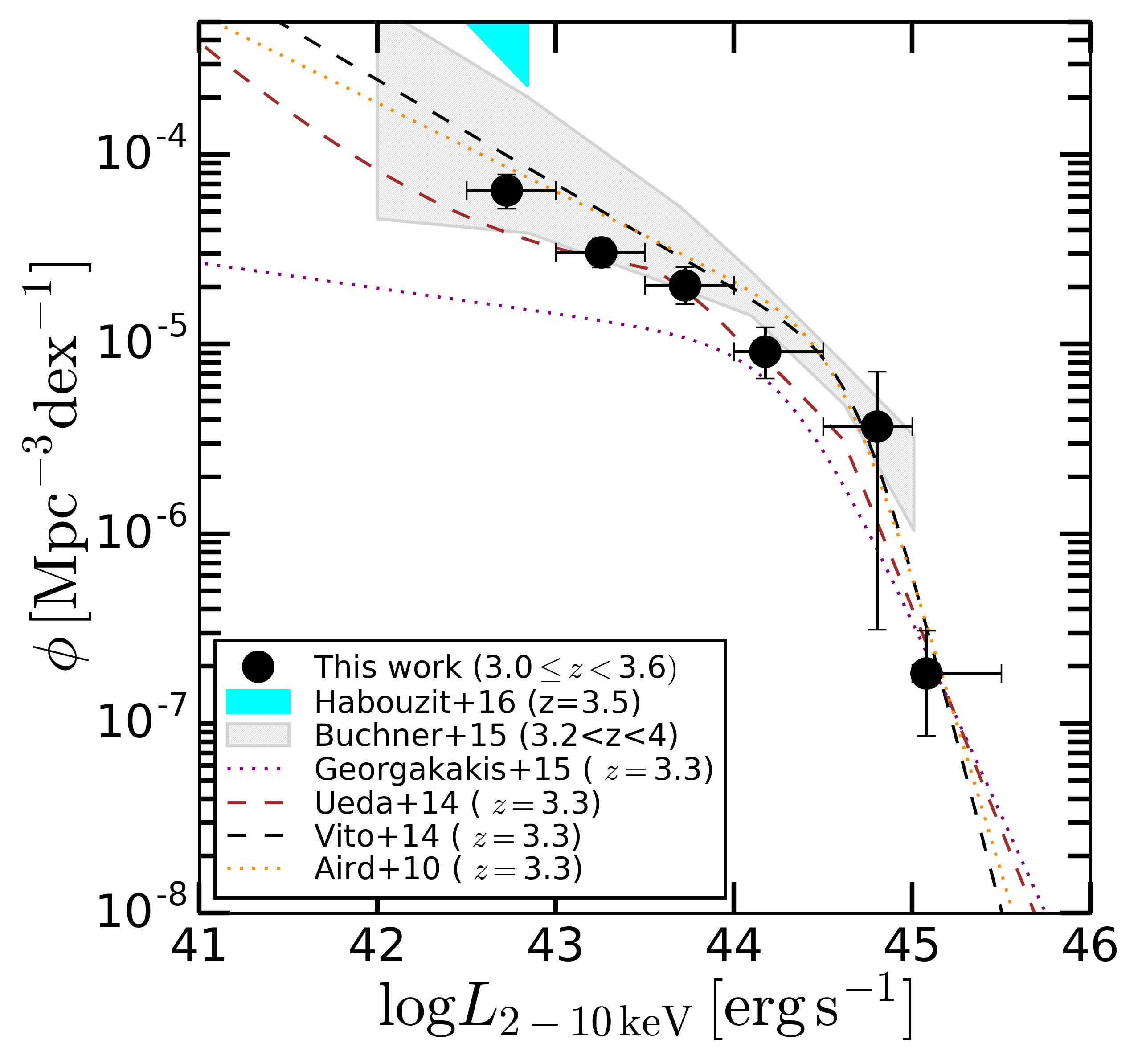}
 	\includegraphics[width=88mm,keepaspectratio]{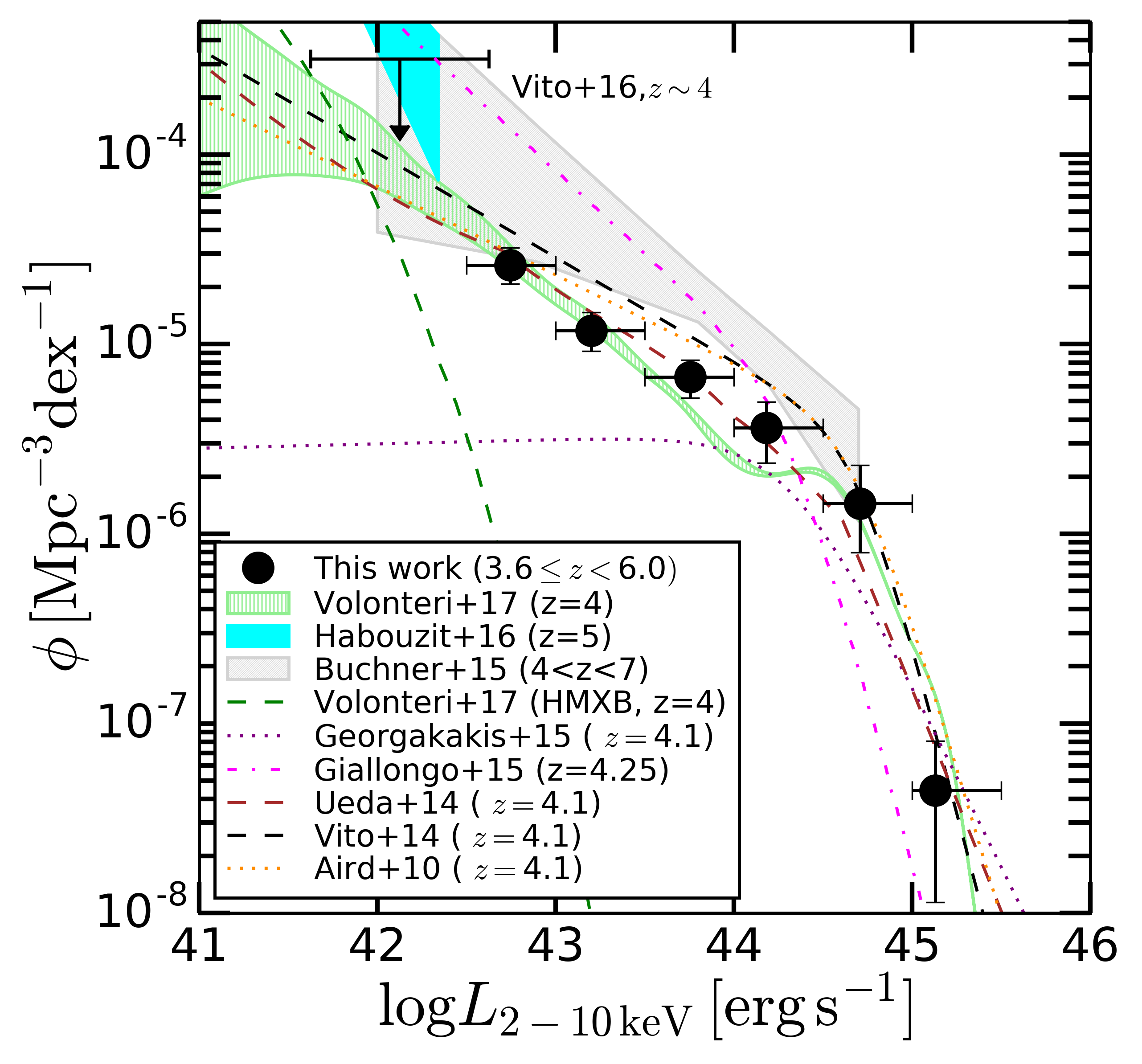}
 	\caption{X-ray luminosity functions of AGN at $z=3-3.6$ (left panel) and $z=3.6-6$ (right panel), compared with results from previous observational (\citealt{Aird10,Ueda14,Vito14,Buchner15,Georgakakis15,Giallongo15}; extrapolated to lower luminosities than those probed by the original works) and theoretical (\citealt{Habouzit16,Volonteri17}) works at similar redshifts. In the right panel, the downward-pointing arrow is the upper limit derived by \citet{Vito16} through a stacking analysis of the 7 Ms CDF-S dataset, and the dashed green line is the expected XLF of HMXBs at $z=4$ (from \citealt{Volonteri17}). Our results exclude extremely steep or flat slopes of the faint end of the AGN XLF.}
 	\label{xlf}
 \end{figure*}

 The AGN X-ray luminosity function can be derived from Eq.~\ref{L_distro_eq} by integrating over narrower redshift intervals and dividing the results by the volume sampled by the surveys in each redshift bin. Fig.~\ref{xlf} presents the X-ray luminosity functions in two redshift bins, chosen to include approximately the same numbers of sources. As stated in \S~\ref{luminosity}, ignoring the correction for Eddington bias does not change significantly the luminosity distribution of our sample, and, as a consequence, the derived XLFs. In particular, the slope of the faint end does not significantly vary when applying or neglecting such a correction. Our XLFs are fairly consistent with previous observational results from \cite{Aird10}, \cite{Ueda14}, and \cite{Vito14, Vito16}. They are also in agreement with \cite{Buchner15} at $z<3.6$, which do not provide an analytical expression for the XLF, but rather confidence regions in redshift intervals.
At higher redshifts the \cite{Buchner15} region exceeds our results. Similarly, the simulated XLFs by \cite{Habouzit16} appear to over-predict the density of low-luminosity AGN.

 \cite{Giallongo15} applied a detection procedure based on searching for clustering of photons in energy, space, and time, at the pre-determined optical positions of CANDELS-detected galaxies (see also \citealt{Fiore12}), which allowed the authors to push the \textit{Chandra} sensitivity beyond the limit reachable by blind detection methods (i.e., with no previous knowledge of the positions of galaxies). The large number of low-luminosity AGN at high redshift detected by \cite{Giallongo15} suggests that AGN could have an important role in cosmic reionization (e.g. \mbox{\citealt{Madau15}; but see also \citealt{Ricci17}}).
 
 \cite{Giallongo15} derived the UV LF of their sample of X-ray detected, high-redshift AGN candidates by deriving the absolute UV magnitude from the apparent optical magnitude in the filter closest to rest-frame $1450\,\angstrom$ at the redshift of each source. 
 We transformed their UV LF into an X-ray LF (see Fig.~\ref{xlf}) by assuming a SED shape $F_\nu\propto\nu^{-0.5}$ between $1450\,\angstrom$ and $2500\,\angstrom$ (as in \citealt{Georgakakis15}), the \cite{Lusso10} $\alpha_{ox}$, and $\Gamma=1.8$ for the X-ray spectrum \citep[see also][]{Vito16}.
 Our results do not support the very steep LFs derived by \cite{Giallongo15}.
 \cite{Parsa17} recently disputed the very detection of the faint $z>4$ AGN in \cite{Giallongo15}. Moreover, \cite{Cappelluti16}, applying a similar procedure which makes use of pre-determined optical positions, did not find such a large number of $z>4$ sources. Other issues plausibly affect the assessment of the population of faint, high-redshift AGN, such as the uncertainties in the photometric redshifts, the expected presence of Eddington bias (which leads to overestimating the number of detected faint sources), and the  XRBs contribution to X-ray emission at faint fluxes \citep[e.g.][]{Lehmer12}. Moreover, the uncertainties related to the assumed UV/X-ray spectral slope (discussed by by \citealt{Giallongo15}) may affect the conversion between UV magnitude and X-ray luminosity, although reasonable values for that conversion cannot produce the tension between their slope of the AGN XLF faint end and our results .

 At the low luminosities probed by the  7 Ms CDF-S ($\mathrm{log}L_{\mathrm{X}}=42-43$), X-ray emission from XRB may provide a non-negligible contribution \citep[e.g.][]{Bauer02, Ranalli03, Ranalli05, Lehmer10, Mineo12}. The green dashed line in Fig.~\ref{xlf} (right panel) is the XLF of star-forming galaxies at $z=4$, where the emission is mostly 
 due to high-mass XRB, from the model by \citet[see in particular their section 2.2 for assumptions and caveats]{Volonteri17}.
 Briefly, they mapped the stellar-mass functions of \cite{Song16} into an XLF through the main-sequence of star-forming galaxies \citep{Salmon15}
 and the SFR-$L_{\mathrm{X}}$ scaling relation by \citet[model 269]{Fragos13}, which \cite{Lehmer16} demonstrated to be the best representation at $z=0-2.5$.
 At log$L_{\mathrm{X}}\gtrsim42.5$, the luminosity limit we applied to our luminosity functions, as the $42\lesssim \mathrm{log}L_{\mathrm{X}}\lesssim42.5$ regime was strongly affected by incompleteness, the density of star-forming galaxies is  a factor of $\gtrsim10$ lower than the AGN density.
 We conclude that at such luminosities our sample is not significantly contaminated by star-forming galaxies. At lower luminosities,
 the XLF of star-forming galaxies is comparable to, or even exceeds, the AGN XLF, in agreement with \cite{Vito16}, who found that the X-ray emission in galaxies individually undetected in the 7 Ms CDF-S is mostly due to XRB.
 Similar conclusions are reached by scaling the star-formation rate functions (SFRF, i.e., the comoving space densities
 of galaxies per unit star-formation rate) from \cite{Smit12} and \cite{Gruppioni15}, at $z\approx3.5$ and 4, respectively, into XLF using 
 the best-fitting, redshift-dependent \cite{Fragos13} and \cite{Lehmer16} relations.

 \subsection{The evolution of the AGN space density, and comparison with the galaxy population}
\begin{figure} 
	\centering
	\includegraphics[width=88mm,keepaspectratio]{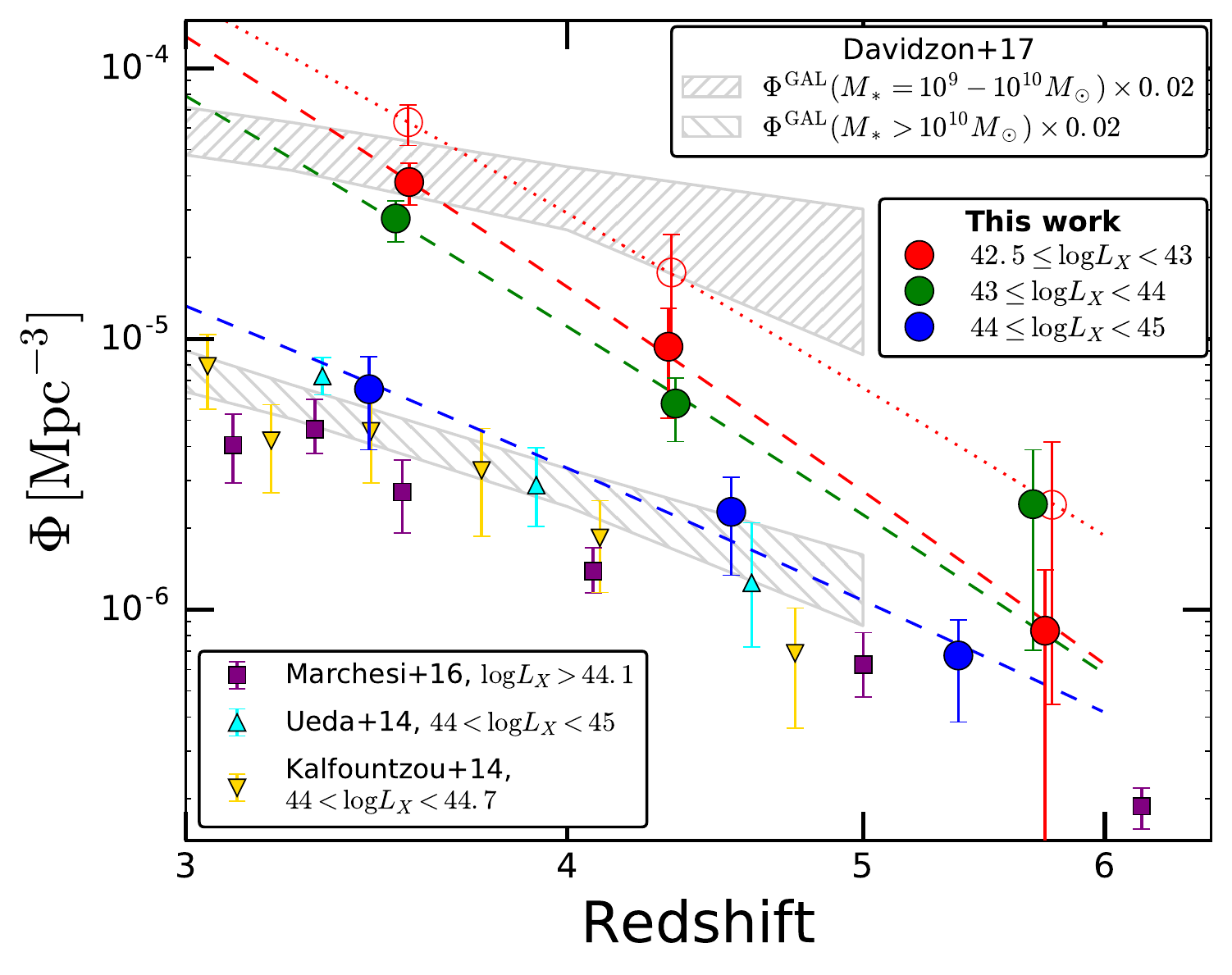}
	\caption{Comoving space density of three luminosity classes of AGN, divided in three redshift bins ($z=3-4$, $4-5$, $5-6$). The points are located at the weighted average redshift of each subsample. For the low-luminosity bin, the empty circles indicate the results derived applying the correction factors described in \chapt{incompl}. At higher luminosities, the corrections have negligible effects, and are not shown for clarity. Since high-luminosity sources are not well sampled by the deep, pencil-beam fields used in this work, we report previous results in similar luminosity bins derived analyzing data from wider fields. The dashed curves are the best-fitting models described in \chapt{xlf_sect}. The dotted curve is the best-fitting model for the low-luminosity bin, after the correction for incompleteness. The space density of galaxies in two different mass regimes from \citet{Davidzon17}, rescaled by an arbitrary factor of 0.02, is shown as a grey stripe. }
	\label{spden}
\end{figure}

 Integrating the XLF over luminosity, we derived the comoving space density of AGN in three luminosity bins (Fig.~\ref{spden}). 
 The decline of the space density in Fig.~\ref{spden}  appears to be steeper for low-luminosity AGN than for luminous sources. However, incompleteness significantly affects the detection of low-luminosity AGN, as can be seen in Fig.~\ref{z_L_compl}. In fact, applying the incompleteness correction discussed in \chapt{incompl} to the density of low-luminosity AGN (empty circles in Fig.~\ref{spden}), a slightly flatter behavior is derived. To quantify the decline of the space density, we fitted a model $\Phi(z)=A((1+z)/(1+z_0))^p$, where $z_0=3$, to the points through a simple $\chi^2$ minimization procedure, before (dashed lines in Fig.~\ref{spden}) and after the correction for incompleteness was applied. Fig.~\ref{spden} displays the best-fitting model corrected for incompleteness only for the low-luminosity bin (dotted line), as at higher luminosities incompleteness is negligible. Tab.~\ref{Phi_par_obs} summarizes the best-fitting values for the total sample and the three luminosity-based subsamples.

The evolution of the space density of luminous (log$L_{\mathrm{X}}\gtrsim44$) AGN in Fig.~\ref{spden} is consistent with previous results by, e.g., \cite{Hiroi12}, \cite{Kalfountzou14}, \cite{Vito14}, and \cite{Marchesi16} (i.e., $\Phi\propto(1+z)^p$ with $p\sim-6$, see Tab.~\ref{Phi_par_obs}). Luminous (log$L_{\mathrm{X}}>44$) AGN are usually hosted by massive ($M_*\gtrsim10^{10}M_{\odot}$) galaxies \citep[e.g.,][]{Xue10, Yang17}. Fig.~\ref{spden} shows that the space densities of luminous AGN and massive galaxies (from \citealt{Davidzon17}) evolve with similar slopes, suggesting that the decline of the space density of log$L_{\mathrm{X}}>44$ AGN is merely driven by the evolution of the galaxy number density. This interpretation is further supported by the recent findings of Yang et al. (submitted), who derive a positive evolution of the \textit{average} black-hole accretion rate in massive galaxies from low redshift up to $z\approx3$, followed by a steady behavior up to $z\approx4$. This trend would be in contrast with the well-established decline of the space density of luminous AGN if the number of massive galaxies itself did not decrease.

Thanks to the unmatched sensitivity of the \textit{Chandra} deep fields, in particular the 7 Ms CDF-S, coupled with the deep multiwavelength coverage, which allows a nearly complete multiwavelength identification of the X-ray sources, we could push the investigation of the evolution of the AGN space density down to log$L_{\mathrm{X}}=42.5$ and up to $z=6$. We found a slightly steeper decline for low-luminosity sources than for high-luminosity AGN, even after applying our fiducial corrections for incompleteness (Fig.~\ref{spden} and Tab.~\ref{Phi_par_obs}). However, such corrections depend on the assumed intrinsic distribution in redshift, column density, and luminosity (see \chapt{incompl}). Moreover, the fit we performed is an approximation (although widely used in literature), as the error bars of the points in Fig.~\ref{spden} are derived through a bootstrap procedure, which formally returns a measurement of the dispersion of the bootstraped distribution. For these reasons, we do not consider the best-fitting values reported in Tab.~\ref{Phi_par_obs} as definitive evidence, but rather a suggestion, for a differential evolution of the AGN space density in different luminosity regimes. However, the strong flattening of the XLF faint-end found by \cite{Georgakakis15} is ruled out by Fig.~\ref{xlf}.
Future X-ray missions, discussed in \chapt{Lynx}, will shed light on the evolution of the population of low-luminosity AGN at higher redshifts than those probed in this work.

Intriguingly, the evolution of the space density of low-luminosity AGN is steeper also than the density evolution of both low-mass and high-mass galaxies (see Fig.~\ref{spden}), as derived by  \cite{Davidzon17}.
This discrepancy requires an evolution of one or more of the physical parameters driving black hole accretion for low-luminosity AGN, other than the mere evolution of the number of galaxies. For instance, a steeper evolution of the AGN density than galaxy density can be expected if the duty cycle and/or occupation fraction decrease at increasing redshift, especially for low-mass galaxies.

\subsection{XLF and space density of obscured and unobscured AGN}
\begin{figure*} 
	\includegraphics[width=88mm,keepaspectratio]{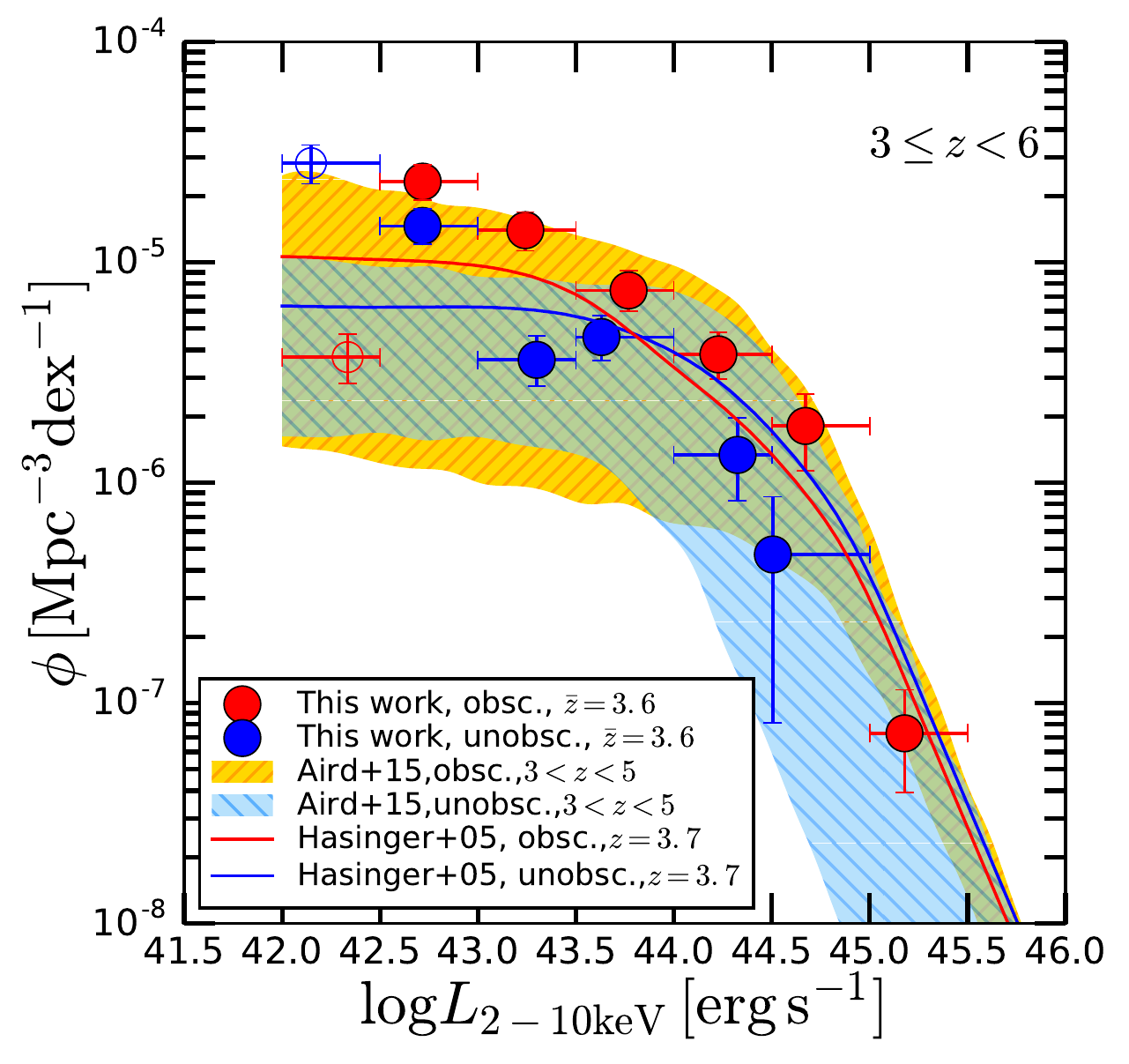}
	\includegraphics[width=88mm,keepaspectratio]{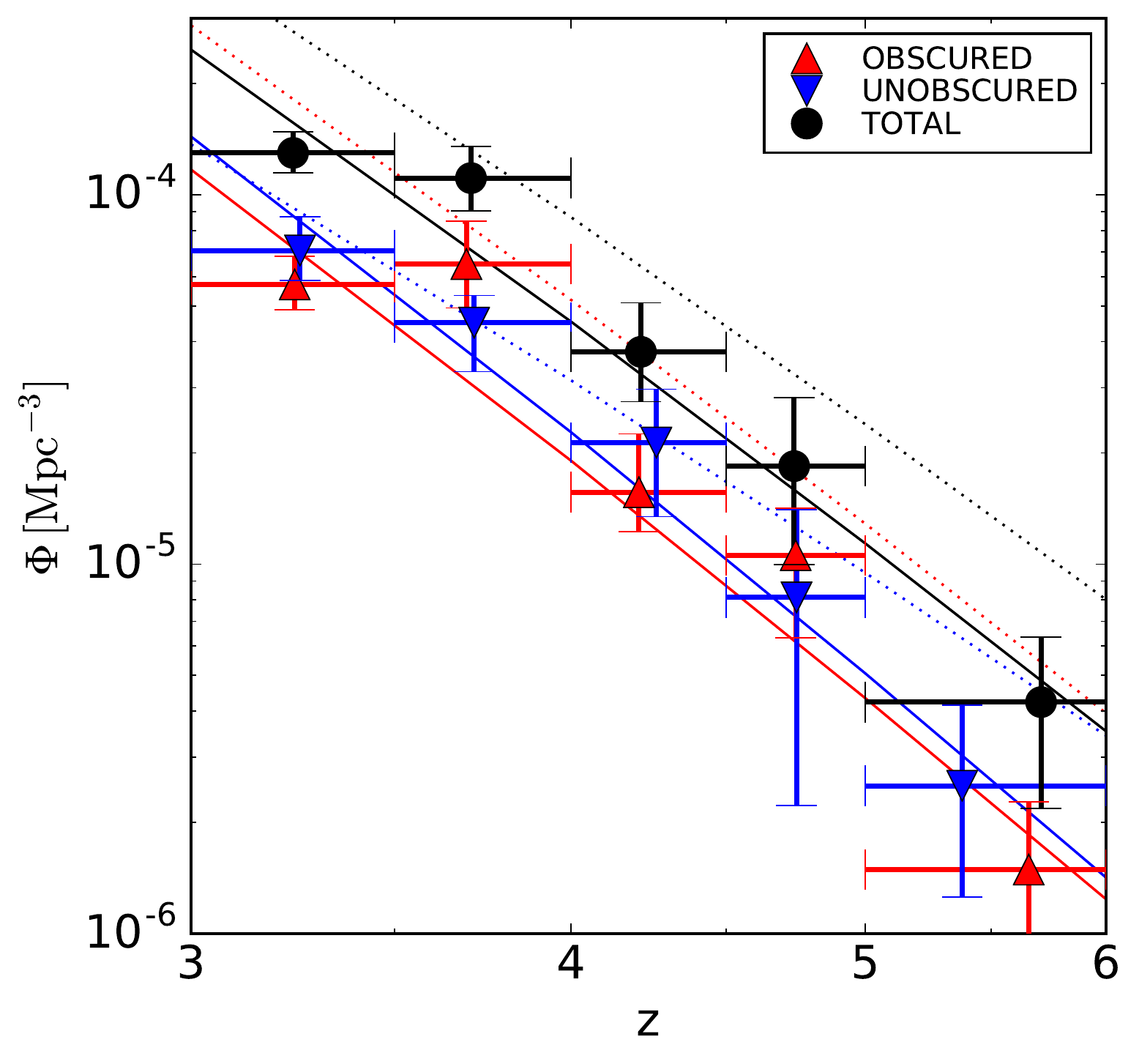}
	\caption{Left panel: X-ray luminosity functions of obscured (red circles) and unobscured (blue circles) AGN at $z=3-6$. The open circles are the binned points in the faintest luminosity bin, and show how incompleteness affects differentially the two subsamples at those luminosities (see \chapt{incompl}). Symbols are centered at the weighted average luminosity of each bin. Red and blue lines are the XLF of obscured (log$N_{\mathrm{H}}>23$) and unobscured (log$N_{\mathrm{H}}<23$) AGN, respectively, by \citet{Hasinger05}.
		Orange and cyan regions represent the areas covered by luminosity functions of log$N_{\mathrm{H}}>22$ and $<22$ AGN, respectively, at  $3\leq z\leq 5$ from \citet{Aird15}. The different column density thresholds assumed to define the obscuration-based subsamples are discussed in the text.
		 Right panel: evolution of the comoving space density with redshift for obscured (red symbols) and unobscured (blue symbols) subsamples. Those symbols are slightly offset for visualization purposes.  Black circles represent the total sample.  Solid lines are the best-fitting linear models obtained by simple $\chi^2$ minimization. To demonstrate the effects of the obscuration-dependent incompleteness (\chapt{incompl}), we present the best-fitting linear models corrected for such incompleteness as dotted lines.  }
	\label{xlf_abs}
\end{figure*}

Fig.~\ref{xlf_abs} (left panel) shows the luminosity functions separately for obscured and unobscured sources, obtained by performing the integral over column density in Eq.~\ref{L_distro_eq} over the ranges $\mathrm{log}N_{\mathrm{H}}=23-25$ and $20-23$, respectively, compared with the X-ray luminosity functions of \citet[used by \citealt{Gilli07} for their X-ray background synthesis model]{Hasinger05} in the same $N_{\mathrm{H}}$ bins. We also report the results from \cite{Aird15}, where obscured and unobscured AGN are defined as those obscured by column densities $\mathrm{log}N_{\mathrm{H}}=22-24$ and $20-22$, respectively.
Points in the first luminosity bin (empty circles) are heavily affected by the obscuration-dependent incompleteness discussed in \chapt{incompl}, as is evident by the drop of the density of obscured AGN with respect to the unobscured subsample. Fig.~\ref{xlf_abs} (right panel) displays the evolution of the space density of obscured and unobscured AGN.  Tab.~\ref{Phi_par_obs} summarizes the best-fitting values for the obscured and unobscured subsamples. 
We found no significant difference in the evolution of the space density of obscured and unobscured AGN. This conclusion is consistent with Fig.~\ref{absfrac_z_fig}, where the obscured AGN fraction is flat with redshift. 
After correcting for incompleteness, a slightly shallower decline of the space density of unobscured AGN is derived, because of the faster decrease of detection completeness with increasing redshift for unobscured sources than fot obscured sources (see Fig.~\ref{z_completeness}). However, the uncertainties due to the procedure of correcting for incompleteness and the relatively small size of the sample prevent us from drawing strong conclusions.

\begin{table}
	\tiny
	\caption{Best-fitting parameters describing the space-density evolution with redshift for the luminosity-based and obscuration-based subsamples (see \chapt{xlf_sect}).}\label{Phi_par_obs}
	\begin{tabular}{cccccccc}
		\hline
		\multicolumn{1}{c}{Subample} &
		\multicolumn{1}{c}{$A\,[\mathrm{Mpc^{-3}}]$} &
		\multicolumn{1}{c}{$p$}\\	
		\hline
All & 	$(2.47\pm0.57)\times10^{-4}$ & $-7.59\pm0.69$\\
$42.5\leq \mathrm{log}L_{\mathrm{X}}<43$ & $(1.31\pm0.08)\times10^{-4}$ & $-9.57\pm0.32$\\
$43\leq \mathrm{log}L_{\mathrm{X}}<44$ & $(7.96\pm2.70)\times10^{-5}$ & $-8.85\pm1.71$\\
$44\leq \mathrm{log}L_{\mathrm{X}}<45$ & $(1.33\pm0.38)\times10^{-5}$ & $-6.17\pm0.87$\\
Obscured & 	$(1.17\pm0.43)\times10^{-4}$ & $-8.12\pm1.11$\\
Unobscured & 	$(1.44\pm0.34)\times10^{-4}$ & $-8.25\pm0.77$\\
		\hline
		\multicolumn{3}{c}{After correction for incompleteness}\\	
All & 	$(4.22\pm1.19)\times10^{-4}$ & $-7.08\pm0.77$\\
$42.5\leq \mathrm{log}L_{\mathrm{X}}<43$ & $(1.80\pm0.02)\times10^{-4}$ & $-8.15\pm0.36$\\
$43\leq \mathrm{log}L_{\mathrm{X}}<44$ & $(8.28\pm2.94)\times10^{-5}$ & $-8.93\pm1.82$\\
$44\leq \mathrm{log}L_{\mathrm{X}}<45$ & $(1.35\pm0.37)\times10^{-5}$ & $-6.16\pm0.84$\\
Obscured & 	$(2.87\pm1.05)\times10^{-4}$ & $-7.66\pm1.14$\\
Unobscured & 	$(1.37\pm0.26)\times10^{-4}$ & $-6.58\pm0.62$\\
		\hline
	\end{tabular}\\
\end{table}

\subsection{Black-hole accretion rate density}\label{bhad}
\begin{figure*} 
	\includegraphics[width=160mm,keepaspectratio]{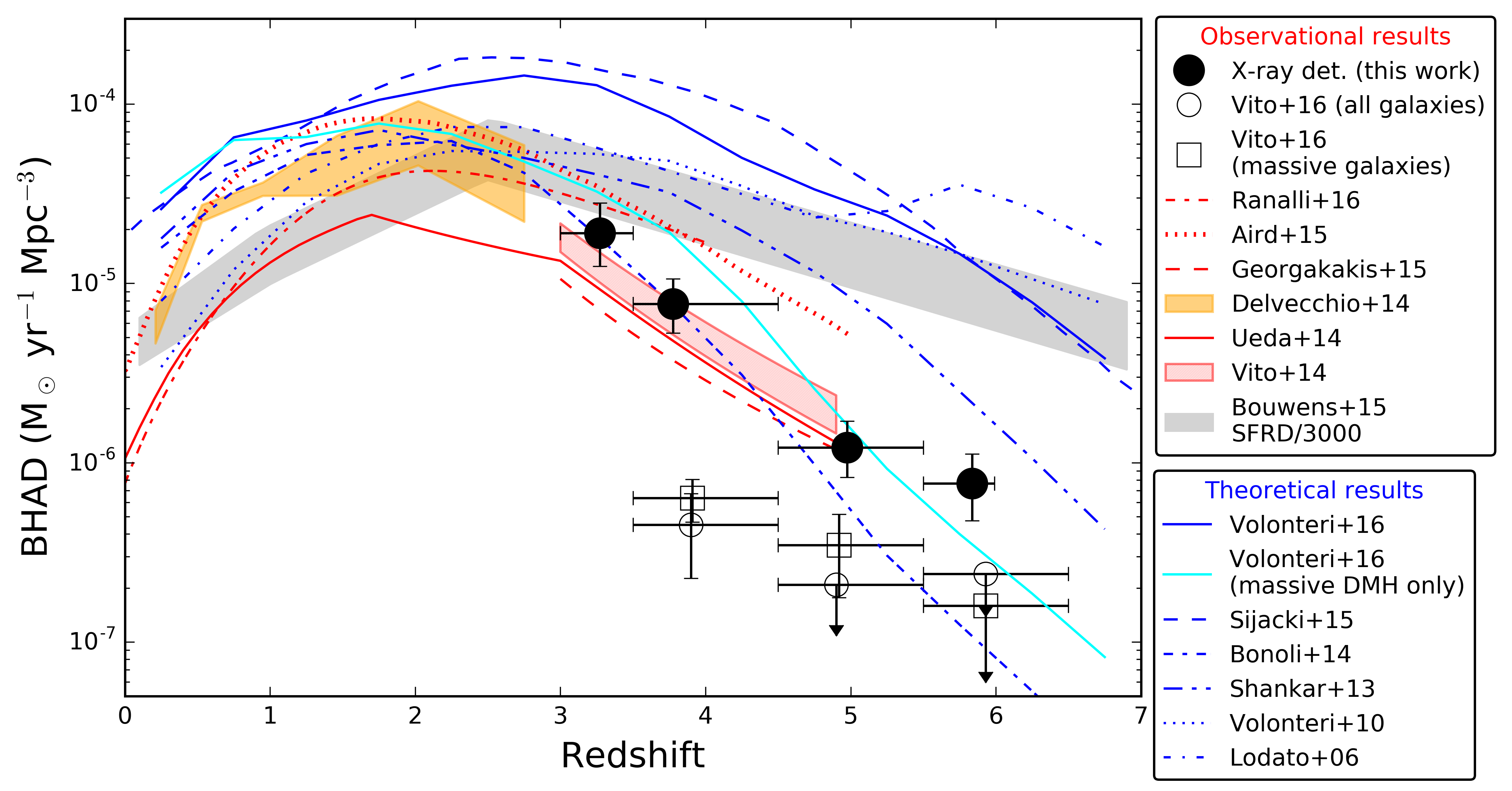}
	
	\caption{BHAD derived from our sample of X-ray detected AGN (filled black symbols), compared with previous observational and theoretical results from literature. We also show the BHAD resulting from the stacking analysis of a sample of individually undetected galaxies \citep{Vito16}, under the most optimistic assumption that all the stacked X-ray signal is due to accretion onto SMBH. Empty circles and squares correspond to the entire sample and the half containing the most massive galaxies, respectively. The grey stripe is the star-formation rate density of \citet{Bouwens15}, rescaled by a factor of 3000.}
	\label{bhad_fig}
\end{figure*}

The black-hole accretion rate density (BHAD) is defined s

\begin{equation}
\Psi(z)=\int \frac{(1-\epsilon)}{(\epsilon c^2)}L_{\mathrm{bol,AGN}}\phi(L_{\mathrm{bol,AGN}},z)\mathrm{dlog}L_{\mathrm{bol,AGN}}
\end{equation}
where $L_{\mathrm{bol,AGN}}=K_{\mathrm{bol}}(L_{\mathrm{X,AGN}})L_{\mathrm{X,AGN}}$ is the AGN bolometric luminosity, $K_{\mathrm{bol}}(L_{\mathrm{X,AGN}})$ is the bolometric correction, $\epsilon$ is the radiative efficiency (fixed here to $\epsilon=0.1$), and $\phi(L_{\mathrm{bol,AGN}},z)$ is the AGN bolometric luminosity function, which can be derived from the AGN XLF through the bolometric correction. We assumed $K_{\mathrm{bol}}(L_{\mathrm{X,AGN}})$ from \cite{Lusso12}. Fig.~\ref{bhad_fig} presents the BHAD derived from our sample as filled black circles. We used the same luminosity bins of \cite{Vito16} in order to facilitate the comparison between the two works. In \cite{Vito16} we derived the BHAD due to X-ray undetected $z=3.5-6.5$ galaxies in the 7 Ms CDF-S through a stacking procedure. Empty circles and squares present the BHAD for the entire sample of galaxies and for the half of most massive ones, respectively, under the most optimistic assumption that all the signal is due to accretion onto SMBH (i.e. the signal due to XRB is negligible). By comparing these points with the BHAD of a preliminary sample of high-redshift, X-ray detected AGN in the 7 Ms CDF-S, we concluded that most of the BH growth at $z=3.5-6.5$ occurs during the bright and fast (relative to galaxy lifespan) AGN phase, with negligible contribution from continuous, low-rate accretion. In this work we refine the sample of X-ray detected, high-redshift AGN and are therefore able to present a more accurate comparison.

Fig.~\ref{bhad_fig} displays the BHAD corresponding to the X-ray luminosity functions of \cite{Ranalli16}, \cite{Aird15}, \cite{Georgakakis15}, \cite{Ueda14}, and \cite{Vito14}, converted into BHAD as described above,\footnote{\cite{Aird15} provided directly their estimate of BHAD, but used the bolometric corrections from \cite{Hopkins07} instead of the \cite{Lusso12} values we assumed. The slightly higher normalization of the \cite{Aird15} curve with respect to the other observational results reflects this difference and can be therefore considered a measure of the uncertainty due to bolometric corrections.} and that derived by \cite{Delvecchio14} exploiting \textit{Herschel} data. We also add the BHAD predicted by simulations (\citealt{Volonteri16}, \citealt{Sijacki15}, \citealt{Bonoli14}, \citealt{Shankar13}, \citealt{Volonteri10}, \mbox{\citealt{Lodato06}}) under different assumptions of BH seed mass and growth mechanisms. Except for \mbox{\cite{Sijacki15}}, who directly reported the BHAD, we derived the curves by differentiating the BH mass densities presented in those works.

Simulations tend to derive BHAD exceeding by about one order of magnitude the observed values at $z>3$ (see Fig.~\ref{bhad_fig}). This discrepancy is usually ascribed to the contribution of continuous, low-rate accretion,  difficult to detect even in deep surveys. In \cite{Vito16} we used a stacking procedure to demonstrate that such contribution is negligible compared to the BHAD due to X-ray detected AGN (see also, e.g., \citealt{Pezzulli17}). With the refined sample of high-redshift AGN of this work, we confirmed this result up to $z\sim6$ and suggest that simulations tend to overestimate the BHAD at high redshift. 

A similar issue is known to affect the faint-end of the AGN LF at high redshift, with simulations finding steeper slopes than the observed ones. \cite{Volonteri16} suggested that this discrepancy could be due to supernovae feedback (i.e. the removal of gas in the galaxy responsible of both nuclear accretion and star formation as a consequence of supernovae activity) being underestimated in low-mass halos in simulations. Considering only massive ($M>5\times10^{11}M_{\odot}$) dark-matter halos, where supernovae feedback has little effect due to the deeper galactic gravitational potential well, improves the agreement with observations. This statement is also true for the BHAD, as can be seen in Fig.~\ref{bhad_fig}, where the results of \cite{Volonteri16} are presented without any mass cut (continuous blue curve) and for massive halos only (cyan curve).

The common evolution of star-formation rate density (SFRD) and the BHAD at $z\lesssim2.5$ is generally considered to be a manifestation of the BH-galaxy coevolution.
However, these two quantities are found to evolve with different slopes at higher redshifts \citep[e.g.][see, e.g., \citealt{Haiman04} for a theoretical interpretation]{Silverman08,Aird15}, mainly through extrapolations of the AGN XLF.
In this work, we can probe this behavior, whose physical causes are difficult to identify, down to the faint flux limits of the \textit{Chandra} deep fields. Most of the BHAD we derived at high redshift is due to luminous (log$L_{\mathrm{X}}>44$) AGN, which contribute $70-90\%$ to the black circles in Fig.~\ref{bhad_fig}; therefore the uncertainties on the density of low-luminosity AGN have no strong impacts in this respect. In particular, the contribution of accretion in individually undetected galaxies is negligible (empty black symbols from \citealt{Vito16} in Fig.~\ref{bhad_fig}). Indeed, the evolution the BHAD is similar to that of the space density of luminous AGN, both declining by a factor of $\gtrsim10$ from $z=3$ to $z=6$.
In contrast, the evolution of the SFRD is more similar to that of low-mass galaxies than massive systems, both declining by a factor of $\sim3$ in the same redshift range (cf. Fig.~\ref{bhad_fig} and \ref{spden}). This discrepancy may mark a break down of the BH-galaxy co-evolution at high redshift.

\section{Future prospects}\label{future}
\subsection{Spectroscopic follow-up of high-redshift AGN candidates}
One of the main sources of uncertainty affecting our knowledge of the demography of high-redshift AGN is the persistently low spectroscopic identification rates of the candidates.
The deep \textit{Chandra} surveys allow sampling of the bulk of the population at high redshift, which is constituted of low luminosity and/or obscured sources. However, these objects are typically faint in the optical/NIR bands (e.g., $\approx80\%$ of our sources located in the CANDELS/GOODS-S area have $H>24$), and are therefore difficult to follow up spectroscopically with current facilities in order to confirm their high-redshift nature. While photometric redshifts of good quality are available in the fields we considered, and despite the careful treatment of their PDF, they may introduce non-negligible uncertainties in the final results (e.g., see \S~4.3 of \citealt{Georgakakis15}). This issue is less problematic for shallower X-ray surveys, which sample AGN with brighter optical/NIR counterparts, but miss the sub-population of low-luminosity AGN.

Current facilities that could be used effectively to increase the spectroscopic completeness of our sample include \textit{ALMA}\footnote{http://www.almaobservatory.org/} \citep[e.g.,][]{Walter16}, the Keck observatory,\footnote{http://www.keckobservatory.org/} and \textit{VLT/MUSE}.\footnote{http://muse-vlt.eu/} The last, in particular, with 27h of integration time in the \textit{Hubble} Ultra Deep Field, has already been able to obtain spectroscopic redshifts for sources below the \textit{Hubble} detection limit \citep{Bacon15}.
In the near future, facilities like \textit{JWST},\footnote{https://jwst.stsci.edu/} and, in the somewhat longer term, the Extremely Large Telescopes (with $\gtrsim30$ m diameter mirrors) will provide the required spectroscopic sensitivity to perform ultradeep spectroscopic follow-up campaigns aimed at identifying high-redshift AGN candidates with extremely high efficiency (see, e.g., the discussion in \citealt{Brandt15}).

\subsection{The case for \textit{Athena} and \textit{Lynx}}\label{Lynx}
The understanding of the SMBH formation and growth mechanisms in the early universe requires the collection of much larger samples of X-ray selected AGN and galaxies at higher redshifts and lower luminosities than those probed by current X-ray surveys. Moreover, while \textit{Chandra} and \textit{XMM-Newton}, jointly with large-area optical/IR surveys, are successfully improving our knowledge of the bright quasar population up to $z\approx7$ \citep[e.g.,][]{Brandt17}, the bulk of the AGN population, constituted by low-luminosity and possibly obscured systems, are currently completely missed beyond $z\approx5-6$ even by the deepest X-ray surveys available, and \textit{Chandra} observations substantially deeper than the 7 Ms CDF-S are not foreseeable. The \textit{Athena} X-ray Observatory \citep{Nandra13}\footnote{http://www.the-athena-x-ray-observatory.eu/}, which will be launched in $\sim2028$, is expected to detect hundreds of $\lesssim L_*$ AGN at $z>6$ \citep{Aird13}, thanks to its survey power a factor of $\sim100$ faster than \textit{Chandra} or \textit{XMM-Newton}, boosting our knowledge of the high-rate accretion phases, which likely dominate SMBH growth at high redshift \citep{Vito16}.

Simulations suggest that different combinations of the physical parameters governing SMBH formation and early growth (e.g., seed-mass and Eddington ratio distributions, occupation fraction, feedback mode, and their dependence on the host-galaxy and environment properties) produce different shapes of the faint-end (log$L_{\mathrm{X}}\lesssim42$) of the AGN LF.
The green-striped region in Fig.~\ref{xlf} (right panel) is the locus of the XLF models of \cite{Volonteri17} at $z=4$, under different assumptions.
In particular, the upper bound is derived using a minimum initial BH mass of $10^2M_{\odot}$ and occupation fraction equal to unity,
while for the lower bound a minimum initial mass of  $10^5M_{\odot}$ and a modified occupation fraction from \cite{Habouzit16},
flagged as ``H17low1" in \cite{Volonteri17}. Different combinations of parameters are mostly encompassed by these boundaries.
The models are normalized to the observed points, which explains the small spread at $\mathrm{log}L_{\mathrm{X}}\gtrsim43$.
However, at lower luminosities, which even the currently deepest surveys cannot sample, different combinations of parameters
produce significantly different XLF shapes. The spread of the XLF shapes in this faint regime is expected
to increase at higher redshifts (see Fig. 12 of \citealt{Volonteri17}). A tremendous step forward toward understanding the early growth of SMBH will be provided by \textit{Lynx} (also known as \textit{X-ray Surveyor}, \citealt{Weisskopf15}), which, with its \textit{Chandra}-like angular resolution and $\approx25$ times higher sensitivity, will reach an expected flux limit of log$F\approx-18.5$ in a blind 4 Ms survey, sufficient to probe the faint end of the XLF down to log$L_{\mathrm{X}}\approx40-42$ at $z\lesssim10$, providing unprecedented constraints on the combinations of parameters driving SMBH formation and early growth. 

The constraints derived by X-ray telescopes will be complemented by deep optical/IR observations provided by future facilities such as \textit{JWST}and \textit{WFIRST}\footnote{https://wfirst.gsfc.nasa.gov/}, necessary to identify the X-ray sources. Recently, several works have focused on the multiwavelength properties of emission from BHs during the early accretion phases. For instance, \cite{Volonteri17} proposed a color-color selection based on \textit{JWST} bands to separate AGN from normal galaxies at high redshift. In particular, the characteristic shape of the spectral energy distribution of accreting direct-collapse BH (with initial $M_{BH}=10^4-10^5\,M_\odot$) has been used by \cite{Natarajan16} to propose a \textit{JWST} color-color method to select such objects. In contrast, optical/IR emission from accreting light-seeds BH ($M_{BH}=10^2-10^3\,M_\odot$) probably will be outshined by stellar emission in the young host galaxies.
 The joint effort of the next generation X-ray observatories and optical/IR telescopes therefore will open a completely new window on the cosmic epoch where SMBH formed.

\section{Conclusions}

We have investigated the population of  $3\leq z < 6$ AGN detected in the two deepest X-ray fields, the \mbox{7 Ms CDF-S} and the 2 Ms CDF-N, selected on the basis of a careful assessment of the photometric-redshift uncertainties for sources lacking spectroscopic identification. For each source, uncertainties on redshift, column density, flux, and luminosity were accounted for by using the full probability distributions of such parameters, and a statistical correction for Eddington bias was applied. Our main results are the following:
\begin{enumerate}
	
	\item We derived the best-constrained column-density distribution of moderate-luminosity obscured AGN at high redshift, finding it to peak at log$N_{\mathrm{H}}\sim23.5-24$. See \chapt{NH}.
	
	\item Comparing the mean photon index of X-ray spectra of high-redshift, $L_X<L_*$ AGN with results derived for luminous quasars, we found no significant variation with redshift and luminosity. See Appendix \chapt{gamma}.
	
	\item The obscured AGN fraction ($F_{\mathrm{obsc}}$), where the column-density threshold to define obscured vs. unobscured AGN is set to log$N_{\mathrm{H}}=23$, is consistent with being flat with redshift at observed values of $\approx0.5-0.6$. Once corrected for incompleteness, we derive $F_{\mathrm{obsc}}\approx0.6-0.7$. This result agrees with previous works finding a saturation of this quantity at $z\gtrsim2$, after a positive evolution at $z\approx0-2$. See \chapt{absfrac_z}..
	
\item 
Thanks to the sensitivity of the \textit{Chandra} deep fields, we could probe the number counts of high-redshift AGN down to fainter fluxes, especially at $z>4$, than those reached by wider and shallower surveys.
Comparing the log$N$-log$S$ of our sample with the predictions from the \cite{Gilli07} X-ray background synthesis model, which was constrained at lower redshifts, separately for obscured and unobscured AGN, we found an excess of obscured sources, likely caused by the positive evolution of $F_{\mathrm{obsc}}$ from the local universe to high redshift. See \chapt{lnls_text}
	
	\item We found $F_{\mathrm{obsc}}\approx0.7-0.8$ at log$L_{\mathrm{X}}\gtrsim43$, somewhat larger, but still in agreement with other works performed at similar redshifts. In contrast with previous results, derived especially at low redshift, there is no significant anti-correlation between  $F_{\mathrm{obsc}}$ and luminosity at log$L_{\mathrm{X}}\gtrsim43$, suggesting a stronger positive evolution of $F_{\mathrm{obsc}}$ towards earlier cosmic times for luminous AGN than for moderate-luminosity sources. See \chapt{absfrac_l}.

	\item At low luminosities (log$L_{\mathrm{X}}\lesssim43$) the observed behavior of  $F_{\mathrm{obsc}}$ is strongly affected by sample incompleteness: low-luminosity, obscured sources are more difficult to detect than unobscured sources, biasing the observed result toward low values of $F_{\mathrm{obsc}}$. We suggest that recent claims of a turn-down of $F_{\mathrm{obsc}}$ toward low luminosities at high redshift may be at least partly due to similar incompleteness issues. By assuming intrinsic distributions of sources in $N_{\mathrm{H}}$ and $L_{\mathrm{X}}$ space, we applied tentative corrections which led $F_{\mathrm{obsc}}$ to be fairly consistent with the values at higher luminosities, although a possible slight decrease down to values of $\approx0.6$ is still visible. See \chapt{absfrac_l} and \chapt{incompl}.
	
\item The space density of luminous (log$L_{\mathrm{X}}\gtrsim44$) AGN declines from $z=3$ to $z=6$, as $(1+z)^d$, with $d\approx-6$, in agreement with previous results. 
 The X-ray luminosity function (XLF) in the redshift range probed by this work does not show significant steepening of the faint end, which has been recently linked with a possible significant contribution of low-luminosity AGN to cosmic reionization. Our results tend to exclude this hypothesis. By contrast, the space density of AGN with low-to-moderate luminosity appears to decline slightly faster than for high-luminosity AGN. However, uncertainties due to incompleteness might affect this result. The evolution of the AGN XLF faint end at high redshift is particularly important for placing constraints on the combination of the physical parameters driving the formation and early growth of supermassive black holes, one of the goals of future X-ray observatories. See \chapt{xlf_sect} and \chapt{Lynx}.

\item The space densities of luminous AGN and massive galaxies show a similar trend with redshift, suggesting that the evolution of luminous AGN is driven by the underlying galaxy population. By contrast, the space density of low-luminosity AGN declines with increasing redshift faster than the space density of both high-mass and low-mass galaxies. This difference suggests an evolution in the physical parameters regulating nuclear activity, such as duty cycle and occupation fraction. We also derived the BHAD at $z=3-6$, and noted that, while the BHAD and SFRD track each other at $z\approx0-2.5$, at higher redshifts the BHAD decreases with a faster rate than the SFRD. See \chapt{xlf_sect}.

\end{enumerate}

\section*{Acknowledgments} \vspace{0.2cm}
We are grateful to the referee, A. Georgakakis, for useful comments that helped in clarifying several aspects of this work.
FV, WNB, and GY acknowledge support from Chandra X-ray Center grant
GO4-15130A, the Penn State ACIS Instrument Team Contract SV4-74018 (issued by the \textit{Chandra} X-ray Center, which is operated by the Smithsonian Astrophysical Observatory for and on behalf of NASA under contract NAS8-03060), and the V.M.
Willaman Endowment. BL acknowledge support from the National Natural Science Foundation of
China grant 11673010 and the Ministry of Science and Technology of China
grant 2016YFA0400702. YQX. acknowledges support from the 973 Program (2015CB857004), NSFC-11473026, NSFC-11421303, the CAS Frontier Science Key Research Program (QYZDJ-SSW-SLH006), and the Fundamental Research Funds for the Central Universities. We warmly thank Harry Ferguson and Dritan Kodra
for providing probability distribution functions of CANDELS photometric-redshifts, and Iary Davidzon, Carlotta Gruppioni, and Melanie Habouzit for providing useful data. The Guaranteed Time Observations (GTO) for the CDF-N included here
were selected by the ACIS Instrument Principal Investigator,
Gordon P. Garmire, currently of the Huntingdon Institute for X-ray
Astronomy, LLC, which is under contract to the Smithsonian
Astrophysical Observatory; Contract SV2-82024.

\bibliography{biblio}

%
%
%
 
%

%
\appendix

\section{Checking the accuracy of the X-ray spectral analysis}\label{check_spec_analysis}

In \chapt{spec_analysis} we assumed a simple spectral model, an absorbed power-law with photon index fixed to $\Gamma=1.8$, as a representation of the X-ray spectra of the sources in our high-redshift sample, and performed a spectral analysis to derive the probability distribution of the intrinsic column density. Although the use of more complex spectral models is generally precluded by the limited number of counts in the X-ray spectra of the typically faint AGN  in our sample, we check in this Appendix for potential biases arising from this assumption using X-ray spectral simulations.

We assumed that the intrinsic X-ray spectra in cases of column densities $\mathrm{log}N_{\mathrm{H}}>22$ are well represented by the MYTorus spectral model\footnote{ http://www.mytorus.com/}\citep{Murphy09}, which fully accounts for the transmission and scattered components, as well as the production of the iron $K$ edge and $K_\alpha$ and $K_\beta$ emission lines. 
We fixed the Galactic absorption to $\mathrm{log}N_{\mathrm{H}}=20$ (the characteristic value toward the CDF-S; \citealt{Kalberla05}), $\Gamma=1.8$, the inclination angle $\Theta=90^\circ$ (i.e. edge-on configuration),  and the normalization of the scattered and line components to unity with respect to the transmitted component. The MYTorus model does not allow lower values of column densities, and therefore for $\mathrm{log}N_{\mathrm{H}}<22$ we used the XSPEC \textit{plcabs} model, which also takes into account Compton scattering. We assumed three values of observed soft-band flux, $\mathrm{log}F=-17$, $-16$ and $-15$, corresponding to sources close to the flux limit of the 7 Ms CDF-S, faint sources, and bright sources (in deep X-ray surveys), respectively. We also considered three different redshifts ($z=3$, 4 and 5) and 8 different values of column density (from $\mathrm{log}N_{\mathrm{H}}=21$ to 24.5 in steps of 0.5 dex). For each of the resulting 72 combinations of these parameters, we simulated 10 X-ray source and background spectra with a 6 Ms effective exposure (typical of the CDF-S), allowing for Poissonian fluctuations, using the \textit{fakeit} command in XSPEC. The background spectrum is sampled from the real CDF-S background. We then fitted the simulated spectra with our simple model and derived the best-fitting column density and its errors at the $90\%$ confidence level.

\begin{figure} 
	\includegraphics[width=80mm,keepaspectratio]{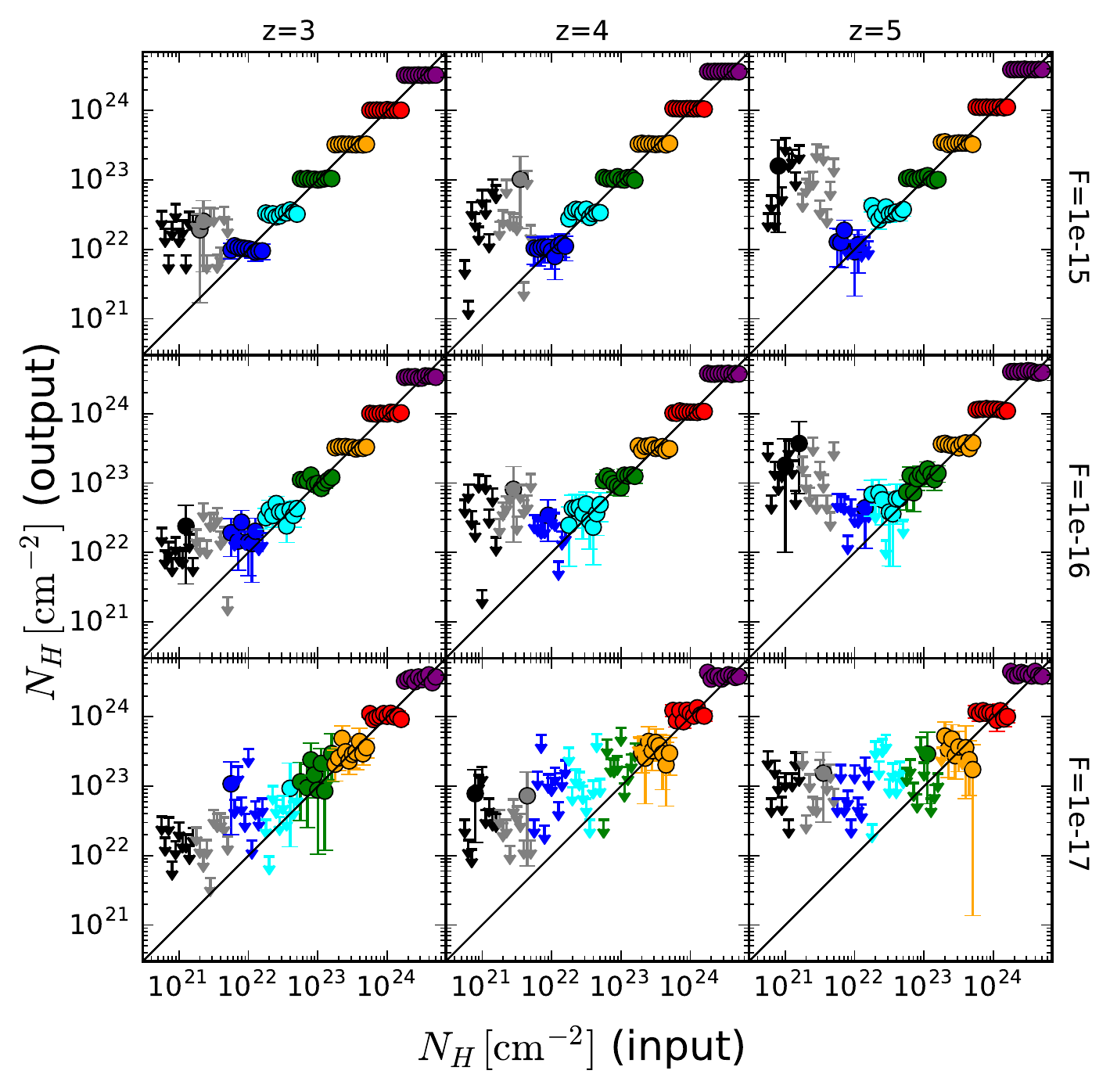}
	\caption{Best-fitting column density and 90\% c.l. errors derived by fitting a simple absorbed power-law model plotted against the simulated value, for different redshifts (columns) and fluxes (rows), as marked in the figure. Spectra are simulated using the MYTorus model and 8 different values of column density (in the range $\mathrm{log}N_{\mathrm{H}}=20-24.5$, in steps of 0.5 dex, color coded in the figure) as described in Appendix \ref{check_spec_analysis}. The $\mathrm{log}N_{\mathrm{H}}$ positions are slightly shifted for presentation purposes. Black lines mark the 1:1 relations.}
	\label{check_spec}
\end{figure}

Fig.~\ref{check_spec} shows the best-fitting column densities against the simulated ones for the different redshift-flux combinations. The output values in cases of large simulated column densities are consistent with the input ones. When decreasing the simulated flux and increasing the redshift, low values of column density ($\mathrm{log}N_{\mathrm{H}}\lesssim22.5$ and, less severely, $\mathrm{log}N_{\mathrm{H}}\sim23$) become increasingly difficult to constrain. In particular, the output column densities of sources simulated with $\mathrm{log}N_{\mathrm{H}}\leq22$ are never constrained to be lower than that value, which is the widely used threshold separating obscured and unobscured AGN at low redshift. This result is due to the photoelectric cut-off shifting close to, and even below, the low-energy spectral limit of the \textit{Chandra}  bandpass. Small differences in the intrinsic photon index (within the expected range $\Gamma=1.7-2.0$) do not affect significantly the output column densities.
We therefore conclude that the spectral model we assumed, despite its simplicity, is a useful and appropriate approximation, at least for the redshift and flux ranges  probed by this work. 

\section{The X-ray power-law photon index of high-redshift, sub-$L_*$ AGN}\label{gamma}
The photon index ($\Gamma$) of the primary power-law emission in AGN X-ray spectra can be used to estimate the Eddington ratio ($\lambda_{\mathrm{Edd}}=L/L_{\mathrm{Edd}}$) of a source through the $\Gamma-\lambda_{\mathrm{Edd}}$ relation \citep[e.g.][but see the caveats discussed in \citealt{Trakhtenbrot17}]{Shemmer06,Shemmer08,Risaliti09,Brightman13,Brightman16,Fanali13}, according to which higher $\lambda_{\mathrm{Edd}}$ would produce steeper $\Gamma$. This relation is likely produced by the coupling between the accretion disk and the hot corona, and can therefore be used to investigate whether the accretion mode depends on the Eddington ratio at different redshifts. 

The spectral analysis presented by \cite{Liu17} on bright ($>80$ net counts) X-ray-selected AGN in the 7 Ms CDF-S reveals a median value of $\Gamma=1.82\pm0.15$, typical of Seyfert galaxies. Evidence for steep photon indexes in a few luminous Type-I QSOs at $z>6$ have been found by \cite{Farrah04}, \cite{Page14} (whose results are disputed by \citealt{Moretti14}) and \cite{Ai16}. However, studies of samples of $z=4-7$ Type-I QSOs did not find a significant evolution of the average $\Gamma$ in luminous QSOs with redshift \citep[e.g.][]{Shemmer05,Vignali05,Nanni17}. Flatter photon indexes ($\Gamma\approx1.1-2.0$) have been derived by \cite{Vignali02} for three spectroscopically-confirmed $z>4$,  moderate-luminosity AGN in the CDF-N, probably due to the presence of intrinsic absorption.

\begin{figure*} 
	\includegraphics[width=160mm,keepaspectratio]{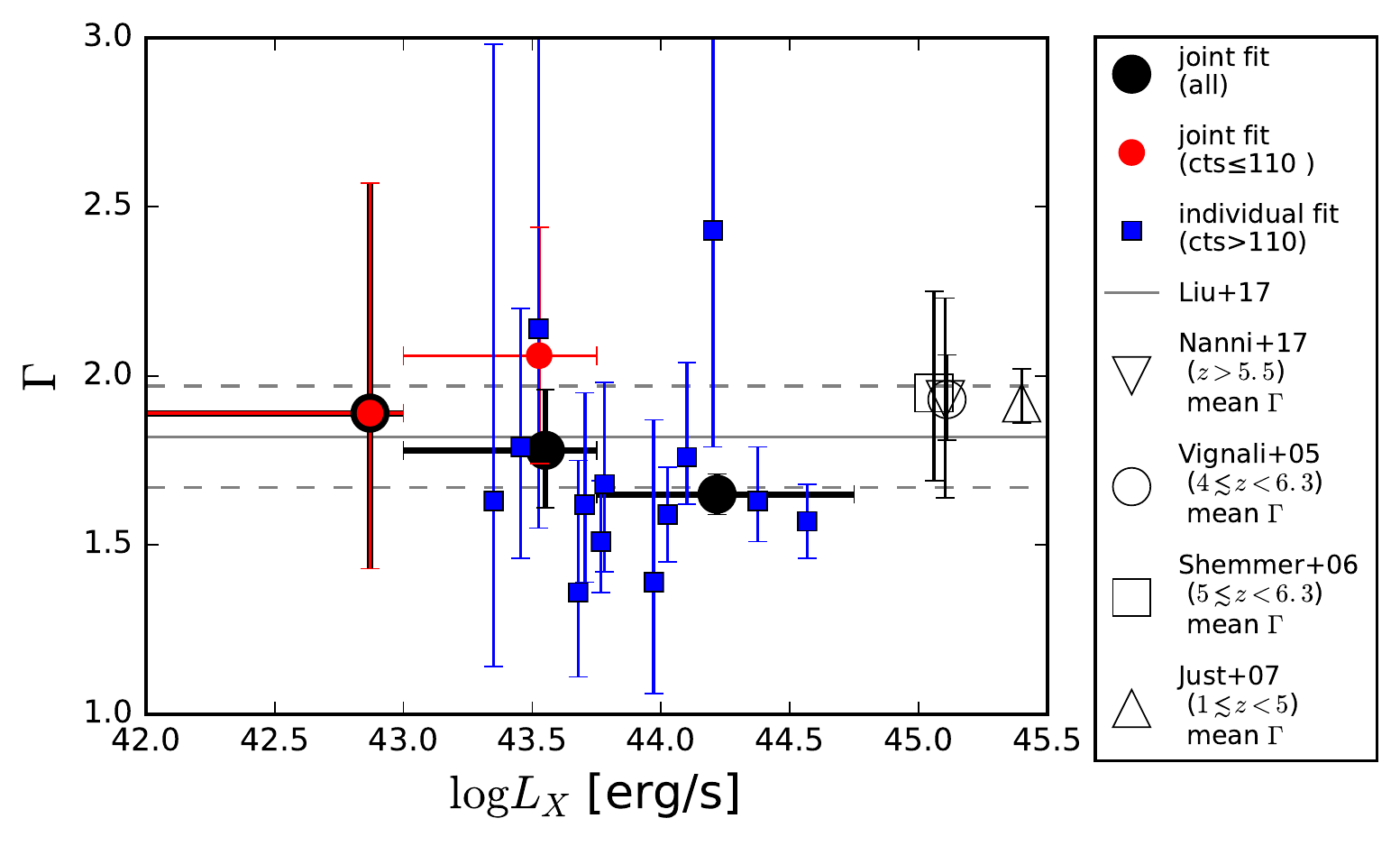}
	\caption{Best-fitting photon indexes for individual, bright sources with $>110$ net counts in the $0.5-7$ keV band (blue squares), and the average values derived from a joint spectral analysis including all sources (black points) and only those with $<110$ net counts (red points), in three luminosity bins. No bright source lies in the first luminosity bin, where therefore the black and red points are coincident. Grey solid and dashed lines mark the median $\Gamma$ derived by \citet{Liu17} in the 7 Ms CDF-S and its standard deviation, respectively. We also show the average $\Gamma$ of samples of luminous, optically-selected Type-I QSOs in different redshift intervals.}
	\label{gamma_Lx_fig}
\end{figure*}

Here we compare the photon indexes of $L\lesssim L_*$ AGN with results derived for much more luminous objects. We selected a subsample of \textit{bona fide} unobscured sources at $z=3-6$ from Table~\ref{sources}, requiring that the best-fitting column density is log$N_{\mathrm{H}}<23$ at the 68\% confidence level and a signal-to-noise ratio SNR$>3$ in the $0.5-7$ keV band. We performed a more detailed spectral analysis than that reported in \chapt{spec_analysis}, by allowing both the intrinsic column density and the photon index to vary. Fig.~\ref{gamma_Lx_fig} (blue squares) shows the resulting best-fitting values of $\Gamma$ as a function of intrinsic luminosity for sources with $>110$ net counts in the $0.5-7$ keV band. This threshold was chosen in order to obtain reasonable uncertainties on the best-fitting $\Gamma$ for individual AGN.

For the fainter sources, we derived the average photon indexes by performing a joint spectral analysis in luminosity bins chosen to include approximately the same number of sources. During the fitting, the column density was left free to vary for each spectrum, while the photon index parameters were linked among all sources, in order to derive an average value. Red circles in Fig.~\ref{gamma_Lx_fig} present the results of this joint analysis for sources with $<110$ net counts. Finally, we performed the joint spectral analysis in three luminosity bins, including both bright and faint objects (black points). Fit results are reported in Tab.~\ref{gamma_tab} for the individual fit of bright sources, and for the joint fit of faint and all (bright and faint) sources.

Fig.~\ref{gamma_Lx_fig} presents as continuous and dashed grey lines the median $\Gamma$ and and the standard deviation of the $\Gamma$ distribution ($\Gamma=1.82\pm0.15$), respectively, derived by \cite{Liu17} for sources with $>80$ net counts in the 7 Ms CDF-S. Our sources are on average broadly consistent with such values, justifying the use of $\Gamma=1.8$ for all sources in \chapt{spec_analysis}. The photon indexes of the luminous sources in our sample appear to be on average slightly flatter. This result is probably driven by the small number of luminous AGN considered here, which is due to the limited area covered by the deep surveys we used. Moreover, the best-fitting average $\Gamma$ in the highest-luminosity bin is quite strongly driven by the $1-2$ brightest sources, which have $\Gamma\approx1.6$ and alone provide the $38-59$\% of the net counts for the joint fitting in this luminosity bin. Neither of these two sources are detected in radio bands, and, in general, discarding the minority of sources (two out of twenty-three) detected in radio surveys does not change the results.

Our findings at $42<\mathrm{log}L_{\mathrm{X}}\lesssim44$ are compared in Fig.~\ref{gamma_Lx_fig} with the average $\Gamma$ of optically selected, luminous (log$L_{\mathrm{X}}>45$) Type-I QSOs at low-to-intermediate redshift ($z=1-5$, \citealt{Just07}) and $z=4-7$ \citep{Vignali05,Shemmer06,Nanni17}. Our results are consistent with both the median value of $\Gamma$ derived by \cite{Liu17} for similar luminosities at all redshifts and the average values of luminous QSOs up to $z\approx7$.

\begin{table}
	\small
	\caption{Best-fitting parameters of the spectral analysis described in \chapt{gamma}.}\label{gamma_tab}
	\begin{tabular}{cccccccc}
		\hline
		\multicolumn{1}{c}{ID} &
		\multicolumn{1}{c}{Sample} &
		\multicolumn{1}{c}{$\Gamma$} &
		\multicolumn{1}{c}{$N_{\mathrm{H}}$}&
		\multicolumn{1}{c}{$\mathrm{log}L_{\mathrm{X}}$}\\
		& size& &$(10^{22}\,\mathrm{cm^{-2}})$& $(\mathrm{erg\, s^{-1}})$ \\		
		\hline
		\multicolumn{5}{c}{Individual fit to bright sources}\\	
		\hline		
		29 & 1&$2.14_{-0.59}^{+0.95}$&$<7$  & 43.52   \\
		207  & 1& $1.76_{-0.14}^{+0.28}$& $<3$ &44.10   \\
		229  & 1& $1.62_{-0.23}^{+0.33}$& $<4$  &  43.70 \\
		293  & 1& $1.39_{-0.33}^{+0.48}$& $<13$ & 43.97  \\
		330  & 1& $2.43_{-0.64}^{+0.83}$& $<19$ & 44.20  \\
		404  & 1& $1.63_{-0.12}^{+0.16}$& $<1$ & 44.37  \\
		617  & 1& $1.79_{-0.33}^{+0.41}$& $<9$ & 43.45  \\	
		774  & 1& $1.57_{-0.11}^{+0.11}$& $3_{-2}^{+2}$ & 44.57  \\
		788  & 1& $1.59_{-0.14}^{+0.14}$& $<5$ & 44.03  \\
		811  & 1& $1.51_{-0.15}^{+0.18}$& $<2$ &  43.77 \\
		921  & 1& $1.68_{-0.26}^{+0.30}$& $<8$ & 43.78  \\
		926  & 1& $1.36_{-0.25}^{+0.39}$& $<11$ &43.68   \\
		965  & 1& $1.63_{-0.49}^{+1.35}$& $<44$ & 43.35  \\
		\hline
		\multicolumn{5}{c}{Joint fit of faint sources}\\	
		\hline
		-- & 3 & $1.89_{-0.46}^{+0.68}$ & -- & $42-43$\\
		-- & 6 & $2.06_{-0.32}^{+0.38}$ & -- & $43-43.75$\\
		\hline
		\multicolumn{5}{c}{Joint fit of all sources}\\	
		\hline
		-- & 3 & $1.89_{-0.46}^{+0.68}$ & -- & $42-43$\\
		-- & 11 & $1.78_{-0.17}^{+0.18}$ & -- & $43-43.75$\\
		-- & 9 & $1.65_{-0.06}^{+0.06}$ & -- & $43.75-44.75$\\
		
		\hline
	\end{tabular}\\
\end{table}

\section{The effect of neglecting the full probability distribution of spectral parameters}\label{nominal_parameters}
In \chapt{spec_analysis} we derived the full probability distribution of flux, and luminosity by convolving the probability distribution of count rate and redshift, and using the proper count-rate-to-flux conversion for each source. Such information has been used in \chapt{absfrac} to derive the dependence of the obscured AGN fraction on redshift, flux and luminosity. 

\begin{figure*} 
	\includegraphics[width=160mm,keepaspectratio]{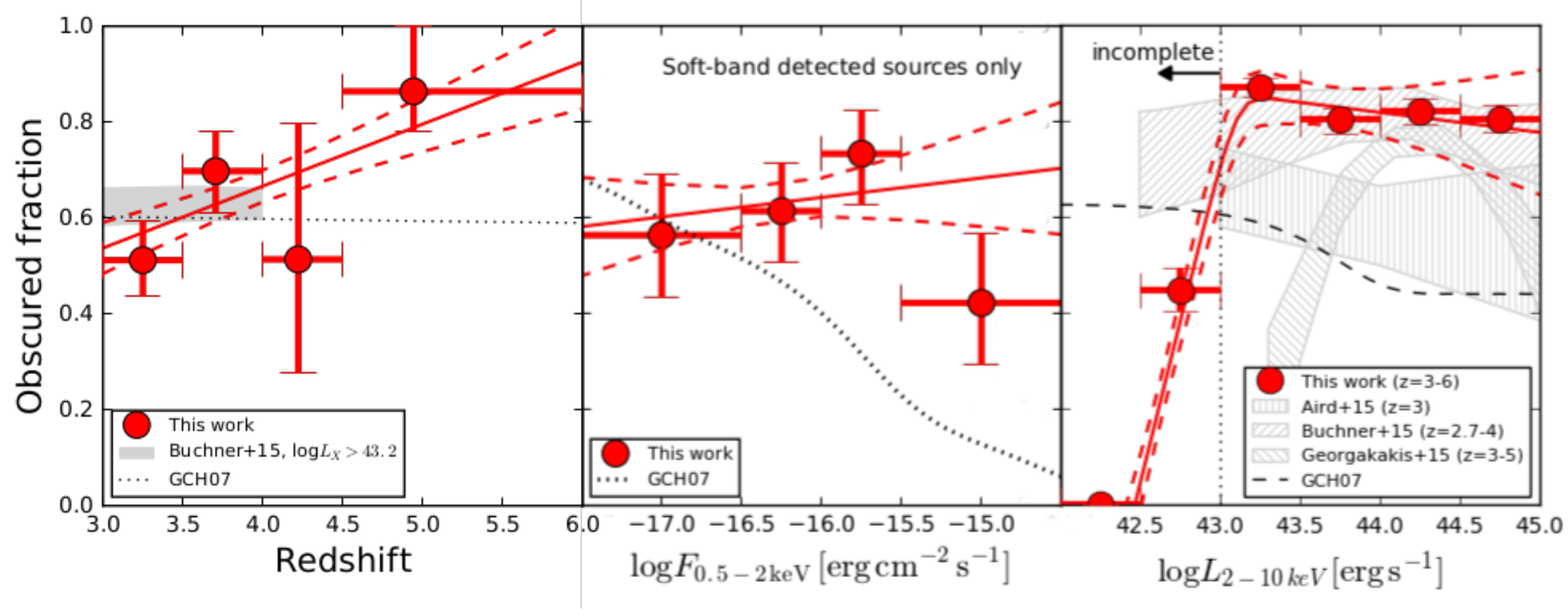}
	\caption{From left to right, obscured AGN fraction as a function or redshift, flux and luminosity, derived applying the procedure described in Appendix~\ref{nominal_parameters}. These figures, derived considering only the nominal values of redshift, count rate and column density for each X-ry source, are to be compared with the results in \chapt{absfrac}.}
	\label{Fobs_nom_par}
\end{figure*}

To check the importance of applying this careful but computationally expensive procedure, we repeat the analysis in \chapt{absfrac} by considering only the nominal value of parameters for each source, according to the following steps. 1) a single redshift ($z^{peak}$), corresponding to the spectroscopic redshift or the peak of PDF($z$), was assigned to each source. We did not include objects with no redshift information. 2) We adopted the values corresponding to the peak of the count-rate ($CR^{peak}$) and column-density ($N_{\mathrm{H}}^{peak}$) distributions. 3) Applying the proper conversion factor for $z^{peak}$ and $CR^{peak}$, we derived single values of flux ($F^{peak}$) and luminosity ($L^{peak}$).
Finally, following the procedures described in \chapt{absfrac}, we derive $F_{\mathrm{obsc}}$ as a function of redshift, flux and luminosity for our sample (Fig.~\ref{Fobs_nom_par}).
Larger fractions of obscured AGN are in general derived by this procedure, affecting in particular the trends with redshift and flux. This may be at least in part due to an overestimation of the best-fitting column density affecting intrinsically unobscured sources, due to statistical fluctuations in low-quality X-ray spectra discussed in \S~3.3 of \cite{Vito13}. When considering the full probability distribution of column density, this issue is greatly alleviated. We also checked that, considering the $F^{peak}$ and $L^{peak}$ values, the number counts resulted not to be strongly different, while the XLF tends to be flatter at low luminosities. We again ascribe this behavior to the tendency of overestimating the column densities, which has the effect of shifting the obscuration-corrected luminosities to larger values.

\end{document}